*Ulysses replaces dynamo with magnetolternator in the Sun and Sun-like stars*

# The Sun as a revolving-field magnetic alternator with a wobbling-core rotator from real data

M. Omerbashich[1*]

[1)] Geophysics Online, 3501 Jack Northrop Ave, Ste. 6172, Los Angeles CA 90250

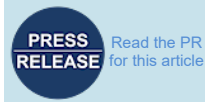

Read the PR for this article

Rather than as a star classically assumed to feature elusive dynamo or a proverbial engine and impulsively alternating polarity, the Sun reveals itself in the 385.8–2.439-nHz (1-month–13-years) band of polar ($\varphi_{Sun}$>|70°|) wind's decadal dynamics, dominated by the fast (>700 km s$^{-1}$) winds, as a globally completely vibrating revolving-field magnetic alternator at work at all times. Thus N̲orth–S̲outh separation of 1994–2008 Ulysses *in situ* <10nT polar-wind samplings reveals Gauss–Vaníček spectral signatures of an entirely ≥99%-significant, Sun-borne global incessant sharp *Alfvén resonance* (AR), $P_i=P_S/i$, $i=2...n$, $i\in\mathbb{Z} \wedge n\in\aleph$, accompanied by a symmetrical sharp antiresonance $P^-$. The *ideal Sun* (slow winds absent) AR imprints to the order u=136 into the fast winds nearly theoretically, with the northerly winds preferentially more so. The spectral peaks' fidelity is very high (≫12) to high (>12) and reaches $\Phi$>2·10$^3$, validating the signatures as a global dynamical process. The fast-wind spectra reveal upward drifting low-frequency trends due to a rigid core and undertones due to a core offset away from the apex. While the consequent core wobble with a 2.2±0.1-yr return period is the AR trigger, the core offset causes northerly preferentiality of Sun magnetism. Multiple total (band-wide) spectral symmetries of solar activity represented by historical solar-cycle lengths and sunspot and calcium numbers expose the solar alternator and core wobble as the moderators of sunspots, nanoflares, and coronal mass ejections that resemble machinery sparking. The *real Sun* (slow winds inclusive) AR resolves to n=100+ and is governed by the $P_S$=~11-yr Schwabe global damping (equilibrium) mode northside, its ~10-yr degeneration equatorially, and ~9-yr southside. The Sun is a typical ~3-dB-attenuated ring system, akin to rotating machinery with a wobbling rotator (core), featuring differentially revolving and contrarily (out-of-phase-) vibrating conveyor belts and layers, as well as a continuous global spectrum with patterns complete in both parities and the >81.3 nHz(S) and 55.6 nHz(N) resolution in lowermost frequencies (≲2 μHz in most modes). The global decadal vibration resonantly (quasi-periodically) flips the core, thus alternating the magnetic polarity of our host star. Unlike a resonating motor restrained from separating its casing, the cageless Sun lacks a stator and vibrates freely, resulting in all-spin and mass release (fast solar winds) in an axial shake-off beyond L1 at discrete wave modes generated highly coherently by the whole Sun. Thus, the northerly and southerly antiresonance tailing harmonic $P^-_{17}$ is the well-known $P_{Rg}$=154-day (or $P_S/3/3/3$ to ±1‰) Rieger period from which the wind's folded *Rieger resonance* (RR) sprouts, governing solar-system (including planetary) dynamics and space weather. AR and its causes were verified against remote data and the experiment, thus instantly replacing the dynamo with a magnetoalternator and advancing basic knowledge on the >100 billion trillions of solar-type stars. Shannon's theory-based Gauss-Vanicek spectral analysis revolutionizes astrophysics and space science by rigorously simulating fleet formations from a single spacecraft and physics by computing nonlinear global dynamics directly (rendering spherical approximation obsolete).

*Key words* — Sun global vibration; Sun engine; Sun core; stellar dynamo; standard stellar models; solar wind; space weather; Rieger period.

**HIGHLIGHTS**
- The first complete recovery of the Sun's global vibration (resonance & antiresonance), thus succeeding where all others had failed repeatedly
- Least-squares spectra of Ulysses polar magnetometer data resemble rotating-machinery operation well-known from mechanical engineering
- Instead of a simplistic dynamo, the Sun is a revolving-field magnetic alternator with machinery-inherent sparking (nanoflares, CMEs, sunspots)
- The core is offset away from the apex (towards the south pole), causing southern interferences & global preferentiality for northern magnetism
- The offset core naturally wobbles, forcing global spins/resonances that flip it (alternate polarity) at the damping equilibrium every ~11 yr
- The Sun exhausts its excess mass globally-resonantly as the solar wind in an axial shake-off and into the heliosphere coherently beyond L1
- The Solar system-permeating Rieger period deciphered as solar in origin decisively (twice) from fast (>700 km s$^{-1}$) and mixed winds
- Now completely known global sun vibration improves fundamental knowledge/standard stellar models of >100 billion trillion solar-type stars
- Accurately computed vibration of the solar wind enables macroscopic space weather forecasting and solar events prediction
- First application of Shannon theory-based, rigorous Gauss-Vaniček Spectral Analysis (GVSA) by least squares in global heliophysics
- GVSA revolutionizes space physics by rigorously simulating multiple spacecraft or fleet formations from a single spacecraft
- GVSA revolutionizes physics by computing nonlinear global dynamics directly (renders spherical approximation obsolete)

1. INTRODUCTION

The Sun is a magnetic star commonly believed to owe its magnetism to *dynamo* — a process occurring in the deep interior, where kinetic energy (of core motion, mostly primordial rotation) gets converted into electric energy that naturally gives rise to magnetic fields maintained by turbulence and other complex motions (Solanki et al., 2006). The mechanism for solar magnetism could also be convectional due to plasma/gas flows — as observed in brown dwarfs that lack a core but still exhibit the same magnetic patterns as our Sun's (Route, 2016) and hemispherical due to the Sun's characteristic latitudinal variations — as manifested in the activity of *sunspots* ("dark" or relatively dimmer and less magnetically active surface regions) occasionally shutting down per hemisphere (Grote & Busse, 2000). Despite our proximity to the Sun, the nature of its dynamo remains elusive (Brown, 2011).

Indeed, one of the main results from Ulysses as the only Space mission that flew above the Sun's polar ($\varphi_{Sun}>|70°|$) regions has revealed latitudinal differentiation of our star's magnetism, discovering significant variations in-between the poles as well as from that of the rest of the Sun, which is dominated equatorially in a "streamer belt" regime. Ulysses also made the first *in situ* observation of solar magnetic polarity reversals — occurring over several months of maximum magnetic activity (electromagnetic radiation and sunspots generation), or once every ~11 yr, which confirmed earlier such indications by the Wilcox Solar Observatory (WSO) (Jones & Balogh, 2003). This quasiperiodicity agrees with the average *Schwabe cycle*, $P_S\in[9\ yr, 13\ yr]$ (Schwabe, 1844), commonly noted since its discovery as a quasiperiodic variation in the number of sunspots. Given our star's size and mass, it is reasonable to assume that this is no coincidence and that the Ulysses result is extendable to mean that the Sun behaves as a magnetic alternator engine, known to normally both vibrate and resonate, i.e., vibrate additionally after its fundamental mode of vibration matched that of another physical system or its subsystem. Specifically, if due to physical coupling with a subsystem, the mechanical resonance (hereafter: resonance) is triggered internally (self-resonating systems), and when it arises in decoupled systems and subsystems, externally (e.g., orbitally).

The father of the widely used magnetohydrodynamics (MHD) theory (Alfvén, 1942) held a view that the Schwabe cycle must be the global resonance period of existing lines of force in the Sun interior (Alfvén, 1943). If this view is correct, the dynamo is a picture too simplistic to hold for the Sun globally. Thus, the *Alfvén waves* are a type of compressional magneto-acoustic waves in the Sun (Campos, 1977), which propagate along magnetic lines of force with a velocity proportional to the magnetic field and can become transverse or perpendicular to the wave motion when they are called kinetic (Alfvén) waves, cf., Alfvén (1942, 1948). The concept of such *Alfvén resonance* (AR) has been opposed, e.g., by Bellan (1994), as well as defended, e.g., by Goedbloed and Lifschitz (1995). While Bellan (1996) found from theoretical considerations that AR is a feature of ideal MHD only and therefore cannot arise in reality, Grant et al. (2018) deduced from observations of a sunspot that, under certain atmospheric conditions, magnetic field-lines flapping, i.e., the Alfvén waves, could form resonantly driven shocks and dissipate into thermal energy. Modeling by Srivastava et al. (2017) suggested that observed high-frequency (~12–42 mHz) torsional oscillations in the quiet Sun are torsional Alfvén waves that transfer ~$10^3$ W·m$^{-2}$ energy into the overlying corona, sufficient to heat it and facilitate the creation of the *solar wind* — the released magnetized plasma (atmosphere's thermally ionized gas; mostly $H^+$–$He^{2+}$ charged particles).

In addition, a linear scaling law has been observed independently from Ulysses mission data, spanning more than two decades and holding across a wide range of scales extending from a few minutes up to 1-day and longer (inertial) scales (Sorriso-Valvo et al., 2007), indicating that the Sun drives solar-wind turbulence itself (Bruno & Carbone, 2013). The unexpected existence of the scaling law in anisotropic, weakly compressible, and inhomogeneous turbulence still needs to be fully understood (Sorriso-Valvo et al., 2007). This linear scaling is the first large-scale evidence that solar-wind turbulence could be describable using the MHD theory.

It is a well-established fact that the Sun and other stars are vibrating bodies (Deubner & Gough, 1984) whose vibrations, magnetism, polarity, and AC (alternating current) propagate via the solar wind into the heliosphere, where the wind's magnetization is then called the *interplanetary magnetic field* (IMF). The Sun's and heliosphere's vibrations occupy various ranges: 4–15-minute short-period and 2-hour–12-day intermediate long-period bands (Thomson et al., 1995), the ~7–30-day long-period band dominated by the Sun's ~27-day surface mean rotational phase (Choi & Lee, 2019), the ~30-day–1-yr long-period band dominated by the widely reported $P_{Rg}$ = ~154-day *Rieger period* (Rieger et al., 1984) and its ⅚$P_{Rg}$, ⅔$P_{Rg}$, ½$P_{Rg}$, ⅓$P_{Rg}$, ⅕$P_{Rg}$ harmonics, i.e., ~128, ~102, ~78, ~51, ~31-days periods referred to as *Rieger-type periodicities* (Dimitropoulou et al., 2008), 1–2-yr intermediate very-long-period ranges (Forgacs-Dajka & Borkovits, 2007), as well as occasionally reported longer-period bands ranging from ~1–11 yr, e.g., Vecchio and Carbone (2009) and Deng et al. (2014).

The first known attempt at recovering the Sun global field-line resonance, as driven by $P_S$ and first proposed by Alfvén (1943), was by Stenflo and Vogel (1986) and Knaack and Stenflo (2005), from the Sun data and using spherical harmonic decomposition modeling with Lomb-Scargle and wavelets spectral techniques. However, their result was dubious as built on inapt techniques applied to regions of deviating turbulence, e.g., Bruno and Carbone (2013), and represented by data of questionable quality so that their alleged recovery turned out to be sparse (in one parity only and without any reasonably discernable patterns) and coarse (both inaccurate and imprecise). In addition, there have been reports of solar-wind resonances, albeit in shorter-period ranges and mostly centered on the Sun rotational frequency, e.g., by Singh and Badruddin (2019). One prominent case of a solar wind's global systematic dynamic is the *Rieger resonance* (RR), encompassing a train of the mentioned Rieger-type harmonics driven by P$_{Rg}$ and for which Bai and Cliver (1990) suggested could be simulated with a damped periodically forced nonlinear oscillator that exhibits both periodic and chaotic behavior. Rieger-type periodicities could be explainable by Rossby-type waves or another effect (Knaack & Stenflo, 2005). RR was reported previously in the IMF, including Earth vicinity (Cane et al., 1998), and in different ranges depending on data, location, epoch, and methodology, as 155–160 days, 160–165 days, 175–188 days, and 180–190 days (Gurgenashvili et al., 2017). Thus, while RR occurs in various ranges, those share the 30–180-day band. Historically, the Rieger period decreased until the middle of the last century and then began to increase again towards the end of the century, opposite to the activity magnitude trend (Zaqarashvili et al., 2010). Likewise, Rieger-type periodicities correlate with solar cycle strength and are shorter during solar cycles with higher magnetic activity (Gurgenashvili et al., 2016). This situation implies that Rieger resonance originates in the Sun engine. However, possible upstream waves in the solar wind from termination shock on the order of a few days and longer were implied from Voyager 2 mission samplings of the solar wind on entry into interstellar space, e.g., (Li et al., 2008), allowing for the classical explanation according to which a mechanical resonance can arise due to waves encountering physical obstacles along propagation paths.

*⁾ Correspondence to: omerbashich@geophysics.online, hm@royalfamily.ba.

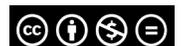

Unlike spectra of spheroidal p- (pressure-force-; compression-restored-) and g-mode (gravity-force-; buoyancy-restored-) vibrations with >1 h periods and extracted in the past, geometrical r-mode (Rossby-waves-like; Coriolis force-restored) months-long periods and toroidal R-mode (hypothetically cavity-confined; electromagnetic-force-restored-) resonances (Dzhalilov et al., 2002) remain intangible and rarely tackled (Knaack & Stenflo, 2005). Crude estimates indicate that R-mode periods are in the range of years and are inexplicable by dynamo theory (Stenflo & Vogel, 1986). The latter two types of long-periodic vibration occupy two adjacent portions of the subrotational frequencies (~30-day to ~11-yr periods), i.e., the entire band of interest in the present study. Therefore, I consider them together and so conjointly term as the a-mode (globally-triggered-and-restored-) resonant vibrations or AR for Alfvén Resonance, with periods $^aP = {}^rP \cup {}^RP$, where AR in the present study includes the accompanying antiresonance unless stated otherwise. In stars with a uniform rotation, very-low subrotational frequency vibrations (like the a-mode vibration) are likely candidates for (global) resonance with the orbital motion (Papaloizou & Pringle, 1978), which due to our Sun's non-uniform rotation excludes external triggering of global resonances. Furthermore, in non-uniformly rotating stars like our Sun, we can expect to see a continuous spectrum of modes that can undergo amplification and thereby lead to enhanced dissipation (*ibid.*). Specifically, since the solar wind is a physical system characterized by multi-scale evolution (Verscharen et al., 2019), global solar wind data should feature a continuous spectrum of modes.

Spacings and orientations of current sheets in the solar wind reveal that the magnetic structure of the heliosphere is a network of braided magnetic-flux tubes of unknown origin (Borovsky, 2018). These hypothetical tubes could arise as propagating modes experience resonances that generate coherent structures in the solar wind, allowing for the solar origin of the convected component of interplanetary MHD turbulence (Bruno & Carbone, 2013). Field-line resonance was invoked previously as the mechanism behind two still unresolved problems of large-scale dynamics: solar abundance (Asplund et al., 2009) and the million-degree corona (Davila, 1987). While the global overheating of the corona and the peculiar intrinsic acceleration of the solar wind to distances of ~10$R_\odot$ or ~0.05 AU likely share the same (unknown) underlying mechanism (Grail et al., 1996), the internal heating of the solar wind on release is due to small-scale ion-cyclotron resonance (Kasper et al., 2013) and even 100–500-s Alfvénic waves are strong enough to power the solar wind (de Pontieu et al., 2007). Completing the knowledge of the Sun on a global scale would have very large implications for our understanding of stars and galaxies in general; for example, such a breakthrough would cast light on how stars lose mass and angular momentum to stellar winds, while if measured, r-modes will become sensitive new probes of stellar physics and structure (Wolff & Blizard, 1986).

The present study then aims at extracting a signature of the Sun's highest-power (monthly-to-decadal) global periodic information as imprinted in the nearby solar wind using an apt methodology to analyze gapped data. Note here the use of the term *vibrations* for dynamical and *oscillations* for kinematical ("without mass") macroscopic considerations. Namely, referring to vibrations indicates focusing on dynamical considerations, i.e., mechanical waves physically discernable as they propagate through (all forms of) matter, which is most useful for practical considerations. Referring to oscillations requires mostly abstract thinking, primarily in terms of kinematics (like orbital), but also other geometric relations and it is useful primarily for theoretical considerations.

## 2. Data & Methodology

### 2.1 *Data*

To extract the Sun periodicities in the 1–13-years band (of magnetic polarity reversals), I first compute frequency spectra of the only type of globally emitted solar wind whose origin is undisputed — the fast (>700 km s$^{-1}$) wind, emitted mainly from polar regions, as sources of the slow wind (~400 km s$^{-1}$) and their locations remain a matter of debate (Brooks et al., 2015). Accordingly, other sources of the slow solar wind (IMF) are ignored, including open field lines from the quiet region. Thus, I regard the fast wind as ideal (matching the original physical conditions at and for the release as closely as possible), and therefore first examine the polar wind since the fast wind dominates polar regions both spatially and temporally. The goal here is to put the often-heard syntagm on the Sun internal engine to the test by examining if how the Sun emits the wind corresponds in any way to a shake-off typical of many engines trying to rid themselves of their casing while experiencing a mechanical resonance. In engines, the resonance is due to structure irregularities and conditions or other factors that introduce jolts into vibration. In my inspection, I tacitly assume that variations in the internal structure and composition of the Sun could produce such mechanical resonances and that the (equatorially dominated) rest of the Sun mainly hosts the sources of the slow wind.

As mentioned, I carry the examination first in the little-explored portion of <~11-yr solar vibrations: the 1–13-years very-long-period (31.71–2.439-nHz) range of the reversals, or in the ~0.01–0.13-ZeV (~$1.6 \cdot 10^7$–$2 \cdot 10^8$ erg) band of solar-wind extreme base energies. Namely, if a systematic physical process could be discerned spectrally from magnetic (MAG-data) variations within the highest system energies — contrary to smooth spectra due to turbulence — then the process is most likely caused by the Sun's internal engine (Gough, 1995). Furthermore, if the mechanism can be shown resonant in the band of interest, as established previously for other bands (see the Introduction), our star behaves like and maintains the state of a magnetic alternator engine. In that case, we can expect the Sun to resonate entirely predictably, which would be usable for probing the interior and modeling heliosphere dynamics, including space weather. Secondly, I investigate if the heliosphere maintains solar a-mode vibrations in the band of interest and as transpired via the solar wind, which, if true, would allow learning more about the inner workings of the Sun and the energy budgets of the heliosphere (Brooks et al., 2015).

Successfully extracting information about vibration could allow learning from modal analysis of resonances (Deubner & Gough, 1984) and to the degree that depends upon the regularity and strength of the resonant process of interest. In addition, any detection of a-modes would provide a substantial new probe of the interior rotation and, eventually, a constraint on convection theory (Wolff & Blizard, 1986). Note Fossat et al. (2017) claimed the absolute value for solar core rotation of ~one week and noted that current understanding and modeling in heliophysics cannot explain such a rapid rotation. Importantly, knowledge on the core from observations is sparse, while a lot is guessed or assumed based on modeling and theory.

To achieve the goal and extract (measure) the desired information in an approach modernized in terms of data and methodology, I use what arguably appear to be the four (for global studies) best data sets of solar and solar-wind magnetic (MAG) field data. Thus, to examine how well the heliosphere overall maintains solar-wind vibrations, I compare the spectra of the global fields (including the mean magnetic field (MMF) and polar fields (PF) from the WSO telescope (Scherrer et al., 1977) and the N-S-polar magnetic fields as reflected in the polar wind from the Ulysses mission) against the spectra of IMF at L1 from the WIND mission. Presumably, dominating in MMF are background/equatorial fields in a ~9:1 ratio to other magnetic features, including local fields, i.e., sunspots (Bose & Nagaraju, 2018), and reflected in the slow wind. Since the MMF and Ulysses data sets depict relatively slow vs. mostly fast winds at or close to the respective sources, successfully matching the spectra of those data against the WIND/L1 data spectra would mean that the heliosphere reflects solar vibrations in the band of interest at planetary distances.

As mentioned earlier and based on previous studies of Ulysses magnetometer measurements, while the Sun emits the fast wind from polar regions, the slow wind emitted from lower heliographic latitudes gets slowed down by rotation and thereby equatorially mixed to the streamer belt regime; see, e.g., Smith and Marsden (2003). Then, from the perspective of the Sun vibrating like a classical magnetically alternating motor, the polar (mainly fast) wind should reflect the Sun's inner workings more faithfully than the equatorially mixed (mainly slow) wind ever could. Importantly, given the southern region's instability under relatively higher turbulence and wandering fields and since the polar regions themselves emit the solar wind at significantly varying rates, resulting in significant N–S asymmetry of solar wind speeds irrespectively of solar activity (Tokumaru et al., 2015), I separate the polar data into northerly and southerly polar solar winds. Also, if the fast solar wind's resonation in the band of polarity reversals indeed permeates the heliosphere, then the Sun should be expelling the wind at those same resonance modes. In brief, after the separation of resonance from heliophysical background noise, the Sun parameters (vibration modes) and thus its structure, composition, and engine could all be tackled from a new perspective based on modal analyses. This new perspective could enable us to gain a more detailed or even complete (globally primarily) insight into our star than previously attainable with tools such as helioseismology, spectroscopy, cosmochemistry, or gravity modes. Finally, as mentioned earlier, such new fundamental knowledge of the Sun's inner workings would also improve our understanding of stars and stellar systems. Data used in the present study are in Fig. 1.

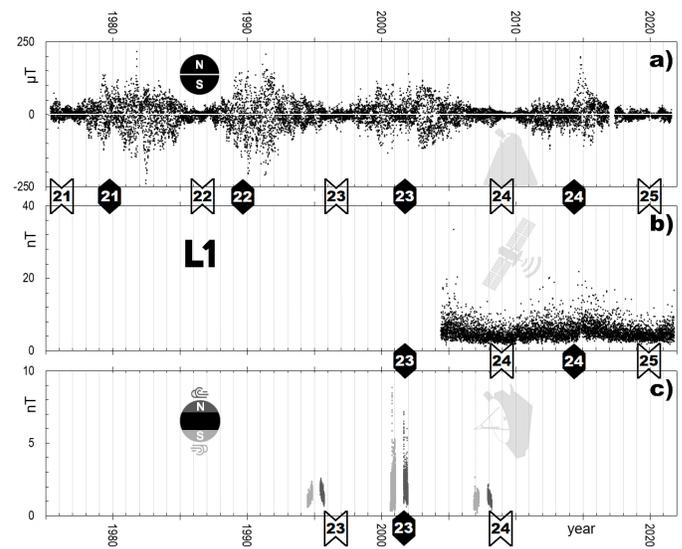

**Figure 1**. Plots of analyzed average strengths, B, of magnetic fields studied. Panel a: Daily averages of the B≲150 µT mean magnetic field (MMF) by the Wilcox Solar Observatory (WSO) telescope from 16 May 1975–03 August 2021. Panel b: Daily averages of the B≲50 nT, 2004–2021 interplanetary magnetic field (IMF) by WIND at the L1 Lagrangian point from 01 June 2004–04 November 2021. Panel c: hourly averages of the B≲10 nT IMF from Ulysses over the Sun's polar regions ($\varphi_{Sun}>|70°|$) in three flybys above the southern (light gray): 26 June–05 November 1994, 6 September 2000–16 January 2001, 17 November 2006–03 April 2007, and the northern polar region (dark gray): 19 June–29 September 1995, 31 August–10 December 2001, 30 November 2007–15 March 2008. As can be seen already in the time domain both at L1 (panel b) and over the Sun's polar regions (panel c), not only does IMF act in unison with solar cycles of magnetic activity's maxima and minima, but it also closely resembles usual year-to-year activity. IMF thus largely preserves yearslong systematic contents, prompting decade-scale spectral analyses of the data down to the solar cycle's Schwabe period, $P_S$= ~11 yr. IMF observations by WIND before assuming the mission target orbit at L1, i.e., while still in the lunar orbit from 01 January 1995–31 May 2004, were omitted from computations and the plot for clarity. Numbered arrows mark the times of respective solar cycles that, by convention, always start at a minimum activity (shown as white arrows) and develop into a maximum solar activity (as black arrows) around midway through the respective cycle. To compensate for a relatively lower data resolution in polar data vs. other datasets, high-resolution (hourly) averages of the Ulysses samplings (panel c), presumably richer in systematic information than daily averages, were used. Data were plotted co-temporally amongst the panels for comparison and are in the Supplement; see the Data sources statement.

51

2.2 *Methodology*

To compute the spectra, I use Shannon theory-based Gauss–Vaníček rigorous method of spectral analysis (GVSA) (Omerbashich, 2006) by Vaníček (1969, 1971), which represents spectral peaks against linear background noise levels. The spectral peaks are expressible in percentages of the respective peak's contribution to data variance (var%) or in decibels, dB (Pagiatakis, 1999). GVSA has many benefits and, in numerous ways and situations, outperforms the Fourier method as merely a special case of GVSA (Craymer, 1998); for example, when analyzing long gapped records, i.e., most natural-data records (Omerbashich, 2022, 2007, 2006; Press et al., 2007; Pagiatakis, 1999; Wells et al., 1985). By discarding unreliable data in the records, such as non-calibrated telescope observations, I also take advantage of this blindness to data gaps as a feature exclusive to the least-squares class of spectral analysis techniques. This property of GVSA becomes a crucial advantage in analyses of scarcely clustering intermittent observations like those of the solar wind above polar regions (which are intermittent because of the Ulysses mission objective to target polar regions as rarely observed, Fig. 1–c). Other GVSA properties of use in the present study include the enforcing of periods (Wells et al., 1985), a procedure performed in parallel with a spectrum computation where periods of choice get mathematically ignored as though nonexistent in the data, thus enabling us to decipher spectral content otherwise buried under noise, as well as underlying systematic processes in a physical system. Tests of GVSA, showing its superiority, have been performed, e.g., by Taylor and Hamilton (1972) and Omerbashich (2003).

GVSA comes integrated with a comprehensive statistical analysis toolbox in the form of a scientific software package LSSA for frequency or periodicity estimates in the strictly least-squares sense, unlike the more popular Lomb-Scargle approximation of GVSA, which no longer can be regarded as least-squares and which underperforms when analyzing noisy and complicated signals such as those of solar activity (Carbonell et al., 1992; Danilović et al., 2005). For instance, noise in turbulence data can create false-alarm peaks in the Lomb periodogram, making it difficult to judge if a given Lomb peak is significant (Zhou & Sornette, 2001). Fourier spectral analysis methods can also report false peaks when analyzing patched-up or data edited otherwise (Omerbashich, 2006), where, unlike in GVSA, editing is necessary for most types of natural data as gapped inherently. GVSA revolutionizes physical sciences by enabling direct computations of global nonlinear dynamics (Omerbashich, 2023a), rendering approximative approaches like spherical decomposition obsolete. As for GVSA limitations, those are not critical when studying global dynamics of closed physical systems like astrophysical bodies, including stars, planets, and moons. Remarkably and unlike in any other method, the minimum number of values GVSA can estimate a spectrum from is just three. This stunning feature stems from the normed fitting of data with trigonometric functions, which enables accurate descriptions of global dynamics even from data widely separated in space and time as though the analyzed data set were complete at a declared sampling rate of interest.

A GVSA spectrum, *s*, is obtained at a spectral resolution *r* (1000 spectral values/lines in the present study throughout), for *r* corresponding periods $T_j$ or frequencies $\omega_j$ and output with spectral magnitudes $M_j$, as:

$$s_j(T_j, M_j); \; j = 1, \ldots r \wedge j \in \mathbb{Z} \wedge r \in \aleph. \tag{1}$$

In its simplest form, i.e., when there is no *a priori* knowledge on data constituents such as datum offsets, linear trends, and instrumental drifts, a GVSA spectrum *s* is computed as (Omerbashich, 2003):

$$\boldsymbol{s}(\omega_j, M_j) = \frac{\boldsymbol{l}^\mathrm{T} \cdot \boldsymbol{p}(\omega_j)}{\boldsymbol{l}^\mathrm{T} \cdot \boldsymbol{l}}, \tag{2}$$

obtained after two orthogonal projections. First, of the vector of *b* observations, *l*, onto the manifold $Z(\boldsymbol{\Psi})$ spanned by different base functions (columns of **A** matrix) at a time instant *t*, $\boldsymbol{\Psi} = [\cos \omega t, \sin \omega t]$, to obtain the best fitting approximant $\boldsymbol{p} = \sum_{i=1}^{b} \hat{c}_i \boldsymbol{\Psi}_i$ to *l* such that the residuals $\hat{\boldsymbol{v}} = \boldsymbol{l} - \boldsymbol{p}$ are minimized in the least-squares sense for $\hat{\boldsymbol{c}} = (\boldsymbol{\Psi}^\mathrm{T} \mathbf{C}_l^{-1} \boldsymbol{\Psi})^{-1} \cdot \boldsymbol{\Psi}^\mathrm{T} \mathbf{C}_l^{-1} \boldsymbol{l}$. The second projection, of *p* onto *l*, enables us to obtain the spectral value, Eq. (2). Vectors $\boldsymbol{u}_j = \boldsymbol{\Psi}^\mathrm{T} \boldsymbol{\Psi}_{NK+1}$ and $\boldsymbol{v}_j = \boldsymbol{\Psi}^\mathrm{T} \boldsymbol{\Psi}_{NK+2}$, *j* = 1, 2…. NK$\in \aleph$, compose columns of the matrix $\mathbf{A}_{NK,NK} = \boldsymbol{\Psi}^\mathrm{T} \boldsymbol{\Psi}$. Note here that the vectors of known constituents compose matrix $\hat{\mathbf{A}}_{b,b} = \hat{\boldsymbol{\Psi}}^\mathrm{T} \hat{\boldsymbol{\Psi}}$, in which case the base functions that span the manifold $Z(\boldsymbol{\Psi})$ get expanded by known-constituent base functions, $\hat{\boldsymbol{\Psi}}$, to $\boldsymbol{\Psi} = [\hat{\boldsymbol{\Psi}}, \cos \omega t, \sin \omega t]$. For a detailed treatment of GVSA with known data constituents, see Wells et al. (1985). Subsequently, the method got simplified into non-rigorous (strictly non-least-squares) formats like the above-mentioned Lomb-Scargle technique created to lower the computational burden of the Vaníček method, but which no longer is an issue.

GVSA is rigorous because it estimates a statistical significance in var% for the desired level, say 95%, in a spectrum from a time series with *p* data values and *c* known constituents as $1-0.95^{2/(p-c-2)}$ (Steeves, 1981) (Wells et al., 1985), as well as imposes an additional constraint of fidelity or "realism" ($\Phi$) to determine the physical validity of each significant spectral peak individually, e.g., if together with other peaks it likely belongs to a dynamical process. As known from advanced statistics, fidelity is a general information measure based on the coordinate-independent cumulative distribution and critical yet previously neglected symmetry considerations (Kinkhabwala, 2013). In communications theory, fidelity measures how undesirable it is (according to some fidelity criterion we devise) to receive one piece of information as another piece is transmitted (Shannon, 1948). Fidelity is then defined in GVSA, i.e., in the theory of spectral analysis, as a measure of how undesirable it is for two frequencies to overlap at (occupy) the same frequency space of a sample. A fidelity value is the time interval (in units of the timescale) by which

52

the period of a significant spectral peak must increase or decrease to be π-phase-shiftable within the time-series length. Φ thus measures the unresolvedness between two consecutive significant spectral peaks (that cannot be π-phase-shifted). Two adjacent peaks are resolvable if their periods differ by more than the fidelity value of the former. This clustering tendency criterion reveals whether a spectral peak can share a systematic nature with another spectral peak, e.g., be part of a batch, an underlying dynamical process like resonance, antiresonance, reflection, overtone, or undertone. The spectral peaks that meet this criterion are in the LSSA software output listed amongst insignificant and the rest amongst significant (hereafter: *physically-statistically significant spectral peaks,* or just (fully) *significant peaks* for short). Subsequently, Omerbashich (2006) deduced empirically an additional criterion of stringency: that GVSA fidelity in prominently periodic time series (with more than just a few periodicities) to reasonable approximation satisfies a Φ>12 common criterion for the individually genuine significance of a systematic process and therefore most of its periodicities as well.

To prevent aliasing, the spectral band's lower end, 13-yr, was selected as not too close to $P_S$ and out of phase with the Ulysses orbital period of 6.2 yr. Thanks to the ability of GVSA to handle spectral/energy leakages methodologically, i.e., simply by slightly adjusting the band limits if needed, this band choice should also prevent any leakages for that band. Also, unlike in the Fourier class of spectral analysis methods, GVSA does not depend on the Nyquist frequency (Craymer, 1998) and so can fit trigonometric functions even in, i.e., extract significant periodicity from data that span intervals slightly (empirically: up to ~50%) shorter than twice the period of interest. The declared uncertainty on least-squares estimates of all periods extracted in the present study is ±5%.

3. RESULTS

Spectra of the Fig. 1 data were computed first in the 1–13-years (31.710–2.439-nHz or 100.00–7.69-cycles per century) band, or the ~1.6·10⁷ –2·10⁸ erg band of solar-wind extreme base energies, and plotted in five panels of Fig. 2, respectively to the panels of Fig. 1 whose panel c also corresponds to Fig. 2–d & e after data separation into the southerly and northerly polar wind, respectively. As seen, notable spectra-wide features in the revealed series of ordered frequency response functions are resonance (sharp peaks) and symmetrical antiresonance (sharp troughs) trains, a ~3-dB attenuation typical of controlled mechanically resonating structures like engines (Ewins, 1995), and a constantly high relative system energy in the lead mode and all the harmonics, of ~25 var%. These features are typical of a classical motor and reflect the previously little understood workings of the nevertheless often invoked Sun internal engine. The signature of a very long-period resonance of solar magnetic fields, extracted from the solar winds at or close to their sources, contains several additional fundamental properties of vibrating/resonating systems, which are addressed in terms of experimental validation of the present study's result later on. One property worth noting is the symmetry between the resonance and antiresonance, which reveals a real engine at work.

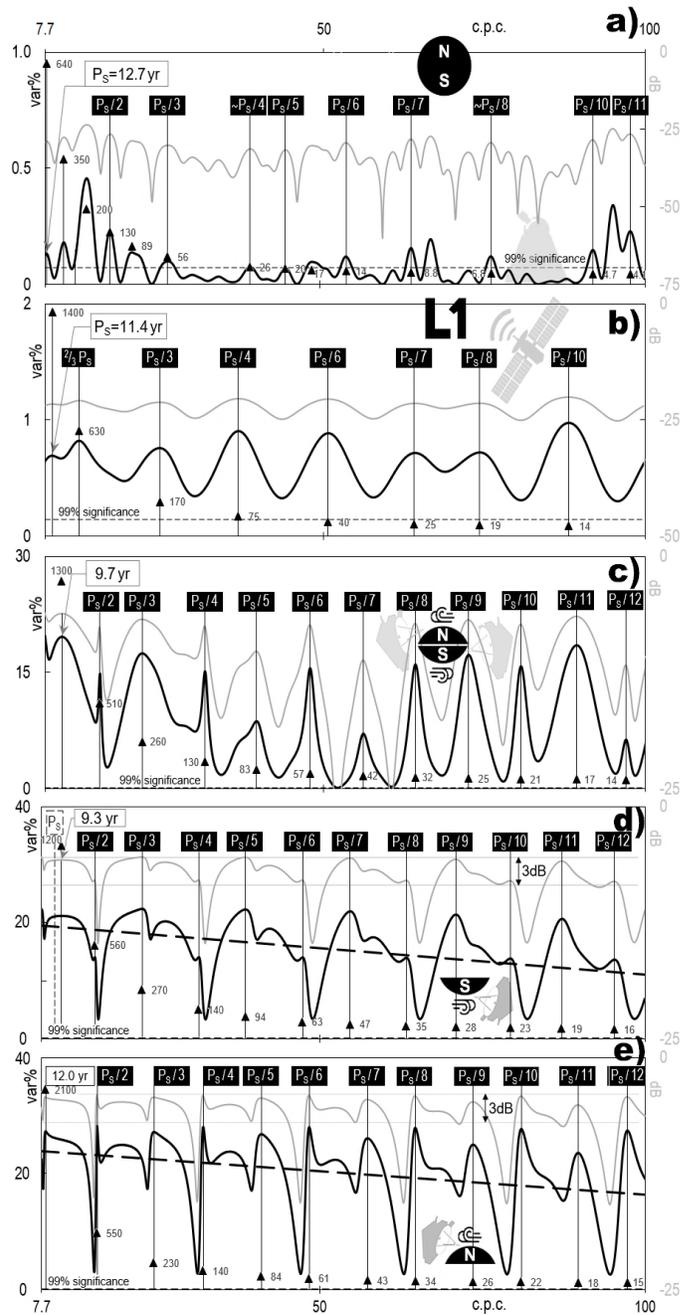

**Figure 2**. GVSA spectra of magnetic fields' strength variations from Fig. 1 reveal the $P_S$=~11-yr (Schwabe) global mode-driven *a-mode resonance* (Alfvén resonance, or AR) in the 100.0–7.69 cycles per century (c.p.c.), i.e., 1–13-years or 365–4745 days (31.710–2.439-nHz) band. Panel a — from WSO-derived daily values of the MMF; panel b — from the WIND spacecraft's quasi-stationary IMF samplings at L1; panel c — from Ulysses flybys over the Sun's polar regions dominated by the fast solar winds (>700 km s⁻¹); panel d — separately from Ulysses magnetometer data over the northern; and panel e — separately from Ulysses magnetometer data over the southern polar region. Dashed lines on panels d & e are linear trends. Numbered triangles mark the fidelity values on respective spectral peaks along a third arbitrary vertical axis (not shown). Fidelity on extracted spectral periodicities stayed well above 12, indicating a genuinely dynamic process (Omerbashich, 2006). Note here that the overall fidelity increase has followed the N–S data separation; also note an increase in system energy occupied by AR, from ~10 var%, panel c, to ~20 var%, panel d, and ~25 var%, panel e. In addition, the mutual resemblance of variance and power spectra, which virtually coincide in shape, indicate that the spectrally described process is happening over far wider bands, which calls for widening spectral bands and re-computing the spectra. Fig. 3 depicts the results from that re-computation in the solar full long-period band of (rotational-to-Schwabe) frequencies. Note here that modes in the present study take order per separate extractions, as shown in Table 1 (marked as m, $m \in \aleph$), Table 2 (as k, $\in \aleph$), Table 3 (as n, $n \in \aleph$), and Table 4 (as w, $\mathbf{w} \in \aleph$), Table 5 (as u, $u \in \aleph$), Table 6 (as q, $q \in \aleph$), rather than imagining a symmetrical (parity-dependent) Sun in crude modeling approaches such as spherical decomposition. Compare to experimental results, Fig. 7.

53

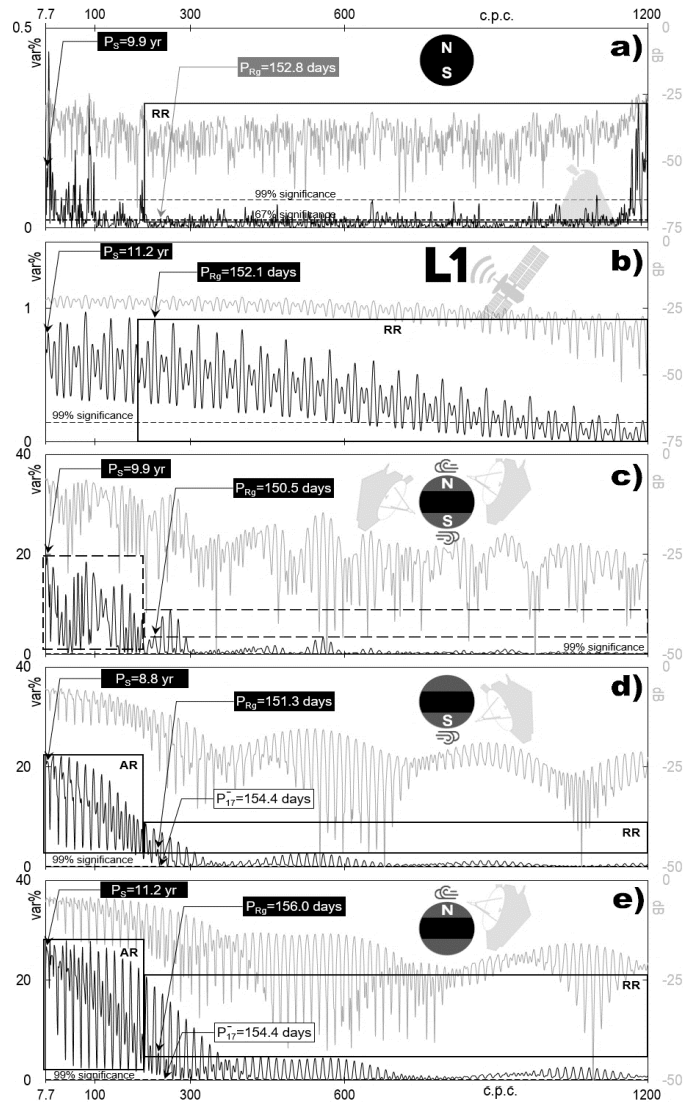

The lead resonance period of the north wind in this condensed band was $P_S$=12.0 yr, panel e. This northerly data spectrum somewhat outperformed the southerly, panel d, due to relatively higher levels of southern turbulence and the temporal proximity of solar cycles' activity extremes to the times of northerly samples, Fig.1. Still, all periods for both polar regions are ≥99%-significant and in terms of resonance and (couplings-triggered) antiresonance trains, were resolved. The difference between the 11- and 9-yr periods itself, as a consequence of polar asymmetry in the solar magnetic activity, can be expected to trigger a theoretical (perfectly integer-ordered) mechanical resonance in the wind's waving, here recovered in both polarities, as shown in Fig. 2 over the 1–13-years (31.71–2.439-nHz) condensed spectral band. Recall that, as opposed to turbulence that exhibits smooth spectra, such a perfect resonance can only be dictated by the Sun internally, e.g., Gough (1995), here especially so since characterized by sharp peaks and troughs, and very high fidelity — all indications of a deterministic driving process. This result agrees even with Knaack and Stenflo (2005), who claimed that ~1-yr and longer Sun periodicities align to solar cycles and are unlikely to arise randomly. Note again that the Ulysses hourly data contained presumably more complete systematic information and have outperformed the WIND and WSO data despite the latter two records featuring relatively higher completeness rates and longer spans. The significant variation between the spectral magnitudes of resonance peaks in panels e vs. d of Fig. 2 reflects the N–S asymmetry in solar-wind/IMF, found affecting wind speeds and confirmed independently of Ulysses data, e.g., by Tokumaru et al. (2015).

For a more detailed insight into the AR and RR, and since AR is driven presumably by the Schwabe cycle (Alfvén, 1943), the spectra of data from Fig. 1 were re-computed in (zoomed out to) the 1-month–13-years (385.802–2.439-nHz or 1200.0–7.69-cycles per century) full band of subrotational frequencies, or the 0.01–1.60 ZeV (~$1.6 \cdot 10^7$ –$2.5 \cdot 10^9$ erg) full band of base energies. This approach is justified further by the results plotted in Fig. 2–d & e matching experimental results already, Fig. 7–a. The results are in Tables 1–3 and on five panels of Fig. 3, again respective to the panels of Fig. 1 and the five panels of Fig. 2. As in Fig. 2, MMF in Fig. 3, panel a, and the N–S combined data, panel c, reveal that the spectra tend to flatness due to turbulence (Gough, 1995).

However, despite a heavy suppression under turbulence, equatorial instability never attenuates the resonance entirely, as seen in the partial and 0.1-var% faint extraction of RR at ≥67% significance, panel a. (Here, well-known physical processes are regarded as significant if extracted with at least 67% significance.) Furthermore, any radical disparity amongst the $P_S$ global resonance modes, Fig. 2, is now gone, Fig. 3, and those periods are now more congruent, as discussed later. Note an improvement in $P_S$ and $P_{Rg}$ going from panels a–e here and from Fig. 2–e to Fig. 3–e, as well as in the spectral resolution of the Rieger resonance (both extraction-wise and magnitude-wise), seen as a relative increase in the area of the left-hand frame with data separation. Also seen is a magnitude-of-order increase in spectral magnitudes on panels c–e compared to a & b, i.e., in system-energy bands occupied by AR, meaning the $P_{Rg}$ from panel a (value grayed out), while ≥67%-significant, is still part of background noise and thus not emitted equatorially. Panel b

**Figure 3**. GVSA spectra of the data from Fig. 1, in the 1200.0–7.69 c.p.c. (1-month–13-years or 30–4745 days or 385.802–2.439-nHz) band of Sun subrotational frequencies, revealing the entire global AR, i.e., a continuous spectrum of modes. Panel a — AR absence from the MMF; panel b — AR detection in the IMF at L1 from WIND; panel c — AR extraction from the northerly- and southerly-polar-wind data combined; panel d — AR extraction from the northerly-polar-wind data; panel e — AR extraction from the southerly-polar-wind data. The left-hand frame, labeled as AR (note the AR proper includes within its 30–180-day subband the RR as well), partly seen in Fig. 2, delimits the main (0.5–13-yr) frequency subband of highest resonantly magnified energies in the Sun vibrations, as seen from intensive and systematic peak-splitting due to coupling but not turbulence anisotropy, which thus turns out to have been moderated by AR. The right-hand frame delimits the 30–180 days band of the Rieger resonance (RR) controlled by the $P_{Rg}$=~154-day Rieger period reported widely in solar indices across the solar system. This band turned out to be a subband of the AR global resonance, Fig. 2. $P_{Rg}$ estimates recover best from the IMF at L1 (WIND), panel b, and northern-polar IMF, panel e, reflecting the known fact that it primarily is the northerly wind that reaches the IMF at L1. Thanks to GVSA spectral magnitudes in var% being directly proportionate to system energy, the resonances are extracted completely (with a continuous spectrum of modes and complete with antiresonances and patterns in both parities) so that frames scale also magnitude-wise, i.e., as energy bands. Fidelity (not shown) satisfied the Φ>12 threshold for periods of up to ~1 yr, implying that the AR spectrum represents a genuine dynamic process, here up to the ~annual cycle(s). This verification exposed RR as a non-self-sustained physical process but a carrier of the Sun's global resonance mode $P_S$ and its (by couplings superimposed) antiresonance harmonics. Thus, the real (gravitationally unperturbed) $P_{Rg}$ is $P^-_{17}$= 154.4-day from both polar regions, panels d & e and Table 2. Note that an antiresonance mode $P^-$ always precedes AR, which for check also holds for $P_o$ when the band's lower limit gets further lowered, say to 5 c.p.y. (not shown). The period labels highlighted in gray mark a (67–99%]-confidence. Frame labels are omitted for clarity where necessary, and the frames are dashed in instances in which a resonance driver is missing. Numerical values of the plotted AR modes are in Table 1, the antiresonance modes in Table 2. Notably, the Sun vibrates incessantly albeit with not just one global mode as Alfvén (1943) surmised, but three global modes of vibration, i.e., the original $P_S$ northwards plus its two supposed degenerations: a ~10-yr equatorially that then transitions into a ~9-yr degeneration southwardly, panels e & b, c & a, d, respectively. This gradual damping of the global vibration already in its highest dynamic energies indicates an uneven mass distribution in the Sun's deep interior, clustering southward. Compare to experimental results, Fig. 7.

| m | WSO MMF $P_m$ | WSO MMF $\Phi_m$ | WIND IMF/L1 $P_m$ | WIND IMF/L1 $\Phi_m$ | Ulysses NS $P_m$ | Ulysses NS $\Phi_m$ | Ulysses S $P_m$ | Ulysses S $\Phi_m$ | Ulysses N $P_m$ | Ulysses N $\Phi_m$ | m | WSO MMF $P_m$ | WSO MMF $\Phi_m$ | WIND IMF/L1 $P_m$ | WIND IMF/L1 $\Phi_m$ | Ulysses NS $P_m$ | Ulysses NS $\Phi_m$ | Ulysses S $P_m$ | Ulysses S $\Phi_m$ | Ulysses N $P_m$ | Ulysses N $\Phi_m$ |
|---|---|---|---|---|---|---|---|---|---|---|---|---|---|---|---|---|---|---|---|---|---|
| 0 | 3609.3 | 390 | 4100.0 | 1300 | 3609.3 | 1300 | 3223.6 | 1100 | 4100.0 | 1800 | 66 | | | 44.1 | 0.2 | 55.2 | 0.3 | 59.4 | 0.4 | 64.7 | 0.5 |
| 1 | 2655.9 | 210 | 2912.3 | 670 | 2258.2 | 510 | 1558.2 | 260 | 2258.2 | 550 | 67 | | | 43.4 | 0.2 | 53.9 | 0.3 | 57.9 | 0.4 | 62.9 | 0.4 |
| 2 | 2100.9 | 130 | 1481.6 | 170 | 1643.1 | 270 | 1349.1 | 200 | 1844.0 | 370 | 68 | | | 42.7 | 0.1 | 53.0 | 0.3 | 56.4 | 0.3 | 61.3 | 0.4 |
| 3 | 1737.7 | 89 | 993.6 | 78 | 1144.3 | 130 | 932.1 | 93 | 1481.6 | 240 | 69 | | | 42.1 | 0.1 | 52.4 | 0.3 | 55.0 | 0.3 | 59.7 | 0.4 |
| 4 | 1349.1 | 54 | 712.1 | 40 | 904.2 | 82 | 853.0 | 78 | 1102.5 | 130 | 70 | | | 41.5 | 0.1 | 51.7 | 0.3 | 53.6 | 0.3 | 58.2 | 0.4 |
| 5 | 932.1 | 26 | 565.3 | 25 | 766.3 | 59 | 766.3 | 63 | 993.6 | 110 | 71 | | | 40.9 | 0.1 | **51.1** | 0.3 | 52.3 | 0.3 | 57.5 | 0.4 |
| 6 | 747.4 | 17 | 491.6 | 19 | 650.6 | 42 | 665.0 | 47 | 877.9 | 83 | 72 | | | 40.3 | 0.1 | 50.5 | 0.3 | **51.1** | 0.3 | 56.8 | 0.4 |
| 7 | 680.0 | 14 | 416.8 | 14 | 565.3 | 32 | 623.7 | 42 | 747.4 | 60 | 73 | | | 39.8 | 0.1 | 50.0 | 0.3 | 50.3 | 0.3 | 56.1 | 0.3 |
| 8 | 565.3 | 10 | 357.5 | 10 | 499.7 | 25 | 576.1 | 36 | 695.6 | 52 | 74 | | | 39.3 | 0.1 | 49.4 | 0.2 | 49.9 | 0.3 | 55.4 | 0.3 |
| 9 | 544.8 | 9 | 323.1 | 8 | 454.5 | 21 | 516.8 | 29 | 636.9 | 44 | 75 | | | 38.7 | 0.1 | 48.8 | 0.2 | 49.3 | 0.3 | 54.8 | 0.3 |
| 10 | 476.0 | 7 | 291.8 | 7 | 411.2 | 17 | 461.5 | 23 | 565.3 | 34 | 76 | | | 38.2 | 0.1 | 48.3 | 0.2 | 48.8 | 0.3 | 54.1 | 0.3 |
| 11 | 395.0 | 5 | 263.7 | 6 | 380.1 | 14 | 416.8 | 19 | 491.6 | 26 | 77 | | | 37.8 | 0.1 | 47.5 | 0.2 | 48.2 | 0.3 | 53.5 | 0.3 |
| 12 | 384.9 | 4 | 240.6 | 5 | 353.4 | 12 | 384.9 | 16 | 447.8 | 22 | 78 | | | 37.3 | 0.1 | 47.1 | 0.2 | 47.8 | 0.2 | 52.9 | 0.3 |
| 13 | 375.3 | 4 | 226.2 | 4 | 306.6 | 9 | 353.4 | 13 | 405.6 | 18 | 79 | | | 36.7 | 0.1 | 46.7 | 0.2 | 47.2 | 0.2 | 52.2 | 0.3 |
| 14 | 349.3 | 4 | 207.5 | 3 | 283.6 | 8 | 326.6 | 11 | 375.3 | 15 | 80 | | | 35.9 | 0.1 | 46.2 | 0.2 | 46.7 | 0.2 | 51.7 | 0.3 |
| 15 | 184.6 | 1 | 192.9 | 3 | 252.7 | 6 | 303.6 | 10 | 345.3 | 13 | 81 | | | 35.4 | 0.1 | 45.3 | 0.2 | 46.2 | 0.2 | **51.1** | 0.3 |
| 16 | 181.3 | 1 | 182.4 | 3 | 238.7 | 6 | 286.3 | 9 | 323.1 | 11 | 82 | | | 34.2 | 0.1 | 44.6 | 0.2 | 45.7 | 0.2 | 50.6 | 0.3 |
| 17 | **55.0** | 0 | 171.0 | 2 | 226.2 | 5 | 268.4 | 8 | 300.5 | 10 | 83 | | | 33.0 | 0.1 | 43.9 | 0.2 | 45.3 | 0.2 | 50.0 | 0.3 |
| 18 | 32.8 | 0 | 161.0 | 2 | 216.4 | 5 | 254.8 | 7 | 280.9 | 9 | 84 | | | 32.6 | 0.1 | 43.1 | 0.2 | 44.8 | 0.2 | 49.5 | 0.3 |
| 19 | 30.9 | 0 | **152.1** | 2 | 206.1 | 4 | 238.7 | 6 | 266.1 | 8 | 85 | | | **31.6** | 0.1 | 42.3 | 0.2 | 44.5 | 0.2 | 48.9 | 0.3 |
| 20 | 30.8 | 0 | 146.2 | 2 | 196.7 | 4 | 229.6 | 6 | 250.6 | 7 | 86 | | | 30.3 | 0.1 | 41.9 | 0.2 | 43.9 | 0.2 | 48.5 | 0.3 |
| 21 | 30.6 | 0 | 138.1 | 2 | 189.3 | 4 | 216.4 | 5 | 236.8 | 6 | 87 | | | | | 41.5 | 0.2 | 43.1 | 0.2 | 47.9 | 0.3 |
| 22 | 30.5 | 0 | 131.5 | 1 | 181.3 | 3 | 207.5 | 5 | 224.5 | 5 | 88 | | | | | 41.1 | 0.2 | 42.2 | 0.2 | 47.5 | 0.2 |
| 23 | 30.3 | 0 | **126.0** | 1 | 169.1 | 3 | 198.0 | 4 | 214.9 | 5 | 89 | | | | | 40.5 | 0.2 | 41.5 | 0.2 | 47.0 | 0.2 |
| 24 | 30.2 | 0 | 121.4 | 1 | 161.0 | 3 | 190.5 | 4 | 204.7 | 5 | 90 | | | | | 39.9 | 0.2 | 40.7 | 0.2 | 46.5 | 0.2 |
| 25 | 30.1 | 0 | 116.3 | 1 | **150.5** | 2 | 182.4 | 4 | 196.7 | 4 | 91 | | | | | 39.3 | 0.2 | 40.0 | 0.2 | 46.0 | 0.2 |
| 26 | | | 111.1 | 1 | 146.2 | 2 | 175.0 | 3 | 188.1 | 4 | 92 | | | | | 38.7 | 0.2 | 39.3 | 0.2 | 45.6 | 0.2 |
| 27 | | | 108.0 | 1 | 141.4 | 2 | 168.2 | 3 | 181.3 | 4 | 93 | | | | | 38.1 | 0.1 | 38.6 | 0.2 | 45.1 | 0.2 |
| 28 | | | **103.9** | 1 | 137.5 | 2 | 162.7 | 3 | 174.0 | 3 | 94 | | | | | 37.5 | 0.1 | 38.0 | 0.2 | 44.7 | 0.2 |
| 29 | | | 100.1 | 1 | 133.2 | 2 | 156.8 | 3 | 168.2 | 3 | 95 | | | | | 37.0 | 0.1 | 37.4 | 0.2 | 44.3 | 0.2 |
| 30 | | | 96.6 | 1 | **129.8** | 2 | 151.3 | 3 | 161.0 | 3 | 96 | | | | | 36.7 | 0.1 | 36.8 | 0.1 | 43.9 | 0.2 |
| 31 | | | 93.9 | 1 | 126.0 | 2 | 146.9 | 2 | **156.0** | 3 | 97 | | | | | 36.4 | 0.1 | 36.2 | 0.1 | 43.5 | 0.2 |
| 32 | | | 90.7 | 1 | 120.0 | 1 | 141.4 | 2 | 150.5 | 2 | 98 | | | | | 36.1 | 0.1 | 36.1 | 0.1 | 43.0 | 0.2 |
| 33 | | | 87.8 | 1 | 117.2 | 1 | 137.5 | 2 | 146.2 | 2 | 99 | | | | | 35.8 | 0.1 | 35.7 | 0.1 | 42.8 | 0.2 |
| 34 | | | 85.6 | 1 | 114.1 | 1 | 133.2 | 2 | 141.4 | 2 | 100 | | | | | 35.6 | 0.1 | 35.4 | 0.1 | 42.2 | 0.2 |
| 35 | | | 83.0 | 1 | 110.7 | 1 | **129.8** | 2 | 137.5 | 2 | 101 | | | | | 35.2 | 0.1 | 35.1 | 0.1 | 42.0 | 0.2 |
| 36 | | | 80.6 | 1 | 108.0 | 1 | 125.5 | 2 | 133.2 | 2 | 102 | | | | | 34.7 | 0.1 | 34.9 | 0.1 | 41.3 | 0.2 |
| 37 | | | **78.5** | 0.5 | 105.3 | 1 | 121.9 | 2 | **129.8** | 2 | 103 | | | | | 34.3 | 0.1 | 34.6 | 0.1 | 40.6 | 0.2 |
| 38 | | | 76.7 | 0.5 | **102.8** | 1 | 119.0 | 2 | 126.0 | 2 | 104 | | | | | 33.8 | 0.1 | 34.3 | 0.1 | 39.9 | 0.2 |
| 39 | | | 74.4 | 0.4 | 100.8 | 1 | 115.4 | 1 | 122.9 | 2 | 105 | | | | | 33.5 | 0.1 | 34.1 | 0.1 | 39.4 | 0.2 |
| 40 | | | 72.4 | 0.4 | 98.1 | 1 | 112.8 | 1 | 119.5 | 2 | 106 | | | | | 33.2 | 0.1 | 33.8 | 0.1 | 39.2 | 0.2 |
| 41 | | | 70.9 | 0.4 | 95.3 | 1 | 109.5 | 1 | 116.7 | 2 | 107 | | | | | 32.7 | 0.1 | 33.5 | 0.1 | 38.8 | 0.2 |
| 42 | | | 69.3 | 0.4 | 93.3 | 1 | 107.2 | 1 | 113.7 | 1 | 108 | | | | | 32.2 | 0.1 | 33.3 | 0.1 | 38.5 | 0.2 |
| 43 | | | 67.6 | 0.4 | 91.3 | 1 | 104.6 | 1 | 111.1 | 1 | 109 | | | | | 31.7 | 0.1 | 33.1 | 0.1 | 38.2 | 0.2 |
| 44 | | | 65.9 | 0.3 | 89.4 | 1 | **102.1** | 1 | 108.4 | 1 | 110 | | | | | 31.4 | 1.0 | 32.8 | 0.1 | 37.9 | 0.2 |
| 45 | | | 64.7 | 0.3 | 87.6 | 1 | 99.8 | 1 | 106.1 | 1 | 111 | | | | | 31.2 | 1.0 | 32.6 | 0.1 | 37.5 | 0.2 |
| 46 | | | 63.2 | 0.3 | 85.8 | 1 | 97.5 | 1 | 103.5 | 1 | 112 | | | | | **31.0** | 1.0 | 32.3 | 0.1 | 37.3 | 0.2 |
| 47 | | | 61.8 | 0.3 | 84.2 | 1 | 95.6 | 1 | **101.4** | 1 | 113 | | | | | 30.8 | 0.9 | 32.1 | 0.1 | 36.9 | 0.2 |
| 48 | | | 60.6 | 0.3 | 83.0 | 1 | 91.6 | 1 | 99.1 | 1 | 114 | | | | | 30.5 | 0.9 | 31.9 | 0.1 | 36.6 | 0.1 |
| 49 | | | 59.3 | 0.3 | 80.8 | 1 | 87.6 | 1 | 97.2 | 1 | 115 | | | | | 30.1 | 0.9 | 31.6 | 0.1 | 36.4 | 0.1 |
| 50 | | | 58.1 | 0.3 | **78.1** | 1 | 85.6 | 1 | 95.0 | 1 | 116 | | | | | | | **31.2** | 0.1 | 36.1 | 0.1 |
| 51 | | | 56.9 | 0.3 | 75.5 | 1 | 84.2 | 1 | 93.3 | 1 | 117 | | | | | | | 30.8 | 0.1 | 35.8 | 0.1 |
| 52 | | | 56.0 | 0.3 | 74.2 | 1 | 82.6 | 1 | 91.3 | 1 | 118 | | | | | | | 30.3 | 0.1 | 35.3 | 0.1 |
| 53 | | | 54.8 | 0.2 | 73.1 | 1 | 81.2 | 1 | 89.4 | 1 | 119 | | | | | | | | | 34.7 | 0.1 |
| 54 | | | 53.7 | 0.2 | 72.1 | 1 | 79.7 | 1 | 87.8 | 1 | 120 | | | | | | | | | 34.2 | 0.1 |
| 55 | | | 52.9 | 0.2 | 70.1 | 0.5 | **78.3** | 1 | 86.1 | 1 | 121 | | | | | | | | | 33.7 | 0.1 |
| 56 | | | 52.0 | 0.2 | 68.3 | 0.5 | 76.9 | 1 | 84.6 | 1 | 122 | | | | | | | | | 33.2 | 0.1 |
| 57 | | | **50.9** | 0.2 | 66.5 | 0.4 | 75.5 | 1 | 83.0 | 1 | 123 | | | | | | | | | 32.7 | 0.1 |
| 58 | | | 50.0 | 0.2 | 64.7 | 0.4 | 74.4 | 1 | 81.7 | 1 | 124 | | | | | | | | | 32.3 | 0.1 |
| 59 | | | 49.3 | 0.2 | 62.9 | 0.4 | 73.1 | 1 | 80.1 | 1 | 125 | | | | | | | | | 31.8 | 0.1 |
| 60 | | | 48.5 | 0.2 | 61.3 | 0.4 | 70.7 | 1 | **77.5** | 1 | 126 | | | | | | | | | 31.4 | 0.1 |
| 61 | | | 47.5 | 0.2 | 59.9 | 0.4 | 68.6 | 1 | 75.0 | 1 | 127 | | | | | | | | | **31.0** | 0.1 |
| 62 | | | 46.9 | 0.2 | 59.2 | 0.4 | 66.5 | 0.5 | 72.8 | 1 | 128 | | | | | | | | | 30.8 | 0.1 |
| 63 | | | 46.2 | 0.2 | 58.2 | 0.3 | 64.7 | 0.5 | 70.6 | 1 | 129 | | | | | | | | | 30.6 | 0.1 |
| 64 | | | 45.4 | 0.2 | 57.9 | 0.3 | 62.8 | 0.4 | 68.5 | 1 | 130 | | | | | | | | | 30.4 | 0.1 |
| 65 | | | 44.6 | 0.2 | 56.7 | 0.3 | 61.1 | 0.4 | 66.5 | 0.5 | 131 | | | | | | | | | 30.2 | 0.1 |

**Table 1**. GVSA-extracted (measured) global-vibrational a-modes of the *real* Sun (i.e., from fast and slow polar winds combined), including Alfvén resonance (AR) in the raw form (anisotropic peak splitting inclusive) and the original Rieger resonance (RR) (as just freed from Sun-internal couplings) of order m, as periods $P_m$ of ≥99%-significant peaks in GVSA spectra of (columns from left to right, respectively): the Sun's MMF (WSO), the IMF at L1 (WIND), the IMF above the Sun's polar ($\varphi_{Sun}>|70°|$) regions combined (Ulysses NS), and separately the IMF above the Sun's southern (Ulysses S) and northern polar regions (Ulysses N). The spectral peaks with values listed compose complete AR trains in some cases, i.e., a continuous spectrum of modes, Fig. 3. Values of order m, corresponding to fidelities $\Phi_m \geq 12$ indicative of a genuine systematic process (Omerbashich, 2006), are highlighted gray, RR a-modes black. Note that the present study succeeded in extracting AR from Ulysses samplings of the IMF to at least the order m=131 in both polar winds, but even a resolution that high might be improved on in the future as new data become available. Periods are in Earth days. Note that the classically termed r- and R-modes are in the range of months and years, respectively. In all tables, all periods are in Earth days unless stated otherwise.

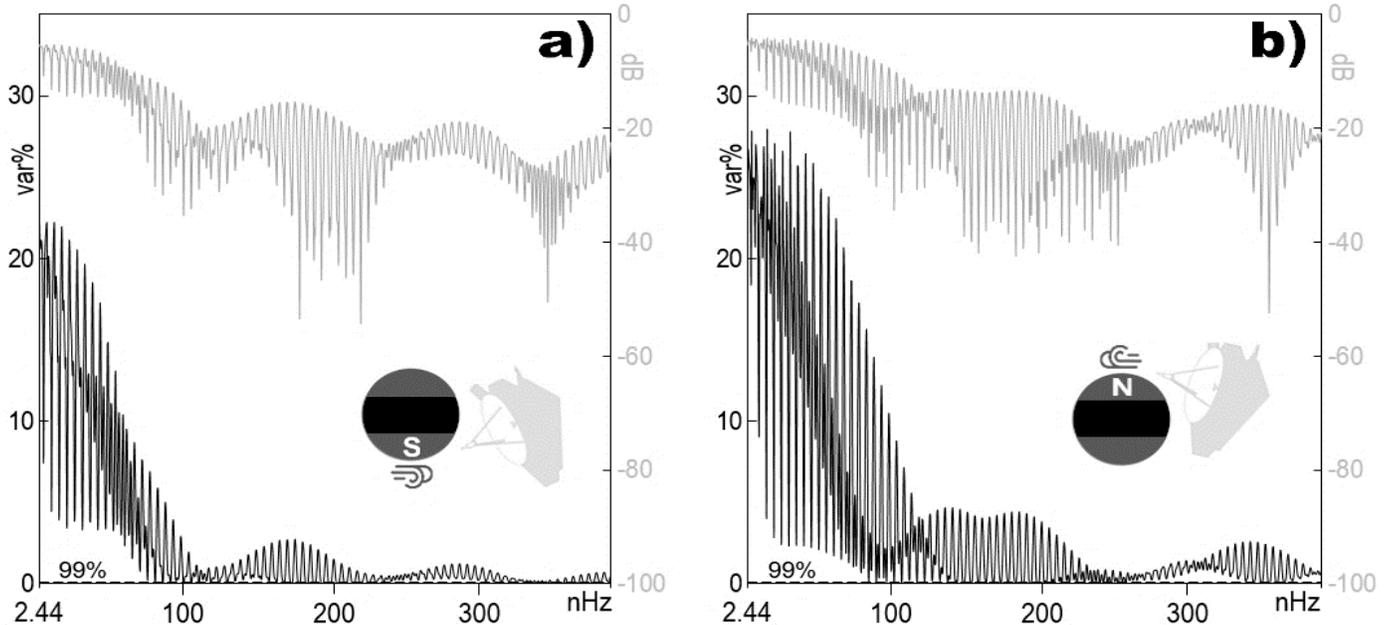

**Figure 4**. AR, compressed view, highlighting the formational regularity in the GVSA-extracted resonances' (spectral) waveform envelopes: in the southerly polar wind from Fig. 3–d (panel a) and the northerly polar wind from Fig. 3–e (panel b). The regularity means that a single secondary (here resonant) dynamical process acts globally at the Sun.

of Fig. 3 has revealed from WIND measurements of IMF at L1 that the heliosphere maintains AR and RR at least to L1 distances, each resonance virtually as a whole. However, this pureness of the signal breaks down temporally beyond/above the termination point of the antiresonance train at ~annual periodicity (at which the global coupling ceases but without vibration relaxation), spatially beyond L1, thus leaving RR open to external influences ranging from planetary constellations and gravitation to increased activity of magnetic fields, all suggested in the past and discussed later.

As seen from Figs. 2–e and 3–e, the couplings-free RR is a re-emitted (offshoot of) AR via the tailing antiresonance mode $P^-_{17}$, and primarily around the extremal phases, i.e., maxima and minima, when conditions least favor turbulence (depicted in the frequency space as anisotropic peak-splitting), making room for a partial reorganization of the fields' activity into resonance and antiresonance trains. Note the original, i.e., system value of $P_{Rg}$=156.5 days as released by the Sun during the solar cycle 23 maximum, Fig. 1. The fluctuation in $P_{Rg}$ throughout the heliosphere is constrained to not more than a few % of $P_{Rg}$, and seen formatted into the Rieger-type harmonics. Namely, this limited variability in RR is an additional indicator of a possible impact of interplanetary constellations, including planetary gravitation and magnetism (since those are geodynamically most significant, as experienced by the solar system's bodies), on $P_{Rg}$ and the couplings-freed RR that thus gets reformatted into Rieger-type periodicities as widely reported. Fidelity stayed well below 12 on all RR peaks (when computed RR alone, i.e., in the 30–180-day band; not shown), revealing that RR as part of the Sun's AR signal is never a solitary, i.e., uncoupled physical process. Thus, as soon as it gets released from the Sun couplings, factors external to the Sun, like the above-mentioned planetary constellations, including planetary gravitation and magnetism, take over the role of additional actors in suppressing the fidelity on peaks down to ~annual termination modes.

| $P^-_k$ | Ulysses IMF-S [days] | [yr] | [var%] | Ulysses IMF-N [days] | [yr] | [var%] |
|---|---|---|---|---|---|---|
| 2 | 2258.2 | 6.19 | 7.44 | 2440.9 | 6.69 | 18.39 |
| 3 | 1102.5 | 3.02 | 4.47 | 1144.3 | 3.14 | 9.37 |
| 4 | 747.4 | 2.05 | 4.19 | 766.3 | 2.10 | 4.62 |
| 5 | 554.9 | 1.52 | 3.45 | 576.1 | 1.58 | 3.37 |
| 6 | 447.8 | 1.23 | 3.94 | 461.5 | 1.26 | 2.91 |
| 7 | 370.7 | 1.02 | 3.33 | 384.9 | 1.05 | 2.68 |
| 8 | 316.3 | 0.87 | 3.81 | 330.1 | 0.90 | 2.53 |
| 9 | 278.3 | 0.76 | 3.34 | 289.0 | 0.79 | 2.40 |
| 10 | 246.5 | 0.68 | 3.43 | 257.0 | 0.70 | 2.26 |
| 11 | 222.8 | 0.61 | 3.28 | 231.4 | 0.63 | 2.10 |
| 12 | 201.9 | 0.55 | 3.04 | 210.4 | 0.58 | 1.92 |
| 13 | 194.1 | 0.53 | 2.35 | 192.9 | 0.53 | 1.71 |
| 14 | 179.1 | 0.49 | 1.61 | 178.1 | 0.49 | 1.50 |
| 15 | 171.0 | 0.47 | 2.47 | 165.4 | 0.45 | 1.29 |
| 16 | 165.4 | 0.45 | 0.82 | 158.4 | 0.43 | 1.12 |
| 17 | 154.4 | 0.42 | 0.24 | 154.4 | 0.42 | 1.11 |
| 18 | 144.8 | 0.40 | 0.09 | 138.8 | 0.38 | 0.14 |
| | 99%@: | | 0.10 | 99%@: | | 0.13 |

**Table 2**. GVSA-extracted (measured) global Alfvén antiresonance of the real Sun (troughs), Figs. 3 & 4, from the Ulysses magnetometer (MAG) data collected over the southern-polar (left-hand side; Ulysses IMF-S) vs. northern-polar (right-hand side; Ulysses IMF-N) regions of the Sun. All extracted periods are ≥99%-significant, with antiresonance $P^-_k$ modes accompanying by always preceding the corresponding $P_m$ (i.e., AR), Table 1, up to the order k=18, at which point the significance drops below 99%, grayed values in the last row. Note the original (as generated under Sun couplings) Rieger period $P^-_{17}$=154.4 days (highlighted in black), which was one of only two modes extracted with the same value from both polar data sets (the other pair $^{NP^-}_{15}$ and $^{SP^-}_{16}$ belonged to different modes, so the pairing likely was coincidental). The Sun couplings give rise to antiresonances, while such a strong folded signal at this mode, where doubling is due to the signal emitted from both the northern and southern polar regions, is the reason behind this antiresonance mode's vast presence in solar indices and its confusion sometimes for a Sun's resonance rather than antiresonance mode.

The successful complete extraction of AR and RR explains why the Rieger period $P_{Rg}$ is so profoundly present throughout our solar system. Namely, $P_{Rg}$ is the final ≥99%-significant antiresonance mode, here found of order k=17, at whose frequency the internal global couplings cease, and the $P^-_{17}$ mode remains the shortest a-mode produced by the Sun alone, allowing this mode as the real, i.e., externally-gravitationally still unperturbed $P_{Rg}$, to effectively escape the inherent ani-sotropy and turbulence of the Sun. Once set free (becoming exposed to planetary fields, mainly gravitational but at gaseous giants' magnetic perturbations as well), dynamic waves in the solar ejecta that propagate under the $P^-_{17}$ mode create at least two reflection trains complete with resonances but without anti-resonances (internal couplings) anymore and up to the rotational frequencies, Figs. 2–4. This composition reveals RR as a powerful actor in interplanetary dynamics within our solar system, and thereby planetary dynamics as well and in unison with planetary gravitational and magnetic fields.

| n | $^SP_n^{meas}$ | $^SP_n^{theor}$ | $^S\Delta$ | n | $^SP_n^{meas}$ | $^SP_n^{theor}$ | $^S\Delta$ | n | $^NP_n^{meas}$ | $^NP_n^{theor}$ | $^S\Delta$ | n | $^NP_n^{meas}$ | $^NP_n^{theor}$ | $^S\Delta$ |
|---|---|---|---|---|---|---|---|---|---|---|---|---|---|---|---|
| 1 | 3223.6 | | | 51 | 62.8 | 63.2 | 0.7% | 1 | 4100.0 | | | 51 | 80.1 | 80.4 | 0.3% |
| 2 | 1558.2 | 1611.8 | 3.3% | 53 | 61.1 | 60.8 | -0.5% | 2 | 2258.2 | 2050.0 | -10.2% | 53 | 77.5 | 77.4 | -0.1% |
| 3 | 932.1 | 1074.5 | 13.3% | 54 | 59.4 | 59.7 | 0.4% | 3 | 1481.6 | 1366.7 | -8.4% | 55 | 75.0 | 74.5 | -0.6% |
| 4 | 766.3 | 805.9 | -5.9% | 56 | 57.9 | 57.6 | -0.5% | 4 | 993.6 | 1025.0 | 3.1% | 56 | 72.8 | 73.2 | 0.6% |
| 5 | 665.0 | 644.7 | -3.1% | 57 | 56.4 | 56.6 | 0.2% | 5 | 877.9 | 820.0 | -7.1% | 58 | 70.6 | 70.7 | 0.2% |
| 6 | 516.8 | 537.3 | 3.8% | 59 | 55.0 | 54.6 | -0.7% | 6 | 695.6 | 683.3 | -1.8% | 60 | 68.5 | 68.3 | -0.2% |
| 7 | 461.5 | 460.5 | -0.2% | 60 | 53.6 | 53.7 | 0.2% | 7 | 565.3 | 585.7 | 3.5% | 62 | 66.5 | 66.1 | -0.6% |
| 8 | 416.8 | 402.9 | -3.4% | 62 | 52.3 | 52.0 | -0.6% | 8 | 491.6 | 512.5 | 4.1% | 64 | 64.7 | 64.1 | -1.0% |
| 9 | 353.4 | 358.2 | 1.3% | 63 | 51.1 | 51.2 | 0.2% | 9 | 447.8 | 455.6 | 1.7% | 65 | 62.9 | 63.1 | 0.2% |
| 10 | 326.6 | 322.4 | -1.3% | 64 | 50.3 | 50.4 | 0.1% | 10 | 405.6 | 410.0 | 1.1% | 67 | 61.3 | 61.2 | -0.1% |
| 11 | 303.6 | 293.1 | -3.6% | 65 | 49.9 | 49.6 | -0.6% | 11 | 375.3 | 372.7 | -0.7% | 69 | 59.7 | 59.4 | -0.4% |
| 12 | 268.4 | 268.6 | 0.1% | 66 | 49.3 | 48.8 | -0.8% | 12 | 345.3 | 341.7 | -1.1% | 70 | 58.2 | 58.6 | 0.7% |
| 13 | 254.8 | 248.0 | -2.8% | 67 | 48.2 | 48.1 | -0.2% | 13 | 323.1 | 315.4 | -2.4% | 71 | 57.5 | 57.7 | 0.4% |
| 14 | 229.6 | 230.3 | 0.3% | 68 | 47.8 | 47.4 | -0.8% | 14 | 300.5 | 292.9 | -2.6% | 72 | 56.8 | 56.9 | 0.3% |
| 15 | 216.4 | 214.9 | -0.7% | 69 | 46.7 | 46.7 | 0.0% | 15 | 266.1 | 273.3 | 2.7% | 73 | 56.1 | 56.2 | 0.1% |
| 16 | 198.0 | 201.5 | 1.7% | 70 | 46.2 | 46.1 | -0.4% | 16 | 250.6 | 256.2 | 2.2% | 74 | 55.4 | 55.4 | 0.0% |
| 17 | 190.5 | 189.6 | -0.4% | 71 | 45.3 | 45.4 | 0.2% | 17 | 236.6 | 241.2 | 1.8% | 75 | 54.8 | 54.7 | -0.2% |
| 18 | 182.4 | 179.1 | -1.8% | 72 | 44.8 | 44.8 | 0.0% | 18 | 224.5 | 227.8 | 1.4% | 76 | 54.1 | 53.9 | -0.3% |
| 19 | 168.2 | 169.7 | 0.9% | 73 | 44.5 | 44.2 | -0.7% | 19 | 214.9 | 215.8 | 0.4% | 77 | 53.5 | 53.2 | -0.5% |
| 20 | 162.7 | 161.2 | -1.0% | 74 | 43.9 | 43.6 | -0.7% | 20 | 204.7 | 205.0 | 0.2% | 78 | 52.2 | 52.6 | 0.6% |
| 21 | 151.3 | 153.5 | 1.4% | 75 | 43.1 | 43.0 | -0.2% | 21 | 196.7 | 195.2 | -0.7% | 79 | 51.7 | 51.9 | 0.4% |
| 22 | 146.9 | 146.5 | -0.2% | 76 | 42.2 | 42.4 | 0.5% | 22 | 188.1 | 186.4 | -0.9% | 80 | 51.1 | 51.2 | 0.3% |
| 23 | 141.4 | 140.2 | -0.9% | 78 | 41.5 | 41.3 | -0.3% | 23 | 181.3 | 178.3 | -1.7% | 81 | 50.6 | 50.6 | 0.1% |
| 24 | 133.2 | 134.3 | 0.8% | 79 | 40.7 | 40.8 | 0.3% | 24 | 168.2 | 170.8 | 1.6% | 82 | 50.0 | 50.0 | 0.0% |
| 25 | 129.8 | 128.9 | -0.7% | 81 | 40.0 | 39.8 | -0.5% | 25 | 161.0 | 164.0 | 1.8% | 83 | 49.5 | 49.4 | -0.2% |
| 26 | 125.5 | 124.0 | -1.2% | 82 | 39.3 | 39.3 | 0.0% | 26 | 156.0 | 157.7 | 1.1% | 84 | 48.9 | 48.8 | -0.3% |
| 27 | 119.0 | 119.4 | 0.3% | 83 | 38.6 | 38.8 | 0.6% | 27 | 150.5 | 151.9 | 0.9% | 85 | 48.5 | 48.2 | -0.5% |
| 28 | 115.4 | 115.1 | -0.2% | 85 | 38.0 | 37.9 | -0.1% | 28 | 146.2 | 146.4 | 0.2% | 86 | 47.5 | 47.7 | 0.4% |
| 29 | 112.8 | 111.2 | -1.5% | 86 | 37.4 | 37.5 | 0.3% | 29 | 141.4 | 141.4 | 0.0% | 87 | 47.0 | 47.1 | 0.4% |
| 30 | 107.2 | 107.5 | 0.2% | 88 | 36.8 | 36.6 | -0.4% | 30 | 137.5 | 136.7 | -0.6% | 88 | 46.5 | 46.6 | 0.2% |
| 31 | 104.6 | 104.0 | -0.6% | 89 | 36.2 | 36.2 | 0.0% | 31 | 133.2 | 132.3 | -0.7% | 89 | 46.0 | 46.1 | 0.1% |
| 32 | 99.8 | 100.7 | 1.0% | 90 | 36.1 | 35.8 | -0.7% | 32 | 129.8 | 128.1 | -1.3% | 90 | 45.6 | 45.6 | -0.1% |
| 33 | 97.5 | 97.7 | 0.2% | 91 | 35.7 | 35.4 | -0.7% | 33 | 122.9 | 124.2 | 1.1% | 91 | 45.1 | 45.1 | -0.2% |
| 34 | 95.6 | 94.8 | -0.9% | 92 | 35.1 | 35.0 | -0.3% | 34 | 119.5 | 120.6 | 0.9% | 92 | 44.7 | 44.6 | -0.4% |
| 35 | 91.6 | 92.1 | 0.6% | 93 | 34.6 | 34.7 | 0.3% | 35 | 116.7 | 117.1 | 0.3% | 93 | 44.3 | 44.1 | -0.4% |
| 37 | 87.6 | 87.1 | -0.5% | 94 | 34.3 | 34.3 | -0.1% | 36 | 113.7 | 113.9 | 0.2% | 94 | 43.5 | 43.6 | 0.3% |
| 38 | 85.6 | 84.8 | -0.9% | 95 | 34.1 | 33.9 | -0.4% | 37 | 111.1 | 110.8 | -0.3% | 95 | 43.0 | 43.2 | 0.4% |
| 39 | 82.6 | 82.7 | 0.1% | 96 | 33.5 | 33.6 | 0.4% | 38 | 108.4 | 107.9 | -0.4% | 96 | 42.8 | 42.7 | -0.1% |
| 40 | 81.2 | 80.6 | -0.8% | 97 | 33.3 | 33.2 | -0.2% | 39 | 106.1 | 105.1 | -0.9% | 97 | 42.2 | 42.3 | 0.1% |
| 41 | 78.3 | 78.6 | 0.5% | 98 | 32.8 | 32.9 | 0.3% | 40 | 101.4 | 102.5 | 1.0% | 98 | 42.0 | 41.8 | -0.5% |
| 42 | 76.9 | 76.8 | -0.2% | 99 | 32.6 | 32.6 | 0.0% | 41 | 99.1 | 100.0 | 0.9% | 99 | 41.3 | 41.4 | 0.3% |
| 43 | 74.4 | 75.0 | 0.7% | 100 | 32.3 | 32.2 | -0.4% | 42 | 97.2 | 97.6 | 0.4% | 100 | 40.6 | 41.0 | 1.0% |
| 44 | 73.1 | 73.3 | 0.2% | | | | | 43 | 95.0 | 95.3 | 0.3% | | | | |
| 46 | 70.7 | 70.1 | -0.9% | | | | | 44 | 93.3 | 93.2 | -0.1% | | | | |
| 47 | 68.6 | 68.6 | -0.1% | | | | | 45 | 91.3 | 91.1 | -0.2% | | | | |
| 48 | 66.5 | 67.2 | 0.9% | | | | | 46 | 89.4 | 89.1 | -0.3% | | | | |
| 50 | 64.7 | 64.5 | -0.3% | | | | | 47 | 87.8 | 87.2 | -0.7% | | | | |
| | | | | | | | | 48 | 86.1 | 85.4 | -0.8% | | | | |
| | | | | | | | | 49 | 83.0 | 83.7 | 0.8% | | | | |
| | | | | | | | | 50 | 81.7 | 82.0 | 0.4% | | | | |

**Table 3**. The matchings of the measured ≥99%-significant AR of the real Sun, from the Ulysses southerly-polar, $^SP^{meas}$, and northerly-polar, $^NP^{meas}$, data, Table 1 and Figs. 3 & 4, against theoretical-resonance periods $P^{theor}=P_1/i$, $i=2…,n$; $n \in \aleph$, per the $P_1$ estimate of $P_S$ from each data set, up to the order n=100. Matchings within ≤1% to respective theoretical resonance periods highlighted light gray, within ≤1‰ dark gray. From southerly-polar data, 11 theoretical-resonance periods did not have a measured match in the ≥99%-significant spectra, and 24 of all ≥99%-significant spectral peaks were not matched by theoretical-resonance periods, as due primarily to large-scale turbulence effects seen in spectra mainly as anisotropic peak splitting. From northerly-polar data, 8 and 12, respectively. All periods are in Earth days throughout and alternatively displayed as years.

The $P^-_{17}$=154.4 days, as one of only two a-modes obtained separately from both the southerly and northerly polar winds as seen by Ulysses, Table 3, thus appears in various types of heliophysical data as a prevailing driver that guides the from-that-point-on released wind's principal mechanism of propagation — the quasiperiodic (locally transient) flapping about the ecliptic. At the same time, $P_{Rg}$ becomes the carrier wave of the power from all the preceding (lower) frequencies below $P_S$, namely the AR train. $P_{Rg}$ is thus firmly locked in between the two dominant global dynamical regimes (internally: that of AR; externally: that of planetary constellations and fields). This lock makes it the most present resonance mode overall in the heliosphere, which is already well-known; see Introduction. As seen from Figs. 2–5, fidelity drops to Φ<12 beyond the antiresonances termination, i.e., in sup-annual frequency bands, allowing further for the possibility that RR becomes modulated by planetary constellations and fields so that, e.g., the ~22-yr (Hale) cycle too could be modulated by heliosphere's magnetism (Thomas et al., 2014). This flexibility means that from $P_{Rg}$ on, planetary fields couple freely with $P_{Rg}$ waves as the otherwise final (antiresonances-termination-) offshoot of AR. The AR's power gets channeled in the heliosphere by a planets-modified RR (macroscopic waves of solar ejecta) pushed outwards resonantly and thereby sped up, which is known to occur at least until the wind reaches ~$10R_\odot$ but has so far lacked explanation (Grail et al., 1996). Thus the Rieger period is a triple (tri-band) resonance response of the Sun to its three significantly contrary (out-of-phase; contrarian) global vibrations (~11, ~10, ~9-yr) around the global mode: $P_{Rg}=P_S/3/3/3$ to within 1‰ from the commonly adopted average value of $P_S$=11.3 yr (or to within 7‰ from the average value of $P_S$=11.1 yr from historical observations of sunspot numbers since the 17th century, as addressed later).

As seen from Table 3, while both the northerly and the southerly polar data performed about the same in terms of perceived statistics and precision, the former performed significantly better from the physics point of view and in terms of accuracy, recovering the theoretical resonance in more detail, which is seen as more of successful matches by order n. Then, the AR spectral signature in the northern-polar IMF resembles the Sun's inner workings more faithfully than in the southern-polar IMF. That northerly data preserved the signature of the global resonance better follows also from the symmetrical parity of theoretical a-modes that were without respective matches in the spectra. So the only periods missing from the northern theoretical resonance were those of mutually consecutive orders: 52, 54, then 57, 59, 61, 63, and then 66, 68 (i.e., even, even–odd, odd, odd, odd–even, even). In the spectra of the southerly-polar data, 36, 45, 49, 52, 55, 58, 61, 77, 80, 84, 87, thus indicating that some large-scale stochastic effect was overwhelmingly destroying parity in this region, most likely explained by the well-known relative instability and more extreme turbulence in the south. The continuous spectrum of modes, Figs. 3 & 4, is extracted with patterns virtually complete in both parities, to the resolution higher than 81.3 nHz (south) and 55.6 nHz (north) in the lowest frequencies and ~2µHz in most modes.

To discern if the main result of the present study — complete recovery of AR (and its RR offshoot) from Ulysses data, Figs. 3 & 4 — was overstated or/and an artifact of data resolution or span, Fig. 1, I next verified the finding against the 1976–2021, |B|<150 µT WSO polar magnetic fields (PF) data, Fig. 6. However, despite orders-of-magnitude differences amongst data resolution and coverage, power ratios maintained well from the southern to the northern polar region, panels b & d vs. c & e, respectively. This verification is remarkable for improve-ment in both the detected resonances' background levels and relative energy estimates (as the area of AR and RR frames). Besides the precision of estimates, their accuracy was maintained as well: the $P_S$ lead (mode) period, seen as degenerated likely due to field instability caused by inner masses distribution, is also in remarkable agreement with (phase-shifted by –1-yr from) the same obtained for the polar regions from the Ulysses data, Fig. 3–d & e. Note that RR dropped below the 67% significance level, thus emphasizing the outcome from the Ulysses data on planetary constellations and fields as a possible cause of RR reformatting to non-formatted trains again, Fig.4, and on to modified harmonics as widely observed. Note also that the fidelity on AR is again very high ($\Phi \gg 12$), which indicates a systematic physical process discernable (although not in such high detail anymore) from the WSO telescope's polar data as from the Ulysses *in situ* mission at IMF above the poles.

The verification has thus turned out positive since both the lead period estimates and the relative change in energy bands, going from southerly to northerly data, were reproduced successfully. This confirmation of the result from independent datasets of different origins and significantly varying resolution and span (WSO MMF; Ulysses; WIND; WSO PF) also corroborates Gough (1995) on spectra/processes under Sun turbulence tending to flatness. Therefore, the global decade-scale AR appears to be not a feature exclusive to the polar solar winds but the one that originates in the Sun's interior and gets maintained by the whole star.

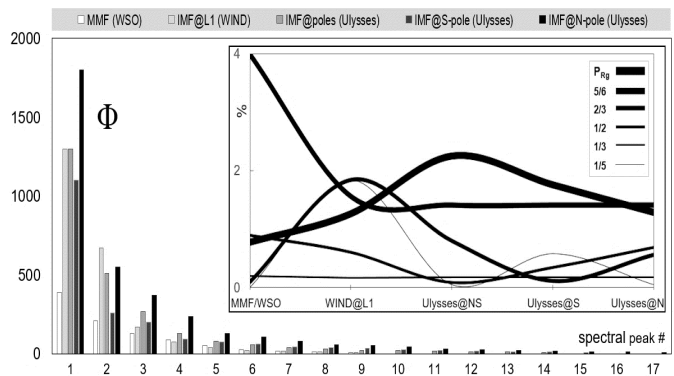

**Figure 5**. Change in significant fidelity, Φ≥12, with a computed spectral peak, up to the first 16 harmonics (highlighted gray in Table 1) plus an additional value, going from the MMF (white bars), IMF at L1 (light gray bars), IMF at polar regions combined (gray bars), IMF at the southern polar region (dark gray bars), to IMF above the northern polar region (black bars). The last shown value, #17, corresponds to ~annual harmonic(s) in the analyzed data sets and marks the antiresonances' (global-couplings) termination point in the frequency space. The plot reveals that the northerly polar wind preserved the resonance signature the best, i.e., 15 out of 16 first (lowest-frequency) a-modes. Note that the period order and grouping are for convenience only, so the clusters as depicted neither have physical meaning nor necessarily refer to the same harmonic. *Callout*: consistency plot, in the matchings of the GVSA-extracted train of RR a-modes ($P_{Rg}$ and Rieger-type periodicities), Table 1, revealing ⅓$P_{Rg}$ = ~51-day as the most consistently recovered (stablest) a-mode of RR, as 51.1-day, at the 0.2%-matching across all data sets (at ≥99%-significance from the four polar-wind data sets). Most of the extracted RR a-modes were well within ~2% of respective Rieger modes; see Introduction for the most commonly reported values of the Rieger train modes, here taken as reference values for computing the matchings (in %). All matchings were for ≥99%-significant periods, except for the MMS WSO data, for which the RR train got extracted with [67%, 89%), [89%, 95%), [89%, 95%), [95%, 99%), [95%, 99%), [89%, 95%), and [95%, 99%) significance going from $P_{Rg}$ to its highest harmonic, respectively. The matchings reflect the known fact that the equatorially mixed wind is overall slower than the polar (fast, mostly) wind and thus unable to maintain the ever-dissipating but never vanishing signature of the self-sustained AR as the overall considerably faster polar wind can where and when dominant, Figs. 2–e and 3–e. See Discussion.

## 4. EXPERIMENTAL AGREEMENT

Commonly in structural and mechanical engineering, one analyzes a structure's response to excitation by studying a response model that consists of a set of frequency response functions defined over the applicable range of frequencies. Functions typically employed are the three pairs of mutual inverses: *accelerance* as the ratio of acceleration and force, *apparent mass* as the ratio of force and acceleration; *mobility* as the ratio of velocity and force, *impedance* as the ratio of force and velocity; *dynamic stiffness* as the ratio of force and displacement, and *receptance* as the ratio of displacement and force; see, e.g., Robson et al. (1971). In physical sciences, the most common way of study is the exact opposite of the engineering way above: from a description of the response properties — most commonly in the form of measured frequency response functions — deduce modal and spatial properties of a physical system. Thus, since the total response of a set of coupled components is expressible in terms of the mobility of individual components, inverse comparisons apply

58

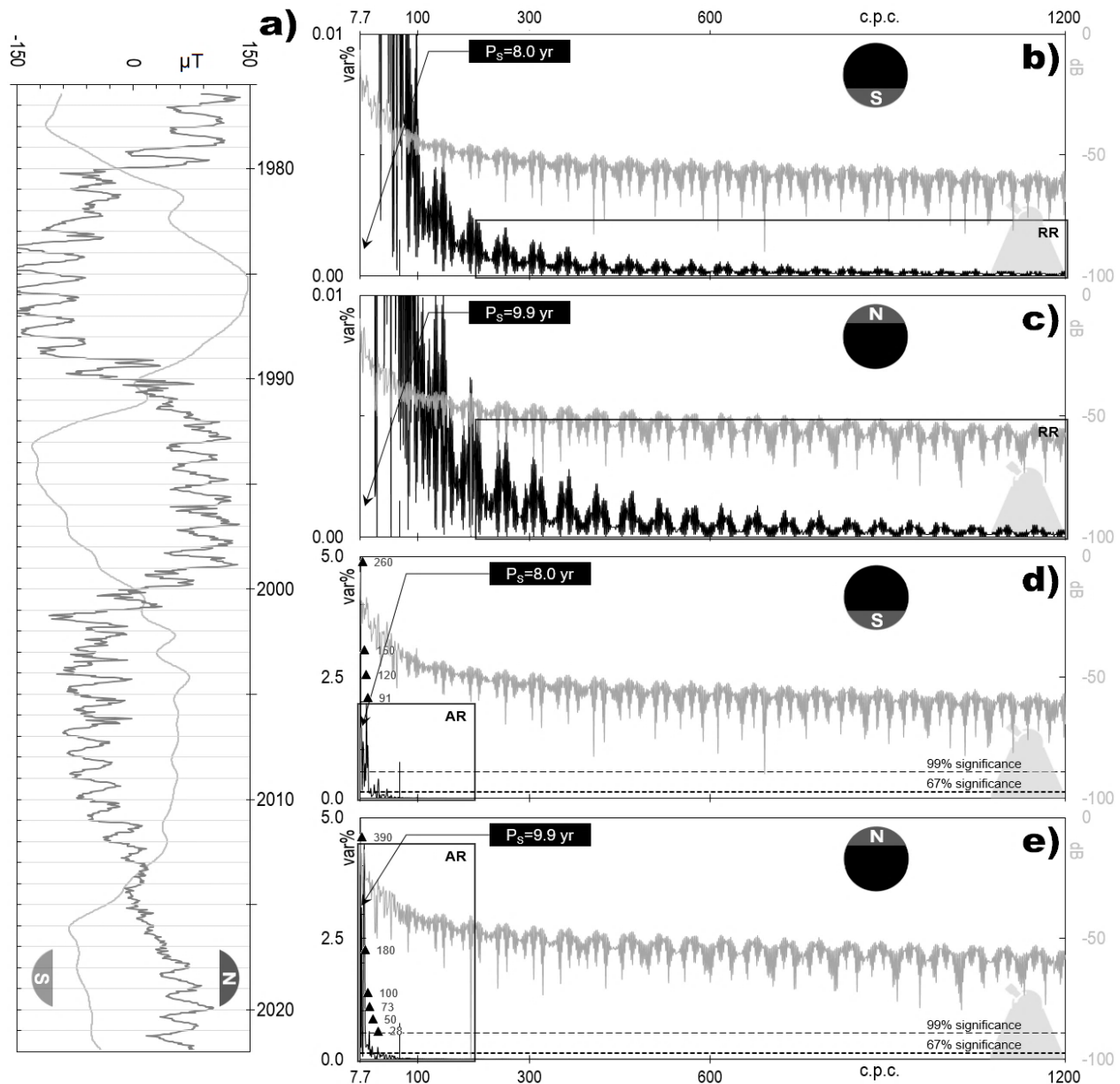

**Figure 6**. Verification of the main result (itself obtained from Ulysses hourly averages, Fig. 3) from ten-day means of the WSO telescope's 31 May 1976–21 November 2021 polar field (PF) observations, panel a, in the same band of interest. Note that panel b depicts the same data as panel d and panel c as panel e, where panels b and c are zoomed in along the left vertical axis to make RR visible.

as well (Ewins, 1995). As we saw from the present study, the Sun is a physical system that vibrates additionally (resonates), most likely due to inner masses distribution and under couplings among its latitudinally and depth-stratified sectors of varying physical and chemical properties. Those sectors are geographically distinct regions that exhibit significantly differential rotation rates, contrarian vibration modes and related resonances, and different mass and point velocities. The resonant response of a physical system most complete to irregular mass distribution and such couplings is constructive–destructive, i.e., the one exhibiting both global resonances (characterized by sharp peaks in the system spectra) and global antiresonances (sharp troughs), Figs. 2 & 3 vs. Fig. 7. Depending on the fundamental properties of the system of interest, antiresonances in a well-behaved system either always immediately precede or always immediately follow resonances.

As expected for the gaseous Sun, a modal comparison of panels a & b in Fig. 7 against panels d & e in Figs. 2 & 3 reveals that the here extracted antiresonance modes always precede the corresponding very-long-period, downwards-drifting (per dashed trends in Fig. 2–d & e) resonance modes as impressed onto the solar wind at or near its source, which corresponds to Fig. 7–a. As seen, this experimental confirmation of the result of extraction of frequency spectra from real *in situ* data correctly reflects the situation with the Sun, in which the genesis of polar (mixed) solar winds is a mass-driven phenomenon and not a stiffness-driven one (which characterizes rigid or solid macroscopic bodies of mass). Furthermore, the solar wind maintains the same downward-drifting regime and a higher order of resonances than respective antiresonances in the heliosphere, e.g., at the L1 point, Fig. 3–b. The data thus confirm that the here revealed global resonation is the way of distributing the solar wind along very low frequencies and that this resonation is the natural state of that distribution due entirely to the Sun engine being a typical revolving-field magnetic alternator. If our Sun was not behaving as a globally structured and therefore well-behaved magnetoalternator, i.e., a real rather than a proverbial engine anymore, and if that property was not fundamental to Sun-like stars, no complete extraction of a-mode vibrations could be possible from system dynamics such as the polar solar wind's dynamics used here for that purpose.

Furthermore, the equal number of corresponding resonance (spectral peak-) and antiresonance (spectral trough-) periods in Fig. 3 reveals *point mobility* (force and velocity are at the same point in frequency space) rather than *transfer mobility* (at different points and thus characterized by more minima than antiresonances) of the excitation source. In addition, the always-matching sharpness of the spectral peak/trough pairs reflects a high quality of data and analysis tools (Ewins, 1995) — which here translates into a praise of the team in charge of designing the Ulysses magnetometer led by André Balogh (1988), and the co-inventors of the GVSA method Carl Friedrich Gauss and Petr Vaníček. In summary, Fig. 7 reveals the variance-spectral signature of the global resonance in the polar wind, Figs. 2–d & e and 3–d & e, as matching best the experimental case II, Fig. 7–d & e, while correctly reflecting the mass-dominated mobility, Fig. 7–a, as seen from the downward trends in Fig. 2–d & e. Therefore, the transient nature of solar resonances, including local transiency of the Rieger's, is due to the classical revolving-field type of alternator our Sun is, of which then specific solar-cycle maxima and minima are merely apparent alternating stages.

5. PROBING THE SUN ENGINE

Thus far, the main concern in the present study was to analyze the *in situ* polar magnetic field variations as felt by the Ulysses spacecraft. As a resulting claim, albeit one based solely on claims by Gough (1995), GVSA of those variations, dominated by the fast (>700 km·s$^{-1}$) solar winds, has indicated a remarkable discovery. Namely, the Sun and, by extension, the Sun-like stars act in the interior and as a whole as well-behaved engines that, in turn, exhaust their stellar winds as a byproduct of those engines at work. Indeed, I have shown above that our star vibrates and resonates globally and along very long (many times the rotational period) temporal scales, as expected from alternating motors as a well-known and understood rotating-machinery concept from mechanical engineering. If validated from disparate data, this discovery replaces the centuries-old notion according to which the Sun acts as a simple and in astrophysics often invoked, yet poorly understood although extensively studied and modeled on, dynamo.

To validate the above discovery, I first examine whether this remarkable result was due to non-physical circumstances like the methodology applied or the degree to which the fast winds dominated the site during data collection. Secondly, to verify that the interior most likely is the originating location of the fast solar winds' complete vibration (i.e., the global solar mechanical resonances and antiresonances), which would corroborate a stellar magnetoalternator at play, I analyze the historical sunspot and solar calcium records as spatiotemporally

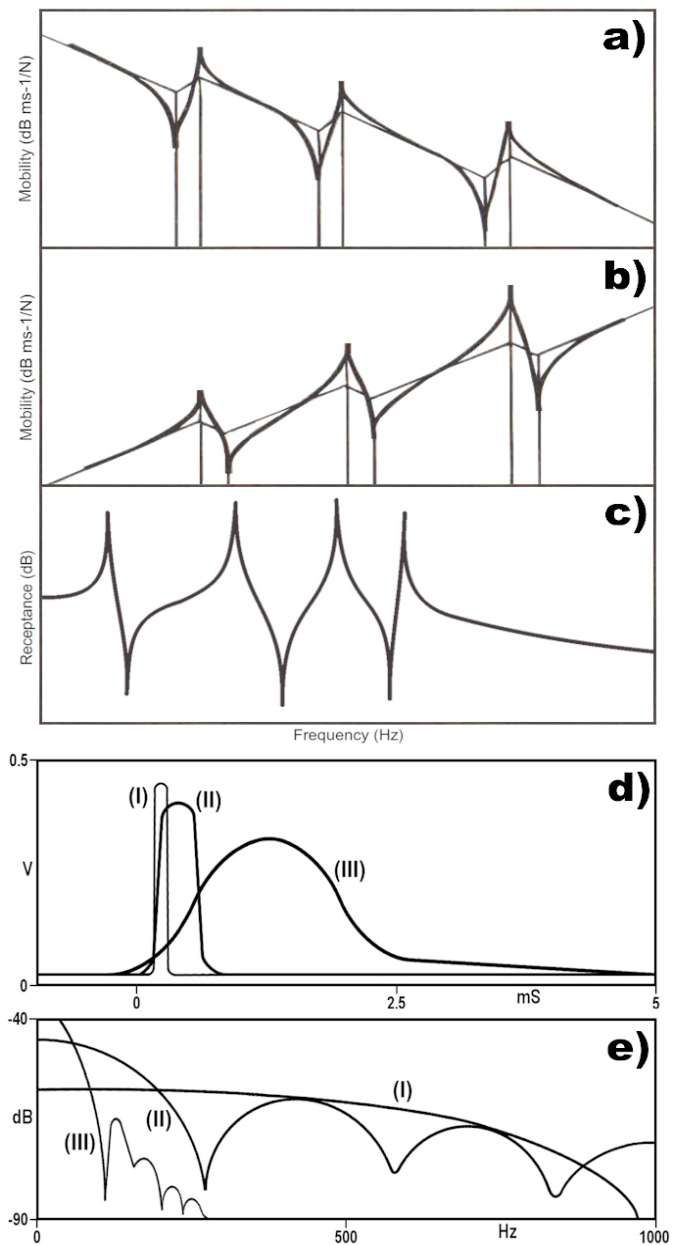

**Figure 7**. Panels a & b: Traced typical dB-scaled plots of *mobility* (ratio of velocity response to force input) in a resonating body: mass-dominated mobility (panel a) that tends to drift downwards with antiresonances occurring immediately before resonances, and stiffness-dominated mobility (panel b) that generally drifts upwards and has antiresonances always immediately above resonances (Ewins, 1995). Skeleton layers respectively represent the additionally constructed ideal mass lines and stiffness lines. These layers are used in experimental studies to trace the frequency response function of resonances vs. antiresonances and thus examine if they are symmetric, i.e., reflective of a genuine physical process. Panel c: An example of *receptance* (ratio of displacement to force input) in a typical (resonating) damped multi-degree-of-freedom system in dB (panel c) (He & Fu, 2001). Panels d & e: The impact type of transient excitations, from experiments with (I) an attached shaker producing the rapid sine sweep (chirp), (II) an attached shaker producing a burst — a short section of signal (random or sine) as a more general version of the rapid sine sweep, but in which the short-duration signal can take any form, and (III) a non-attached impactor such as hammer, inflicting blows upon a test structure. The response data are collected using accelerometers attached to the test structure and processed with an *analyzer* — a voltmeter-type instrument for sampling the system dynamics in volts (V) and computing a frequency response or spectrum in dB (Ewins, 1995). Compare to real-data results, Figs. 2 & 3.

independent indicators of global magnetic activity. We know that sunspot records somehow yet uniquely represent solar activity, so they can here serve as a means for avoiding bias if relying on fast winds data alone, as in the above thus far. Of

60

added interest here is that the chromospheric fine structure around sunspots is marked by bright areas called flocculi or calcium plages when viewed using the K line of calcium, with the plages intensity measured on a scale of 1 (faint) to 5 (very bright). Also worth noting here is that the polar field and active-region field (activity or sunspot) are two faces of magnetism on the Sun. For example, the polar field at the solar minimum is known to herald the activity of the next solar maximum, while the activity (or activity complex) results in the poleward flux streamer about to reverse the polarity of the polar field.

5.1 *Examining the possibility of bias*
So far, I looked into the dynamics of the real Sun — a real-world scenario that reflects the macroscale dynamics of all polar winds while acknowledging that the fast winds spatially and temporally dominate the polar regions. To verify that the agreement of the above-discerned solar magnetoalternator engine with the experiment was not a byproduct of data trends or analysis methodology (a coincidence basically), I next examine the ideal case, i.e., discard the slow polar winds data in the reanalysis of the Ulysses data and designate the remaining fast-wind data as the sole representative of the ideal Sun. If correct, this approach should also enable us to learn more about global solar dynamics.

To obtain a dataset best representing the ideal case, I again utilize the blindness to data gaps as a GVSA feature and separate the polar winds into the fast, tacitly presumed in the above to solely reflect the theoretical Sun as a magnetic alternator, from the slow (here 400–700 km s$^{-1}$) winds, supposedly reflecting the (also resonantly) arising turbulence. If the above-claimed discovery of the global engine type for our Sun is genuine, then such separation ought to systematically improve the real-data-based estimates of global vibration modes, Figs. 2 & 3, to significantly approach respective theoretical forms. At the same time, this separation should also amplify the GVSA spectral magnitudes on the modes (in %var) towards purely theoretical contributions of each respective mode to variance, i.e., towards 100%-var. Furthermore, if they represent the ideal Sun, antiresonance inflections as a supposed byproduct of turbulences must vanish from the spectra of fast winds (i.e., with slow winds discarded).

Indeed, as seen from the results of this separation, Figs. 8 & 9, all estimates of the ideal Sun's modes (periods) of global vibration have now attained their new values more congruently, with both longer and shorter southerly periods shifting, and longer northerly periods remaining the same, with shorter northerly periods shifting somewhat, Table 4. The Rieger period thus remained a part of the antiresonance and retained practically the same value, going from 154.4-day, Table 2, to 153.6-day, Table 6. As before the above separation, $P_{Rg}$ is again equal in the winds emanating both from the north and south. Since the real and the ideal Sun in the present study are macrodynamic spaces, meaning neither is a model space, it is physically plausible to declare the simple mean, 154.0-day, as the final extracted data-based value of the Rieger period. This average agrees absolutely with the original estimate by Rieger et al. (1984). The match is remarkable in itself, given the very high temporal resolution of the Ulysses data and their very long (multi-annual) separation. In turn, this successful and conclusive demonstration of the solar origin of $P_{Rg}$ serves as proof of the validity of GVSA and the spectra.

At the same time, all the spectral magnitudes have significantly approached both the ideal (100%-var) case and the shapes of theoretical spectral envelopes. This stunning and all-encompassing increase in spectral magnitudes has given physical soundness to AR estimates as the vibrational modes of the ideal Sun (slow winds absent), here presumed at the outset to act as a magnetoalternator. In addition, antiresonances lost inflections already by $P_S/7$.

Another evidence that the above results depict the ideal Sun lies in the inverse identity between the northerly vs. southerly results, i.e., the former spectra are virtually mirror-images of the latter and vice-versa, correctly reflecting the hemispherical (ideal) directionality of the global vibration of a nearly-perfectly spherical body like the Sun. At the same time, such mirroring requires that its hemispheres share a common surface (or a mechanism resembling one) necessary for our star to maintain AR over decades, i.e., a rigid or solid core as the most natural such surface. Finally, extracting the well-known Rieger period made the whole AR trains credible. The significance of the main result is now absolute, as the highly congruent (virtually theoretically perfect) spectra of the Sun's global decadal dynamics are a result of analyzing *in situ* magnetometer data rather than modeling, where the northerly results resemble a model field.

| w | $^SP_w$ | $^S\Phi_w$ | $^NP_w$ | $^N\Phi_w$ |
|---|---|---|---|---|
| 1 | 4580.0 | 2200 | 4476.2 | 2200 |
| 2 | 2293.3 | 560 | 2241.3 | 540 |
| 3 | 1529.6 | 250 | 1494.9 | 240 |
| 4 | 1147.5 | 140 | 1124.6 | 140 |
| 5 | 918.1 | 90 | 899.3 | 87 |
| 6 | 765.2 | 63 | 749.2 | 60 |
| 7 | 655.9 | 46 | 642.0 | 44 |
| 8 | 574.0 | 35 | 561.7 | 34 |
| 9 | 510.2 | 28 | 499.2 | 27 |
| 10 | 459.2 | 23 | 449.3 | 22 |
| 11 | 417.5 | 19 | 408.8 | 18 |
| 12 | 382.7 | 16 | 374.7 | 16 |

**Table 4**. Ideal Sun AR, extracted from the fast polar winds in the 1–13-years band, Fig. 8–c & d. Listed are all w=12, ≥99%-significant GVSA spectral periods in Earth days, from the southerly, $^SP_w$, and northerly fast winds, $^NP_w$. Also listed are respective fidelity values, Φ, that in all cases stayed ≫12 to >12, indicating a genuine resonance, i.e., one caused by a single secondary dynamic process at play at (on and inside) the Sun.

When compared to the previous respective results from the fast and slow polar winds combined (the more realistic scenario; real Sun), Figs. 2 & 3, now Fig. 8 reveals that all the even modes in the northerly fast-wind results (the more idealistic scenario; ideal Sun) got progressively somewhat shortened, and all the odd modes got progressively somewhat elongated (about twice as much relative to the neighboring mode). In the southerly fast-wind results, all the modes got progressively somewhat elongated; the change in the odd modes was again about twice that seen in the even modes, Table 4. This uniform non-symmetry in the frequency space when going from solar winds of one speed to those of a significantly different speed indicates a clear separation of the very nature of physical causes of the two dynamics (residing in lowermost frequencies or highest dynamical energies) that thus get spectrally revealed. In other words, different physical but same-kind (primarily dynamical) processes give rise to fast vs. slow (polar) winds.

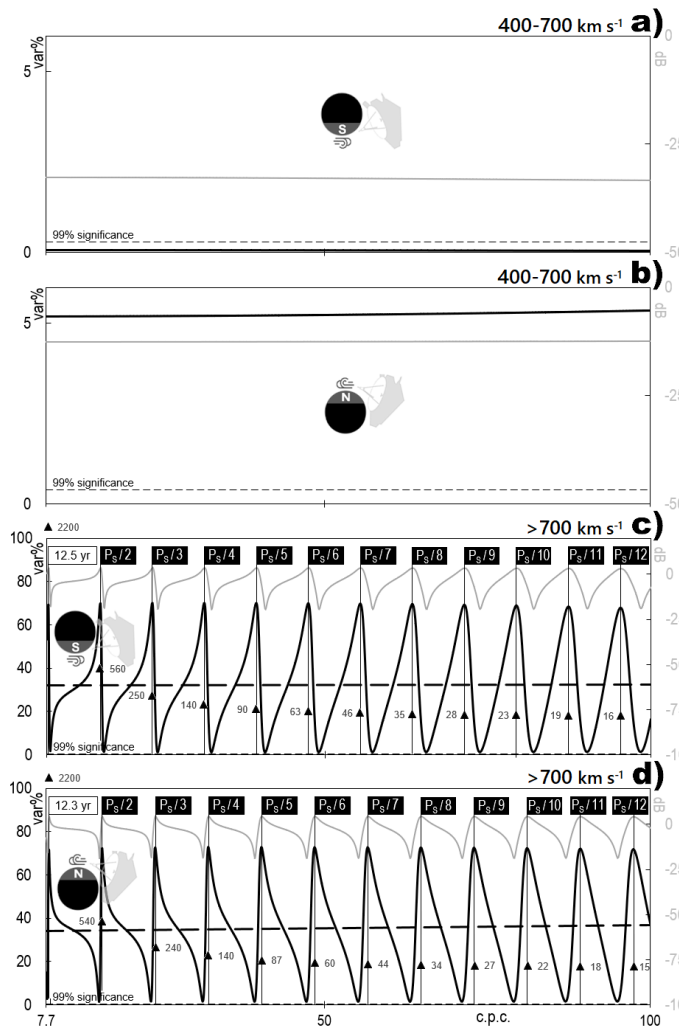

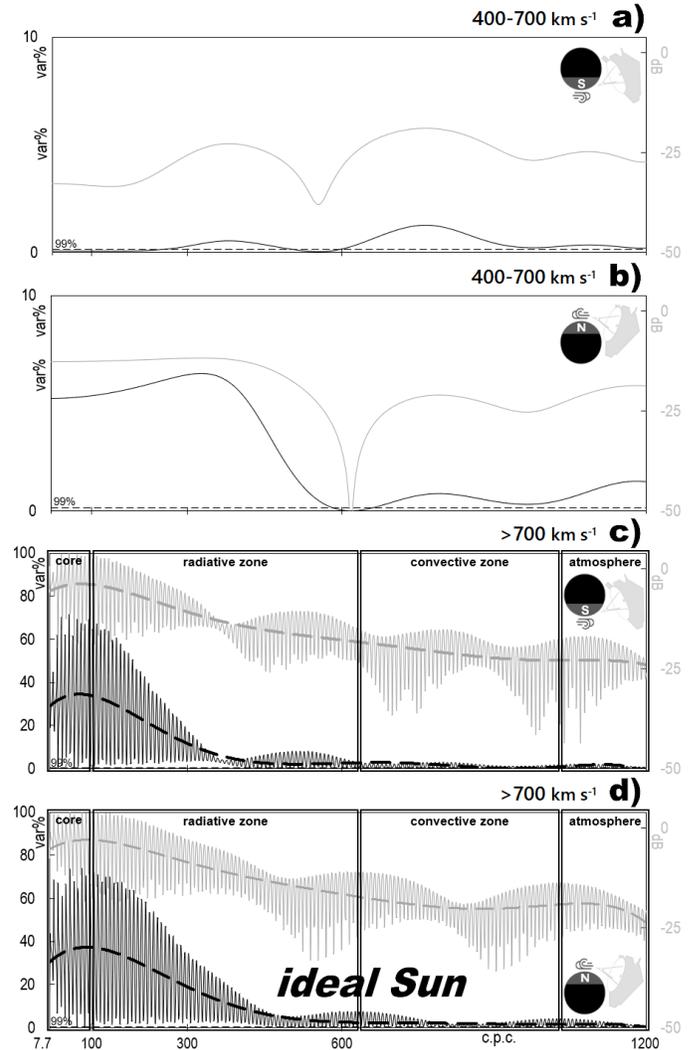

**Figure 8.** Results of repeated analyses from Figs. 2 & 3–d & e after data separation of the polar winds into slow, 400–700 km·s$^{-1}$ (panels a & b) vs. fast, >700 km·s$^{-1}$ (panels c & d) (based on the butterfly map by Fujiki et al., 2019, their Fig. 2), spectral band 1–13 years. Flat spectra of the slow winds (panels a & b) correctly reflect turbulence and, to a smaller extent, a relatively lower temporal resolution of the slow winds data on decadal scales. Importantly, fast-wind signatures of global vibration modes (panels c & d) now exhibit an upwards drift (stiffness-driven rather than mass-driven dynamics, Fig. 7–b), thus revealing the presence of a rigid or solid inner core in the Sun. The same is indicated even from the slow winds of the northerly polar region (panel b), where a ≥99%-significant and relatively strong upward drifting trend at above 5%var is also noticeable. At the same time, the slow winds of the southerly polar region (panel a) are overwhelmed by turbulence that dampens any signatures to below 1var% and below the 99%-significance level, making the slightly downward trending seen in there a less reliable source of information on global dynamics than the strongly upward trending from the northerly polar data (panel b). Furthermore, compared to Fig. 2, fidelity of the results in panels c & d again stayed well above 12, with a drastic increase (practically doubling) on the lowest frequencies (highest macrodynamic energies) from $1.2\cdot10^3$ to $2.2\cdot10^3$ in the southerly (panel c), and a minor increase from $2.1\cdot10^3$ to $2.2\cdot10^3$ in the northerly polar region (panel d), which correctly reflects the fact that the southerly polar region is less dominated by the fast wind than its antipode is. At the same time, the main (global) mode estimate became more consistent, at 12.5-yr southerly vs. 12.3-yr northerly, revealing ~12.4-yr as the global vibration mode of the *ideal Sun* (i.e., of what the Sun would dynamically be in the absence of slow winds and, by extension, turbulence). The north-south difference of ~0.2-yr is mainly due to the above-deduced southward eccentricity of the core. This *ideal global mode* then gets naturally overexcited by global vibrational dynamics, including resonance and antiresonance, to its commonly observed Schwabe value of ~11.3-yr. Also, the fact that AR (resonance and antiresonance) were extracted virtually theoretically perfectly from the IMF carried by northerly and southerly polar fast winds alone confirms that AR is the feature of the interior rather than those winds alone, as that alternative explanation would require an impossible set of coincidences for it to arise and be maintained for a decade or over a solar cycle. Finally, when slow winds of the southern polar region were considered alone (panel a), the significance dropped to nearly 67%, likely due to prominent turbulences in this region. The extracted periods' values are in Table 4.

**Figure 9.** GVSA spectra of the fast- vs. slow-wind separated data, Fig. 1, here in the 1-month–13-years spectral band. Panels a & b again depict turbulence correctly as virtually flat spectra as expected (Gough, 1995). Note that there is a drastic improvement compared to Fig. 3, especially from the northerly fast winds (panel d) for which the spectral envelope attained a virtually perfect (theoretical) shape, even though the present study involves no physical modeling. Panel d thus depicts *the ideal Sun*, i.e., Sun dynamics in the absence of slow polar winds and, by extension, polar turbulences. The Sun zones in panels c & d, ending with the surface rotational frequencies of the photosphere, were extrapolated to in-between the consecutive crests of the northerly fast-wind spectral envelope and based on the core's guidance of the fast polar wind as surmised by comparing Fig. 8–c & d vs. Fig. 2–c & d vis-à-vis Fig. 7–a & b, thus enabling most precise delimitation of the zones. The fact that there exists a single function (here 6$^{th}$ order polynomial, dashed lines) with which one can represent global macroscopic dynamics of the Sun, even if for one hemisphere better than the other, is remarkable in itself. The fast and slow polar winds are the same as in Table 8.

### 5.2 Demonstration of the solar magnetoalternator at work

Spectra of solar activity, sunspot numbers, solar calcium, and other records that directly reflect the interior operation should be consistent with each other and those of solar-wind magnetic variations in the same band of interest and using the same methodology. Therefore, I look next into solar calcium and sunspot activity, expecting to obtain spectra that identify the global Sun as a rotating mechanical engine at work. Of course, one cannot expect both temporally and spatially sparse data (basically local, far less densely distributed than the high-resolution IMF variations sensed by Ulysses) would reveal the exact modes of global action as seen from far better globally distributed Ulysses data. However, the solar calcium and sunspot data should exhibit at least the same trending reflective of such action.

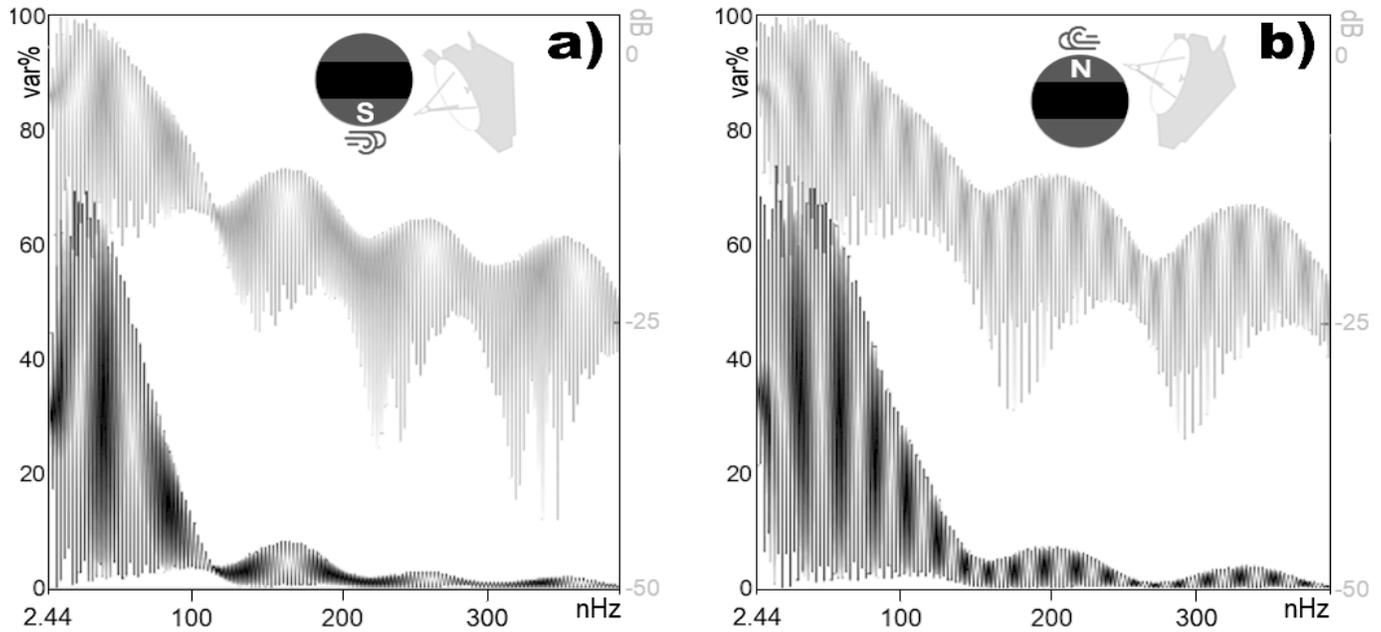

**Figure 10**. AR seen in a compressed view of the GVSA spectra of (magnetic variations in) the fast polar winds, highlighting the formational regularity in the extracted resonances' (spectral) waveform envelopes: in the southerly polar wind from Fig. 9–c (panel a) and the northerly polar wind from Fig. 9–d (panel b). The regularity means that a single secondary (here resonant) dynamical process acts globally at the Sun. The high detail in northerly vs. southerly spectral contents reveals that the former projects the process better, i.e., is less obstructed by its destructive nature (for resonant processes). The preferential resolution indicates an apparent spatial shift of that process towards the south. The spectral envelopes are even more regular now vs. those in Fig. 4, being virtually theoretically perfect (especially so for the northerly polar region), thus indicating that the secondary (here resonant) dynamical process that acts globally at the Sun is related to the fast winds primarily. Therefore, the fast polar wind represents the *ideal Sun* (slow winds absent) most genuinely. Image contrast was enhanced equally in both panels for their better distinction in detail richness. Note that the outstanding results on panels a & b, especially the significantly detail-richer panel b, do not depict any physical models but results of GVSA analysis on real (magnetometer) data collected *in situ*.

If the behavior of the Sun as a magnetoalternator can indeed be verified in an above-described way from independent data, i.e., if the whole Sun from the interior up acts as that specific type of engine rather than just the atmosphere or the convective zone or that from the core to that limited by Alfven radius, spectra of such verification data ought to be symmetric in the frequency band of any systemwide-symmetrical action considered globally in a sphere of mass. For instance, in somewhat viscously dumped systems, mobility plots exhibit symmetry about the resonant frequency (Ewins, 1995). Such behavior must arise here since such a signal in a global-solar magnetoalternator would oscillate on the real axis and be composed of symmetric complex conjugate parts. In mathematics, this is akin to the spin of two phasors rotating in opposite directions around the center of the complex plane. In engineering, the rotating vector directly translates to the rotating magnetic field as the operational principle of an AC generator. Note here that the type of electrical current carried throughout the heliosphere by the solar wind is indeed AC, a preference that still lacks a global source.

Indeed, as seen in Fig. 11 (panel a), the GVSA frequency spectrum (both variance- and power-) in the 1-month–13-years band of group sunspot numbers from 12/1611-03/1995 and their (2$^{nd}$ order polynomial) trends depict pronounced symmetry over the band of interest, which is remarkable given the relatively low temporal resolution of (monthly) averages, and even more so given that the variance spectrum of a closed physical system such as the Sun measures global relative dynamics (total dynamic-energy budget) of that system. The result from panel a improves after discarding the record portion known to be less reliable, panel b. Finally, panel c (panel d is the same as c but after zooming in along the var% axis) and panel e show the GVSA spectra (in the same band) of U.S. NOAA's solar calcium historical monthly averages from 02/1916–04/1999, which are all also pronouncedly symmetric. Note that the spectra are significant in their lowest and highest-frequency ends, so the whole process these and any such spectra depict is statistically significant, in addition to being significant physically (and thus 100%-significant statistically). Besides, even if the above results are judged on the merits of statistics alone, densifying of data improves statistical significance drastically, as seen from panel (e) that depicts the GVSA spectrum of the U.S. NOAA's daily solar calcium values from 06.08.1915-30.12.1984. Finally, since both spectral symmetries — namely the alternating trending of the spectral envelope from the fast polar winds alone (the core vs. zones of Fig. 9–c & d) revealing a rigid or solid mass in the interior; and the solar calcium/sunspot numbers variations (Fig. 11) revealing an alternating mode of operation of the global magnetism — are in the same band, that is the interval of both the core flipping and polarity reversals. Therefore, the Sun's magnetic polarity reverses by the core flipping.

Omerbashich, M. (2023) The Sun as a revolving-field magnetic alternator with a wobbling-core rotator from real data. *J. Geophys.* 65(1):48-77

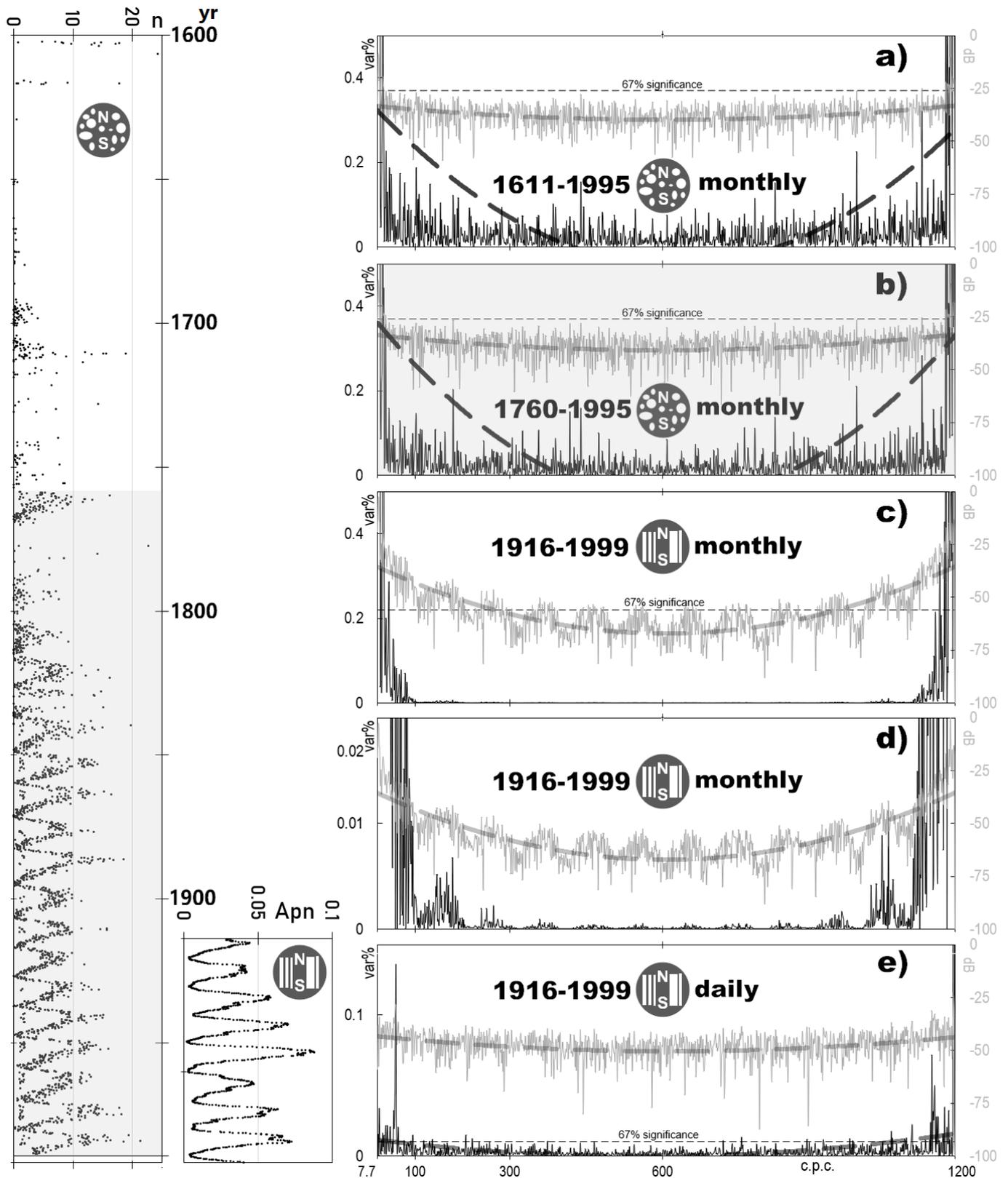

**Figure 11**. GVSA frequency spectra of common indicators of global solar activity, here in the 1-month–13-years spectral band of interest (the entire band of the Sun's macroscopic dynamics, which covers suprotational frequencies down to the polarity flip cycle): group sunspot numbers (panels a & b), and solar calcium (panels c–e). One striking global feature is the total spectral symmetry, seen in all the examined data, as a phenomenon known from mechanical engineering to commonly reflect the work of a revolving-field (alternating) motor. Note that the spectra' ends exceed the 67%-significance level well, meaning that all of the examined spectra can in their entirety be regarded as statistically significant since depicting a typical revolving-field engine as a closed physical system with well-known characteristics of operation (with statistical significance equating 100% due to completely known physics, i.e., governing dynamics). Panel b contains the spectrum of the shaded portion of the sunspots data (left vertical panel), widely held to be the more reliable part of the record. As seen by comparing panel b vs. panel a, this shortening has improved the symmetry, which supports the above claim that the whole Sun behaves as a revolving-field magnetic alternator. Areas of calcium plages and networks (Apn) are in units of a fractional area of the solar disk, here millionths of the hemisphere (right vertical panel). The above historical data are from the U.S. NOAA solar indices database (see Data statement).

64

5.3 *The origin of the (fast) solar winds: wobbling solar core*
Fig. 12 shows the 1-13-yr GVSA spectra of Sun activity as represented by the alternating (maximum–minimum) smoothed 1750-2008 monthly mean sunspot number. As seen on panel b, the ≥99%-significant spectral peaks are those of northerly global resonance mode and its equatorial and southerly degenerations extracted from the Ulysses magnetometer *in situ* data, Fig. 3. Also noticeable is the pronounced dominance of just one (northerly) of the three global modes, whose signal is so strong and clear that even its critical harmonics got extracted, with its spectral magnitude going from the ≥99%-significance down to virtually equally significant ≥95% level, i.e., the northerly global signal is so strong that it could not be overly affected by the combined effect of the equatorial and southerly-polar region dynamics. This sturdiness of the northerly global mode is additional direct evidence that the above presumption that the northerly polar wind is the genuine solar wind (of the ideal Sun) was correct. Most importantly, by applying the well-known concepts from mechanical engineering, namely the vibrational analyses of rotating machinery, one readily establishes that the high spectral peak at the fundamental frequency and high peaks at the harmonics, especially at the second, reveal the Sun core's eccentricity. Furthermore, one can expect that the solar core wobbles in 3D due to the observed eccentricity, as shown by 2D experiments for a circular levitated object when its center gets offset from the center of the circular source of levitation (Kim & Ih, 2007). Vibrational spectral analysis of eccentric multiple-rotor systems, such as the Sun and its differentially rotating layers, particularly the core, has been worked out in mechanical engineering; see, e.g., Zhang et al. (2020) for an example that utilizes subharmonic analysis.

Namely, as components of a vibrational spectrum from the analyses of rotating machinery in mechanical engineering, harmonics are multiples of the fundamental note (here of the three Schwabe global modes: northerly, equatorially, and southerly). Their presence indicates the presence and severity of specific faults and defects in machinery, such as bents, eccentricity, damage, cracked rotors, and wear. The same kind of machinery analysis applies to the Sun. From the spectra of solar activity, Fig. 12, the northerly fundamental note's harmonics are ≥99%-significant, with the three global modes of vibration found to be the most pronounced spectral peaks, based on their significance levels, variance-spectral peaks widths, and power-spectral peaks lobing.

Thus, as well understood from mechanical engineering, the high peak at the fundamental note and high peaks at the harmonics, especially at the second harmonic as is the case here, reveal the rotor's (here: core's) eccentricity; the offset's signal is strong (as characterized by both high and wide spectral peak) since both ½$P_{northerly}$ and ½$P_{equatorially}$ were found to be significant, again correctly and as deteriorating when going away from the apex so that no significant ½$P_{southerly}$ was detected. This internal check, and the fact that the equatorial (as $P_{equatorially}$) and southern turbulence (as $P_{southerly}$) succeeded in somewhat (from ≥99% to ≥95%, i.e., still not drastically) suppressing the $P_{northerly}$ significance, strongly suggest a core offset in the general direction away from the apex, i.e., a core lag on the Sun's way about the Galactic Center. This lagging position of the core relative to the Sun as a whole was expected since the Sun's mass is about thrice the core's. Such a position is also strongly suggested by

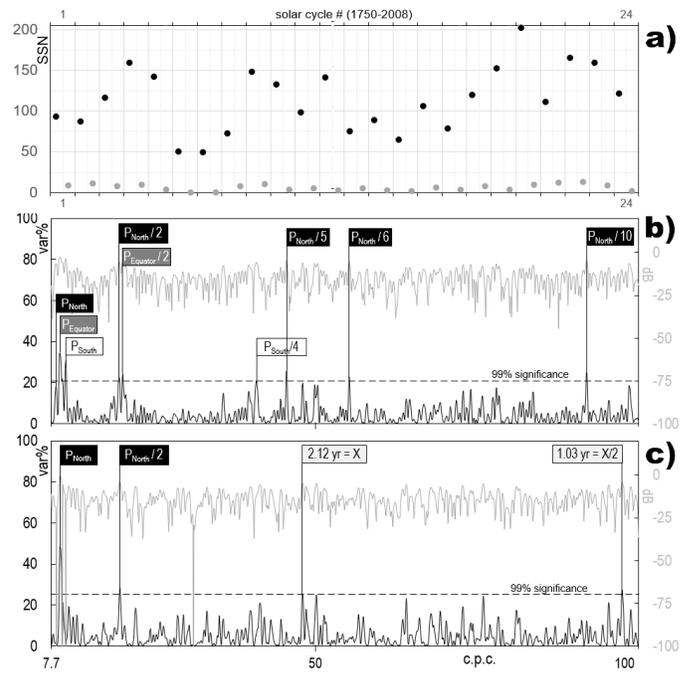

**Figure 12.** GVSA spectra (panels b & c) of 258 years of historical solar activity (panel a) in the 1–13-years band (the core's macroscopic dynamics band, as surmised in the above), represented by the group number of sunspots per solar maximum (black dots) and minimum (gray dots). Panel a: solar activity as the alternating (maximum-minimum) smoothed monthly mean sunspot number (SSN) observed between 1750-2008 (arithmetic average of two sequential 12-month running means of monthly mean numbers), widely held to be the best record of solar engine (interior) operation. Panel b: GVSA was successful at extracting the Schwabe global mode correctly, i.e., as split into three spectral peaks with values: $P_{northerly}$=11.0-yr at ≥95%-significance, $P_{equatorially}$=10.6-yr at ≥99%-significance, and $P_{southerly}$=10.0-yr at ≥99%-significance, all three at a staggering ~20%var (global dynamics energy budget). These values agree with those from Ulysses, Fig. 3: 11.2-yr, 9.9-yr, and 8.8, respectively, to within |0.2-yr|, |0.7-yr|, and |1.2-yr| respectively, deteriorating in absolute terms progressively away from the apex by |0.5-yr| per hemisphere. This good general agreement from one sparse data record in the spectral band of macroscopic dynamics of the solar core confirms that the core offset towards the south. It also reveals that the internal workings of the Sun project unto the fast polar winds, whose global decadal resonance gets exposed as originating incessantly in the interior, as surmised by Gough (1995). The fast-wind AR is maintained and carried on coherently throughout the heliosphere beyond L1, Fig. 3–b. As shown above, the solar global dynamics can now be tackled using the principles of vibrational analysis of rotating machinery from mechanical engineering. Therefore, let us look into the extracted ≥99%-significant harmonics: $P_{northerly}$/2=5.4-yr (to within 1% of twice the extracted $P_{northerly}$), $P_{northerly}$/5=2.2-yr (the same as five times the extracted $P_{northerly}$), $P_{northerly}$/6=1.8-yr (the same as six times the extracted $P_{northerly}$), and $P_{northerly}$/10=1.1-yr (the same as ten times the extracted $P_{northerly}$). Other than the extracted $P_{equatorially}$/2=5.3-yr (the same as twice the extracted $P_{equatorially}$) and $P_{southerly}$/4=2.5-yr (the same as four times the extracted $P_{southerly}$), no significant harmonics of any of the three global modes are present. While the equatorial and southerly peak degenerations together by way of their combined partake in overall system variance took away from the same partake of the northerly peak (so its significance went from ≥99% to ≥95%), it is the northerly harmonics which nonetheless dominate the spectral band of core dynamics. This northerly magnetism preference of the solar activity (as represented by group sunspot numbers) confirms that the northern polar magnetic field represents the Sun magnetism most faithfully. As deduced above from the agreement of extraction of the three global modes with those from the Ulysses data, Fig. 3, the northerly preference also means that the high-resolution Ulysses scans of polar winds snapped the Sun's inner macroscopic dynamics down to the solar core, as proxied unto the fast wind. Panel c: spectrum from panel b recomputed with $P_{equatorially}$ and $P_{southerly}$ enforced (mathematically ignored; see Section 2) as 10.58435-yr and 9.99769-yr respectively, along with their reflections that arise as the above periods get enforced, as 11.73171-yr and 3.32744-yr. The ≥99%-significant spectral peaks remaining are $P_{northerly}$ at a stunning 48%var (whose great global dominance follows also from the enormous lobes in the corresponding power spectrum, gray line) along with a core offset-revealing harmonic ½$P_{northerly}$. Also present are a new peculiar spectral peak X at 2.12-yr and its halved reflection ½X at 1.03-yr. The above historical data are from the U.S. NOAA solar indices database (see Data statement).

the fact that the second southerly harmonic is the only missing one of the three possible ones, correctly reflecting a south-wardly offset because such an offset relative to the northern polar and equatorial regions can only reflect on the northerly and equatorial data. Finally, the presence of the fourth harmonic

65

here found for the southern hemisphere only (and therefore highly indicative of the solar core found to sit in relatively closest proximity to the south pole), reveals a cracked rotor if solid, or a density variation including a possible layer in the highly-dense rotating plasma as the solar core's.

Likewise, the absence of the significant third harmonic correctly reveals no bents, i.e., as expected from a gaseous object (a spectral peak with the value close to this harmonic appears in the above spectrum, but as noise: at a negligible <5%var and below any statistical significance). This insignificance indicates that the Sun's core is of a regular geometric shape globally, with perhaps some minor deformations.

Variance-spectral magnitudes represent a relative measure of global field dynamics over a spectral band of interest (Omerbashich, 2003) and can measure variations in global energy budgets of gaseous astronomical bodies (Omerbashich, 2023c). One can thus readily obtain the solar core's wobble period from the normalized spectrum of the 1615–2000 historical global activity in solar-cycle lengths, Fig. 13, as the repetitive change in the variance spectral envelope over the macroscopic band of the solar core dynamics. As seen along the var% axis, the core undergoes about five such highest variations possible (marked with Roman numerals along the dotted line) over one Schwabe cycle. Then the solar core wobbles with a pseudoperiod of about 1/5 the 11.1-yr cycle, or once every ~2.2 yr. The solar-cycle-to-solar-cycle variation in the global energy budget (spectral magnitudes along the var% axis) reveals that it takes about five wobbles (marked I–V) for the core to flip once, where the wobble period lasts on average ~2.2 yr, which agrees well with the unknown period X = 2.12 yr extracted in the historical record of group sunspot numbers, Fig. 12. The here extracted value of ~2.2 yr agrees with the previously established global observation (confirmed over the solar cycles 21–23) that the maximum rate of CMEs lags ~2 yr behind the peak occurrence of sunspots (Ramesh, 2010), meaning that the core wobble moderates the occurrence of both sunspots and CMEs.

We saw thus far that AR is due to the Sun's lowest-frequencies (highest-energies) global dynamic, Figs. 8–10. The only mechanism capable of generating and maintaining AR at those energy levels is an excentral core via its consequent wobble. While unspecified (whether prograde or retrograde), the wobbling is simple because of the Sun's dominance in its stellar system. Multiple centrally slightly offset yet band-wide spectral symmetries of solar activity, represented by sunspot and calcium numbers, Figs. 11 & 12, confirmed the existence of the core wobble and a global revolving-field magnetic alternator at work at all times, rather than just a star whose polarity alternates impulsively. Figs. 13 helped deduce the wobble period as ~2.2-yr. At the same, the solar core turned out to be a double nuisance for global dynamics: first, by its wobbling giving rise to AR, and secondly, by preferential proximity of the core to the south pole and consequently adding to turbulence in the magnetism there. In addition, unlike automobile motors or jet engines, the Sun lacks a cage and thereby freely resonates. Then, instead of classically believed primordial rotation of the core, the core-wobble-driven incessant AR likely gives rise to the global spin of the rest of the Sun from the core itself (as its weeklong rapid rotation) and up to the photosphere (as its monthlong rotation). The solar wind (correctly identified as the fast wind, with the 11.2-yr northerly global mode of ejection,

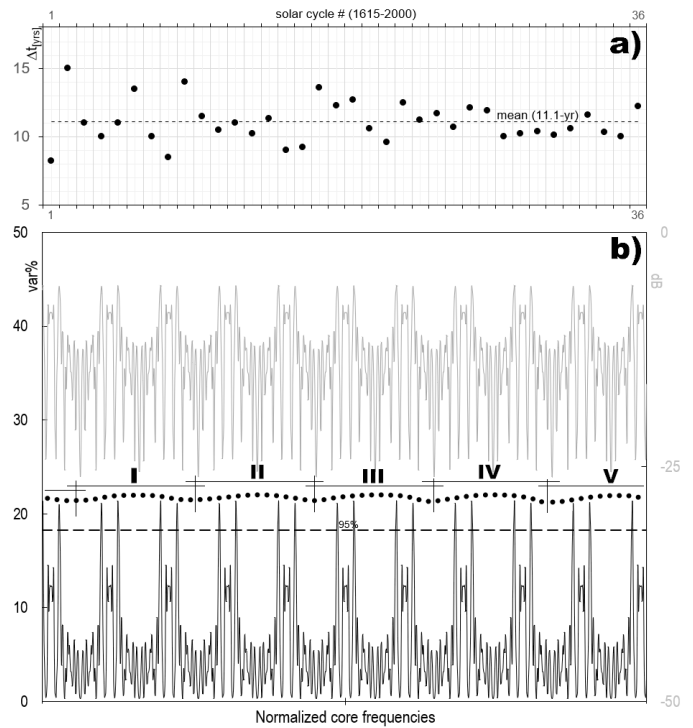

**Figure 13**. GVSA spectra (panel b) of 385 years of historical solar activity (panel a), represented by variation in the length of the solar cycle Δt (in Earth years), taken by year of solar minimum, panel a. By convention, the solar-cycle count usually starts at the year 1755 solar maximum, i.e., after discarding the data from before 1750 as supposedly less reliable (which was an old view, here now refuted for the record of solar cycles durations, as basically due to a lack of an apt spectral analysis technique before GVSA). The data spanned the 36 consecutive solar cycles between 1615–2000. Respective sunspot counts with time are plotted in Fig. 11. Frequencies normalized to the solar-cycles space, i.e., expressed in cycles per solar cycle (spectral band 0.09009–1.17117 c.p.s.c., as the 1–13-years band normalized to the 11.1-yr mean cycle length or one Schwabe period), and represented in cycles per century (c.p.c.) for consistency. A comparison with the rotating-machinery vibrations analysis confirms that the Sun acts as a magnetic alternator (revolving-field engine), e.g., the spectra of open-rotor engine noise at normalized tones, Figs. 16 & 21 of Czech and Thomas (2013). Here, only the extraction of repetitions in spectral magnitude variation (the number of times the spectral envelope protruded out by one full mid-ordinate) is of interest, so the frequency normalization had no bearing on the extraction of periodicities. The above historical data are from the U.S. NOAA solar indices database (see Data statement).

Fig. 3-b) is subsequently released into the heliosphere beyond L1, also at waves guided highly (flappingly) coherently by the whole Sun, Figs. 9 & 10.

Fig. 14 and Table 5 depict the success in spectrally detecting AR and the solar core together with the telling undertones and eccentricity, all from the fast polar winds alone. Thus, a comparison of Table 5 vs. Table 3 reveals that the real-Sun data are characterized by significantly more 1%-matches (Table 3) to their theoretical counterparts than the ideal-Sun data are (Table 5), while the situation reverses for 1‰-matches in favor of the ideal-Sun data. This swapping of places between accuracy and precision in the resemblance of theoretical resonances was as expected, given that the real-Sun data depict our star more realistically and are more accurate overall. On the other hand, the ideal-Sun data paint our star less realistically (they ignore the slow polar winds and, by extension, turbulence) but more precisely, i.e., more correctly to what the Sun is trying to be while struggling with AR as the byproduct of the core wobbling.

66

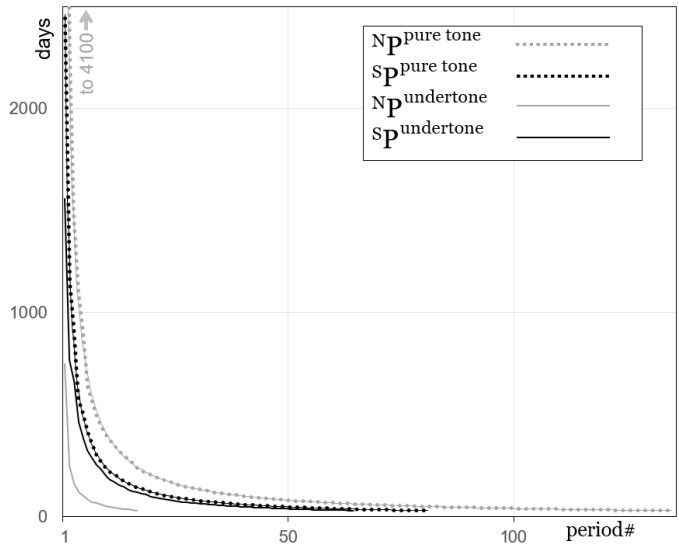

**Figure 14**. Success plot of detecting the solar core (eccentricity) and ≥99%-significant AR periods P from the Northerly and Southerly fast polar winds, Table 5, as relative completeness vs. incompleteness of extracted *pure tones* (the detected signal purified by omitting undertones). As noticed from the fast northerly polar wind, the 4100-day (11.2-yr) Schwabe global equilibrium mode guides AR harmonics extracted to the order u=136 (dotted gray line). With just 17 undertone periods (standalone solid gray line), these superharmonics compose a virtually complete (unobstructed) AR train, which is also represented nearly perfectly with the power-law trend (solid gray line over dotted gray line). On the other hand, AR from the fast southerly polar wind has no guiding global mode and got extracted to u=81 (dotted black line), with a staggering 65 undertone periods (standalone solid black line), i.e., virtually in a 1:1 ratio of tones vs. undertones. This AR train is also represented nearly perfectly with the power-law trend (solid black line over dotted black line). As undertone resonances reveal the presence of a solid or rigid surface in the deep interior, which was deduced in the present study to be the solar core, the above S–N coupling is another confirmation of the core's eccentricity towards the south. Furthermore, the southerly undertone developed in the close vicinity of the tone itself, indicating its relative proximity to the south geographical pole, and revealing the extent (significance) of interference that the solar core causes to the southern polar region.

Finally, Table 6 unveils progressively offset yet persistently (i.e., regardless of different resolutions of the two antipodal trains) mutually matched entire segments of the southerly and northerly antiresonance trains. This general mutual matching of the two antipodal data sets requires the highest dynamic-energy levels to maintain it on decadal time scales, thus revealing that both polar antiresonance trains share a common spatial origin (a surface or a mechanism acting as a surface). Due to the mutually antipodal geographical origin of the data those trains reflect, that shared interface can only be the solar core and its macroscopic dynamics. This global matching means that AR is primarily due to the core wobble. Note here that the higher resolution in resolving the northerly vs. southerly antiresonance trains, or u=153 vs. u=147 in favor of the former (where the same preference is evident from Table 1 as well), is yet another confirmation of the starting physical hypothesis on the former being a more faithful representative of the Sun's global dynamics than the latter. The same conclusion also follows from the signal improvement (purification) after discarding slow polar winds, as seen from comparing the increase in the northerly resonance's resolution from n=131 to n=136 and the (then expected) drop in the southerly resonance's resolution from n=118 to n=81, Table 5 vs. Table 1. Likewise, a significant increase in both antiresonances' resolution after the signal purification (see Table 6 vs. Table 2), confirmed that the fast winds represent the Sun more faithfully than slow winds alone or when part of mixed winds.

In addition to the wobbling period agreeing with it, the shortest global mode period from the Ulysses data, 3224-day (8.83-yr), Table 1 (real Sun), and the longest, 4745-day (12.99-yr), Table 6 (ideal Sun) stand in excellent agreement with the limits of one *Schwabe cycle*, of $P_S \in [9\text{ yr}, 13\text{ yr}]$ (Schwabe, 1844). Limit-cycle oscillations at resonances are a nonlinear phenomenon used in mechanical engineering to model the self-damping of self-excited systems (Hagedorn et al., 2014), such as the Sun under the wobbling core scenario. Limit cycles are inherent to sustained (neither decaying nor growing) vibrations, such as those observed here for the Sun (Figs. 2–4, 8–10), and can be observed directly in the phase plane, e.g., Hellevik and Gudmestad (2017). This global agreement reveals thus that ~13-yr is the primary global natural mode of vibration the Sun attempts to vibrate at (and would succeed if the solar core were perfectly centered). Then, ~11-yr is the vibration equilibrium mode attained due to the global decadal vibration incessantly getting damped under the core wobble and the resulting AR, but never dying out. The rest of the Sun acts as the propagation medium for AR and thereby as a (to the core external) restoring forcer. Note that because the Ulysses data spanned about 13 years and nine months, it was not possible to plausibly extend the band of interest to periods >13-yr to test the estimated (least-squares-normed) 12.99-yr period additionally. However, in their reanalysis of historical records of the Earth's polar motion, An et al. (2023) recently reported both the ~13-yr lower-limit of one Schwabe cycle's interval and its ~11-yr equilibrium period along with the upper-limit 8.5±0.2 yr but which they interpreted by elimination as the Earth inner core's wobble period. The apparently reliable computations of *ibid*. thus provide an independent confirmation of the above extraction of the ~12.99-yr mode as indeed correct, but also expose their interpretation of the ~8.5-yr period for the Earth's inner core wobble as most likely incorrect since the latter is just a natural damping limit of the Sun's global incessant vibration, observed first as the well-known solar activity interval by Schwabe (1844).

This section concludes the proof based on various spatially and temporally independent real-data sets that the Sun acts as a classical revolving-field magnetic-alternator motor instead of a dynamo or proverbial engine, as believed previously and simplistically by some. The global solar vibration, which naturally varies (both resonates and antiresonates) due to the general southward eccentricity of the core combined with differential rotations and properties of the northern vs. southern hemisphere vs. equatorial belt, was shown to give rise to cyclic flipping of the solar core at a period from the Schwabe interval. The core flipping, mostly occurring at the ~11-yr equilibrium mode of the self-sustained and self-damped global decadal vibration, makes the Sun a classical revolving-field magnetic alternator well known from mechanical engineering. But unlike in machinery engines, the Sun lacks a restraining cage, resulting in a constant emission of its excess mass as high-speed magnetized solar winds whose highly structured dynamics dramatically affect the entire solar system and its objects' dynamics.

| u | $^SP_u^{pure}$ | $^SP_u^{pure'}$ | $^S\Delta_u[\%]$ | $^NP_u^{pure}$ | $^NP_u^{pure'}$ | $^N\Delta_u[\%]$ | $^SP_u^{und}$ | $^NP_u^{und}$ | u | $^SP_u^{pure}$ | $^SP_u^{pure'}$ | $^S\Delta_u[\%]$ | $^NP_u^{pure}$ | $^NP_u^{pure'}$ | $^N\Delta_u[\%]$ | $^SP_u^{und}$ | u | $^NP_u^{pure}$ | $^NP_u^{pure'}$ | $^N\Delta_u[\%]$ |
|---|---|---|---|---|---|---|---|---|---|---|---|---|---|---|---|---|---|---|---|---|
| 1 | 2440.9 |  |  | 4100.0 |  |  | 1558.2 | 747.4 | 51 | 47.8 | 47.9 | 0.2 | 80.1 | 80.4 | 0.3 | 38.6 | 101 | 40.5 | 40.6 | 0.1 |
| 2 | 1144.3 | 1220.5 | 6.2 | 2258.2 | 2050.0 | -10.2 | 766.3 | 250.6 | 52 | 46.7 | 46.9 | 0.4 | 78.9 | 78.8 | 0.0 | 37.9 | 102 | 40.1 | 40.2 | 0.1 |
| 3 | 932.1 | 813.6 | -14.6 | 1481.6 | 1366.7 | -8.4 | 665.0 | 161.0 | 53 | 46.2 | 46.1 | -0.4 | 77.7 | 77.4 | -0.4 | 37.2 | 103 | 39.8 | 39.8 | 0.0 |
| 4 | 576.1 | 610.2 | 5.6 | 1102.5 | 1025.0 | -7.6 | 461.5 | 122.4 | 54 | 45.3 | 45.2 | -0.3 | 76.3 | 75.9 | -0.5 | 36.6 | 104 | 39.5 | 39.4 | 0.0 |
| 5 | 508.1 | 488.2 | -4.1 | 904.2 | 820.0 | -10.3 | 384.9 | 101.1 | 55 | 44.4 | 44.4 | 0.0 | 75.1 | 74.5 | -0.8 | 35.7 | 105 | 39.2 | 39.0 | -0.1 |
| 6 | 416.8 | 406.8 | -2.5 | 636.9 | 683.3 | 6.8 | 326.6 | 84.6 | 56 | 43.6 | 43.6 | 0.1 | 73.9 | 73.2 | -0.9 | 35.1 | 106 | 38.8 | 38.7 | -0.1 |
| 7 | 353.4 | 348.7 | -1.3 | 565.3 | 585.7 | 3.5 | 286.3 | 72.8 | 57 | 42.7 | 42.8 | 0.3 | 71.8 | 71.9 | 0.2 | 34.6 | 107 | 38.2 | 38.3 | 0.2 |
| 8 | 306.6 | 305.1 | -0.5 | 499.7 | 512.5 | 2.5 | 254.8 | 65.7 | 58 | 42.3 | 42.1 | -0.5 | 70.6 | 70.7 | 0.2 | 34.1 | 108 | 37.9 | 38.0 | 0.1 |
| 9 | 270.8 | 271.2 | 0.1 | 447.8 | 455.6 | 1.7 | 229.6 | 58.2 | 59 | 41.5 | 41.4 | -0.4 | 69.6 | 69.5 | -0.2 | 33.6 | 109 | 37.5 | 37.6 | 0.1 |
| 10 | 240.6 | 244.1 | 1.4 | 405.6 | 410.0 | 1.1 | 199.3 | 52.2 | 60 | 40.7 | 40.7 | -0.1 | 68.5 | 68.3 | -0.2 | 32.8 | 110 | 37.3 | 37.3 | 0.0 |
| 11 | 218.0 | 221.9 | 1.8 | 375.3 | 372.7 | -0.7 | 182.4 | 47.4 | 61 | 40.0 | 40.0 | 0.1 | 67.6 | 67.2 | -0.5 | 32.3 | 111 | 36.9 | 36.9 | 0.0 |
| 12 | 207.5 | 203.4 | -2.0 | 345.3 | 341.7 | -1.1 | 169.1 | 43.9 | 62 | 39.6 | 39.4 | -0.6 | 66.5 | 66.1 | -0.6 | 31.9 | 112 | 36.7 | 36.6 | -0.1 |
| 13 | 190.5 | 187.8 | -1.4 | 319.6 | 315.4 | -1.4 | 156.8 | 40.8 | 63 | 38.9 | 38.7 | -0.4 | 64.7 | 65.1 | 0.6 | 31.4 | 113 | 36.4 | 36.3 | -0.1 |
| 14 | 175.0 | 174.4 | -0.4 | 300.5 | 292.9 | -2.6 | 146.2 | 38.5 | 64 | 38.2 | 38.1 | -0.2 | 63.9 | 64.1 | 0.3 | 31.0 | 114 | 36.1 | 36.0 | -0.1 |
| 15 | 162.7 | 162.7 | 0.0 | 280.9 | 273.3 | -2.8 | 133.2 | 35.5 | 65 | 37.5 | 37.6 | 0.0 | 62.9 | 63.1 | 0.2 | 30.5 | 115 | 35.8 | 35.7 | -0.2 |
| 16 | 151.3 | 152.6 | 0.8 | 263.7 | 256.2 | -2.9 | 126.0 | 33.7 | 66 | 36.9 | 37.0 | 0.2 | 62.1 | 62.1 | 0.0 |  | 116 | 35.3 | 35.3 | 0.1 |
| 17 | 142.0 | 143.6 | 1.1 | 236.8 | 241.2 | 1.8 | 119.0 | 31.0 | 67 | 36.3 | 36.4 | 0.3 | 61.3 | 61.2 | -0.1 |  | 117 | 35.0 | 35.0 | 0.1 |
| 18 | 137.5 | 135.6 | -1.4 | 224.5 | 227.8 | 1.4 | 112.8 |  | 68 | 36.0 | 35.9 | -0.3 | 60.5 | 60.3 | -0.4 |  | 118 | 34.7 | 34.7 | 0.0 |
| 19 | 129.2 | 128.5 | -0.6 | 214.9 | 215.8 | 0.4 | 107.6 |  | 69 | 35.4 | 35.4 | -0.1 | 59.7 | 59.4 | -0.4 |  | 119 | 34.5 | 34.5 | 0.0 |
| 20 | 122.4 | 122.0 | -0.3 | 204.7 | 205.0 | 0.2 | 99.4 |  | 70 | 34.9 | 34.9 | 0.0 | 59.0 | 58.6 | -0.7 |  | 120 | 34.2 | 34.2 | -0.1 |
| 21 | 115.8 | 116.2 | 0.3 | 195.4 | 195.3 | -0.1 | 95.0 |  | 71 | 34.3 | 34.4 | 0.1 | 57.4 | 57.7 | 0.6 |  | 121 | 34.0 | 33.9 | -0.1 |
| 22 | 110.3 | 111.0 | 0.6 | 188.1 | 186.4 | -0.9 | 91.0 |  | 72 | 33.8 | 33.9 | 0.3 | 56.8 | 56.9 | 0.3 |  | 122 | 33.5 | 33.6 | 0.1 |
| 23 | 105.0 | 106.1 | 1.1 | 180.2 | 178.3 | -1.1 | 85.8 |  | 73 | 33.3 | 33.4 | 0.4 | 56.0 | 56.2 | 0.2 |  | 123 | 33.2 | 33.3 | 0.1 |
| 24 | 102.1 | 101.7 | -0.4 | 174.0 | 170.8 | -1.8 | 82.6 |  | 74 | 33.1 | 33.0 | -0.2 | 55.4 | 55.4 | 0.0 |  | 124 | 33.0 | 33.1 | 0.1 |
| 25 | 97.2 | 97.6 | 0.5 | 167.2 | 164.0 | -2.0 | 79.7 |  | 75 | 32.6 | 32.5 | -0.1 | 54.7 | 54.7 | -0.1 |  | 125 | 32.7 | 32.8 | 0.1 |
| 26 | 93.0 | 93.9 | 1.0 | 156.0 | 157.7 | 1.1 | 75.5 |  | 76 | 32.1 | 32.1 | 0.0 | 54.1 | 53.9 | -0.3 |  | 126 | 32.5 | 32.5 | 0.0 |
| 27 | 89.4 | 90.4 | 1.1 | 150.5 | 151.9 | 0.9 | 73.1 |  | 77 | 31.7 | 31.7 | 0.1 | 53.4 | 53.2 | -0.4 |  | 127 | 32.3 | 32.3 | 0.0 |
| 28 | 87.6 | 87.2 | -0.5 | 146.2 | 146.4 | 0.2 | 70.7 |  | 78 | 31.2 | 31.3 | 0.3 | 52.9 | 52.6 | -0.6 |  | 128 | 32.1 | 32.0 | 0.0 |
| 29 | 84.2 | 84.2 | 0.0 | 141.4 | 141.4 | 0.0 | 68.6 |  | 79 | 30.8 | 30.9 | 0.4 | 51.7 | 51.9 | 0.4 |  | 129 | 31.8 | 31.8 | -0.1 |
| 30 | 81.2 | 81.4 | 0.2 | 137.5 | 136.7 | -0.6 | 66.5 |  | 80 | 30.4 | 30.5 | 0.5 | 51.1 | 51.2 | 0.3 |  | 130 | 31.6 | 31.5 | -0.1 |
| 31 | 78.3 | 78.7 | 0.6 | 133.2 | 132.3 | -0.7 | 63.7 |  | 81 | 30.1 | 30.1 | -0.1 | 50.6 | 50.6 | 0.1 |  | 131 | 31.4 | 31.3 | -0.1 |
| 32 | 76.9 | 76.3 | -0.8 | 129.2 | 128.1 | -0.9 | 61.9 |  | 82 |  |  |  | 50.0 | 50.0 | 0.0 |  | 132 | 31.2 | 31.1 | -0.1 |
| 33 | 74.4 | 74.0 | -0.6 | 126.0 | 124.2 | -1.4 | 60.3 |  | 83 |  |  |  | 49.5 | 49.4 | -0.2 |  | 133 | 30.8 | 30.8 | 0.0 |
| 34 | 71.9 | 71.8 | -0.2 | 119.5 | 120.6 | 0.9 | 58.6 |  | 84 |  |  |  | 48.9 | 48.8 | -0.3 |  | 134 | 30.6 | 30.6 | 0.0 |
| 35 | 69.8 | 69.7 | 0.0 | 116.3 | 117.1 | 0.7 | 56.3 |  | 85 |  |  |  | 48.4 | 48.2 | -0.3 |  | 135 | 30.4 | 30.4 | 0.0 |
| 36 | 67.6 | 67.8 | 0.3 | 113.7 | 113.9 | 0.2 | 54.9 |  | 86 |  |  |  | 47.9 | 47.7 | -0.5 |  | 136 | 30.2 | 30.1 | 0.0 |
| 37 | 65.7 | 66.0 | 0.5 | 110.7 | 110.8 | 0.1 | 53.5 |  | 87 |  |  |  | 47.0 | 47.1 | 0.4 |  |  |  |  |  |
| 38 | 64.7 | 64.2 | -0.7 | 108.4 | 107.9 | -0.4 | 52.3 |  | 88 |  |  |  | 46.4 | 46.6 | 0.3 |  |  |  |  |  |
| 39 | 62.8 | 62.6 | -0.3 | 105.7 | 105.1 | -0.5 | 50.5 |  | 89 |  |  |  | 46.0 | 46.1 | 0.1 |  |  |  |  |  |
| 40 | 61.1 | 61.0 | -0.2 | 103.5 | 102.5 | -1.0 | 49.3 |  | 90 |  |  |  | 45.6 | 45.6 | -0.1 |  |  |  |  |  |
| 41 | 59.4 | 59.5 | 0.1 | 99.1 | 100.0 | 0.9 | 48.3 |  | 91 |  |  |  | 45.1 | 45.1 | -0.2 |  |  |  |  |  |
| 42 | 57.9 | 58.1 | 0.5 | 96.9 | 97.6 | 0.8 | 47.2 |  | 92 |  |  |  | 44.7 | 44.6 | -0.4 |  |  |  |  |  |
| 43 | 57.1 | 56.8 | -0.6 | 95.0 | 95.3 | 0.3 | 45.8 |  | 93 |  |  |  | 44.3 | 44.1 | -0.6 |  |  |  |  |  |
| 44 | 55.6 | 55.5 | -0.2 | 93.0 | 93.2 | 0.2 | 44.9 |  | 94 |  |  |  | 43.5 | 43.6 | 0.3 |  |  |  |  |  |
| 45 | 54.2 | 54.2 | 0.1 | 91.3 | 91.1 | -0.2 | 43.9 |  | 95 |  |  |  | 43.1 | 43.2 | 0.1 |  |  |  |  |  |
| 46 | 53.0 | 53.1 | 0.2 | 89.4 | 89.1 | -0.3 | 43.1 |  | 96 |  |  |  | 42.7 | 42.7 | 0.0 |  |  |  |  |  |
| 47 | 51.7 | 51.9 | 0.4 | 87.8 | 87.2 | -0.7 | 41.9 |  | 97 |  |  |  | 42.3 | 42.3 | -0.2 |  |  |  |  |  |
| 48 | 51.1 | 50.9 | -0.5 | 86.1 | 85.4 | -0.8 | 41.1 |  | 98 |  |  |  | 41.9 | 41.8 | -0.2 |  |  |  |  |  |
| 49 | 49.9 | 49.8 | -0.2 | 83.0 | 83.7 | 0.8 | 40.4 |  | 99 |  |  |  | 41.6 | 41.4 | -0.4 |  |  |  |  |  |
| 50 | 48.8 | 48.8 | 0.1 | 81.7 | 82.0 | 0.4 | 39.3 |  | 100 |  |  |  | 41.2 | 41.0 | -0.6 |  |  |  |  |  |

**Table 5**. Extracted ≥99%-significant *pure tone* periods $P_{pure}$ (AR purified by omitting undertones) and ≥99%-significant undertone periods $P_{und}$, Fig. 14, and their matchings against theoretical-resonance periods $P_i'=P_1/i$, i=2…u; u∈$\varkappa$, per the $P_1$ estimate of $P_S$ from each data set, up to the order u=136. Matchings within ≤1% to respective theoretical resonance periods highlighted light gray, within ≤1‰ dark gray. From southerly-polar data, 11 theoretical-resonance periods did not have a measured match in the ≥99%-significant spectra, and 24 of all ≥99%-significant spectral peaks were not matched by theoretical-resonance periods, as due primarily to large-scale turbulence effects seen in spectra largely as anisotropic peak splitting. From northerly-polar data, 8 and 12, respectively. As seen, 35% of northerly pure-signal periods matched their theoretical counterparts to 1‰ and 48% to 1%, vs. 12% and 65%, respectively, in the real Sun case, Table 3. Southerly, 22% to 1‰ and 59% to 1%, vs. 11% and 70%, respectively, in Table 3.

| $^sP^-_q$ | | | $^sP^-_q$ | | | $^nP^-_q$ | | | $^nP^-_q$ | | |
|---|---|---|---|---|---|---|---|---|---|---|---|
| q | [days] | [var%] | q | [days] | [var%] | q | [days] | [var%] | q | [days] | [var%] |
| 1 | 4100.0 | 19.74 | 76 | 58.2 | 0.53 | 1 | 4745.0 | 14.89 | 76 | 60.0 | 0.51 |
| 2 | 2100.9 | 17.27 | 77 | 57.5 | 0.46 | 2 | 2440.9 | 22.60 | 77 | 59.3 | 0.12 |
| 3 | 1481.6 | 1.13 | 78 | 56.8 | 0.35 | 3 | 1558.2 | 7.95 | 78 | 58.5 | 0.44 |
| 4 | 1102.5 | 3.04 | 79 | 56.0 | 0.26 | 4 | 932.1 | 6.49 | 79 | 57.9 | 0.19 |
| 5 | 877.9 | 5.40 | 80 | 55.3 | 0.19 | 5 | 766.3 | 2.47 | 80 | 57.1 | 0.38 |
| 6 | 747.4 | 7.04 | 81 | 54.6 | 0.13 | 6 | 665.0 | 6.14 | 81 | 56.4 | 0.29 |
| 7 | 636.9 | 1.64 | 82 | 53.9 | 0.08 | 7 | 576.1 | 1.39 | 82 | 55.7 | 0.32 |
| 8 | 554.9 | 1.36 | 83 | 53.3 | 0.05 | 8 | 516.8 | 6.29 | 83 | 55.1 | 0.42 |
| 9 | 491.6 | 2.72 | 84 | 52.6 | 0.03 | 9 | 461.5 | 1.41 | 84 | 54.4 | 0.31 |
| 10 | 441.2 | 4.87 | 85 | 52.0 | 0.02 | 10 | 416.8 | 6.61 | 85 | 53.8 | 0.58 |
| 11 | 405.6 | 2.00 | 86 | 51.3 | 0.03 | 11 | 384.9 | 1.76 | 86 | 53.2 | 0.33 |
| 12 | 370.7 | 1.29 | 87 | 50.7 | 0.05 | 12 | 353.4 | 3.61 | 87 | 52.5 | 0.68 |
| 13 | 341.4 | 2.13 | 88 | 50.2 | 0.10 | 13 | 330.1 | 2.31 | 88 | 52.0 | 0.38 |
| 14 | 316.3 | 4.22 | 89 | 49.7 | 0.10 | 14 | 306.6 | 2.23 | 89 | 51.3 | 0.59 |
| 15 | 297.6 | 2.24 | 90 | 49.1 | 0.05 | 15 | 289.0 | 3.07 | 90 | 50.8 | 0.45 |
| 16 | 278.3 | 1.44 | 91 | 48.5 | 0.02 | 16 | 270.8 | 1.63 | 91 | 50.2 | 0.54 |
| 17 | 261.5 | 2.16 | 92 | 48.0 | 0.03 | 17 | 257.0 | 4.12 | 92 | 49.7 | 0.51 |
| 18 | 246.5 | 4.46 | 93 | 47.5 | 0.13 | 18 | 242.5 | 1.48 | 93 | 49.2 | 0.53 |
| 19 | 235.0 | 2.41 | 94 | 47.0 | 0.21 | 19 | 229.6 | 4.35 | 94 | 48.7 | 0.54 |
| 20 | 222.8 | 1.67 | 95 | 46.5 | 0.13 | 20 | 219.6 | 1.70 | 95 | 48.2 | 0.52 |
| 21 | 211.9 | 2.50 | 96 | 46.0 | 0.10 | 21 | 208.9 | 3.03 | 96 | 47.7 | 0.53 |
| 22 | 203.3 | 4.43 | 97 | 45.5 | 0.13 | 22 | 200.6 | 2.30 | 97 | 47.2 | 0.51 |
| 23 | 194.1 | 2.53 | 98 | 45.1 | 0.22 | 23 | 191.7 | 2.19 | 98 | 46.7 | 0.49 |
| 24 | 185.8 | 2.00 | 99 | 44.7 | 0.29 | 24 | 184.6 | 3.35 | 99 | 46.2 | 0.50 |
| 25 | 178.1 | 3.00 | 100 | 44.2 | 0.24 | 25 | 177.0 | 1.76 | 100 | 45.8 | 0.43 |
| 26 | 172.0 | 4.11 | 101 | 43.8 | 0.25 | 26 | 171.0 | 4.89 | 101 | 45.4 | 0.48 |
| 27 | 165.4 | 2.65 | 102 | 43.3 | 0.30 | 27 | 164.5 | 1.73 | 102 | 44.9 | 0.36 |
| 28 | 159.3 | 2.48 | 103 | 42.9 | 0.37 | 28 | 158.4 | 3.74 | 103 | 44.5 | 0.33 |
| 29 | **153.6** | 3.45 | 104 | 42.5 | 0.41 | 29 | **153.6** | 2.13 | 104 | 44.1 | 0.32 |
| 30 | 149.0 | 3.73 | 105 | 42.1 | 0.38 | 30 | 148.3 | 2.77 | 105 | 43.7 | 0.22 |
| 31 | **144.1** | 2.92 | 106 | 41.7 | 0.37 | 31 | **144.1** | 2.93 | 106 | 43.3 | 0.20 |
| 32 | **139.4** | 3.04 | 107 | 41.3 | 0.35 | 32 | **139.4** | 2.19 | 107 | 42.9 | 0.18 |
| 33 | 135.0 | 3.73 | 108 | 40.9 | 0.34 | 33 | 135.6 | 3.99 | 108 | 42.5 | 0.11 |
| 34 | **131.5** | 3.61 | 109 | 40.5 | 0.31 | 34 | **131.5** | 2.00 | 109 | 42.2 | 0.12 |
| 35 | **127.6** | 3.30 | 110 | 40.1 | 0.27 | 35 | **127.6** | 4.16 | 110 | 41.8 | 0.08 |
| 36 | 123.9 | 3.41 | 111 | 39.8 | 0.22 | 36 | 124.4 | 2.14 | 111 | 41.4 | 0.05 |
| 37 | 120.5 | 3.65 | 112 | 39.4 | 0.17 | 37 | 121.0 | 3.19 | 112 | 41.0 | 0.11 |
| 38 | 117.7 | 3.65 | 113 | 39.1 | 0.12 | 38 | 118.1 | 2.49 | 113 | 40.7 | 0.03 |
| 39 | 114.5 | 3.48 | 114 | 38.7 | 0.09 | 39 | 115.0 | 2.62 | 114 | 40.4 | 0.12 |
| 40 | 111.6 | 3.40 | 115 | 38.4 | 0.07 | 40 | 112.2 | 2.85 | 115 | 40.0 | 0.03 |
| 41 | 108.7 | 3.28 | 116 | 38.0 | 0.07 | 41 | 109.5 | 2.34 | 116 | 39.7 | 0.11 |
| 42 | 106.1 | 3.11 | 117 | 37.7 | 0.04 | 42 | 107.2 | 3.10 | 117 | 39.3 | 0.03 |
| 43 | 103.2 | 2.69 | 118 | 37.4 | 0.02 | 43 | 104.6 | 2.24 | 118 | 39.0 | 0.13 |
| 44 | 100.8 | 2.23 | 119 | 37.1 | 0.01 | 44 | 102.5 | 3.15 | 119 | 38.6 | 0.04 |
| 45 | 98.4 | 1.78 | 120 | 36.8 | 0.01 | 45 | 100.1 | 2.16 | 120 | 38.4 | 0.18 |
| 46 | 96.3 | 1.38 | 121 | 36.5 | 0.01 | 46 | 97.8 | 2.93 | 121 | 38.0 | 0.04 |
| 47 | 94.1 | 1.03 | 122 | 36.1 | 0.02 | 47 | 95.9 | 2.02 | 122 | 37.7 | 0.23 |
| 48 | 92.1 | 0.77 | 123 | 35.8 | 0.05 | 48 | 93.9 | 2.57 | 123 | 37.4 | 0.06 |
| 49 | 90.2 | 0.58 | 124 | 35.6 | 0.09 | 49 | 92.1 | 1.80 | 124 | 37.1 | 0.20 |
| 50 | 88.4 | 0.48 | 125 | 35.3 | 0.05 | 50 | 90.5 | 2.23 | 125 | 36.8 | 0.11 |
| 51 | 86.6 | 0.49 | 126 | 35.0 | 0.01 | 51 | 88.6 | 1.51 | 126 | 36.5 | 0.17 |
| 52 | 85.1 | 0.59 | 127 | 34.7 | 0.00 | 52 | 87.1 | 1.73 | 127 | 36.2 | 0.18 |
| 53 | 83.5 | 0.39 | 128 | 34.5 | 0.03 | 53 | 85.4 | 1.20 | 128 | 35.9 | 0.15 |
| 54 | 81.9 | 0.27 | 129 | 34.2 | 0.11 | 54 | 83.9 | 1.24 | 129 | 35.7 | 0.28 |
| 55 | 80.4 | 0.26 | 130 | 34.0 | 0.13 | 55 | 82.3 | 0.93 | 130 | 35.4 | 0.16 |
| 56 | 78.9 | 0.41 | 131 | 33.7 | 0.00 | 56 | 81.0 | 0.80 | 131 | 35.1 | 0.38 |
| 57 | 77.7 | 0.63 | 132 | 33.4 | 0.09 | 57 | 79.5 | 0.75 | 132 | 34.9 | 0.19 |
| 58 | 76.3 | 0.38 | 133 | 33.2 | 0.09 | 58 | 78.3 | 0.44 | 133 | 34.6 | 0.31 |
| 59 | 75.0 | 0.29 | 134 | 32.9 | 0.07 | 59 | 76.9 | 0.68 | 134 | 34.3 | 0.25 |
| 60 | 73.7 | 0.41 | 135 | 32.7 | 0.04 | 60 | 75.7 | 0.29 | 135 | 34.1 | 0.28 |
| 61 | 72.4 | 0.74 | 136 | 32.5 | 0.05 | 61 | 74.6 | 0.40 | 136 | 33.8 | 0.33 |
| 62 | 71.4 | 0.56 | 137 | 32.2 | 0.11 | 62 | 73.3 | 0.22 | 137 | 33.6 | 0.27 |
| 63 | 70.2 | 0.42 | 138 | 32.0 | 0.12 | 63 | 72.3 | 0.18 | 138 | 33.4 | 0.41 |
| 64 | 69.1 | 0.50 | 139 | 31.8 | 0.10 | 64 | 71.1 | 0.25 | 139 | 33.1 | 0.28 |
| 65 | 68.0 | 0.74 | 140 | 31.5 | 0.10 | 65 | 70.1 | 0.08 | 140 | 32.8 | 0.41 |
| 66 | 67.1 | 0.73 | 141 | 31.3 | 0.13 | 66 | 69.0 | 0.34 | 141 | 32.6 | 0.30 |
| 67 | 66.1 | 0.58 | 142 | 31.1 | 0.17 | 67 | 68.0 | 0.05 | 142 | 32.4 | 0.37 |
| 68 | 65.1 | 0.58 | 143 | 30.9 | 0.16 | 68 | 67.1 | 0.34 | 143 | 32.2 | 0.31 |
| 69 | 64.1 | 0.68 | 144 | 30.7 | 0.16 | 69 | 66.1 | 0.05 | 144 | 31.9 | 0.33 |
| 70 | 63.2 | 0.85 | 145 | 30.5 | 0.16 | 70 | 65.2 | 0.07 | 145 | 31.7 | 0.31 |
| 71 | 62.4 | 0.69 | 146 | 30.2 | 0.17 | 71 | 64.3 | 0.06 | 146 | 31.5 | 0.31 |
| 72 | 61.5 | 0.59 | 147 | 30.0 | 0.17 | 72 | 63.4 | 0.35 | 147 | 31.3 | 0.29 |
| 73 | 60.6 | 0.55 | | | | 73 | 62.5 | 0.07 | 148 | 31.1 | 0.28 |
| 74 | 59.8 | 0.53 | | | | 74 | 61.8 | 0.43 | 149 | 30.9 | 0.25 |
| 75 | 59.0 | 0.53 | | | | 75 | 60.9 | 0.09 | 150 | 30.7 | 0.25 |
| | | | | | | | | | 151 | 30.5 | 0.20 |
| | | | | | | | | | 152 | 30.3 | 0.22 |
| | | | | | | | | | 153 | 30.1 | 0.15 |

**Table 6**. GVSA-extracted (measured) ≥99%-significant periods (troughs) of the ideal Sun's global Alfvén antiresonance, Figs. 9 & 10. Compared to Table 2, the Rieger period $P_{Rg}$ after discarding the slow polar winds went from 154.4-day to 153.6-day (simple mean: 154.0-day). The analysis sensitivity increased, so $P_{Rg}$ of the ideal Sun is part of a batch with four more identical or twin periods (highlighted black) from the northerly and southerly fast winds. Furthermore, additional twin periods (highlighted gray) appear from there on in the two antipodal antiresonances trains, and mostly also in batches albeit at progressively shifted orders: the second batch shifted by one order, the third by two, and so on until the seventh batch that shifted by six orders. Note here that twin periods were also obtained from various data sets in Table 1, but only in pairs, i.e., never in batches. The batching only from the fast winds reveals the true nature of the Rieger period, i.e., as solar in origin; in turn, this is another confirmation that the fast winds indeed represent the Sun as if it were a well-tuned, balanced revolving-field magnetoalternator with a perfectly centered rotator (core).

## 6. DISCUSSION

The traditionally favored model for the genesis of the Sun magnetism is that of the dynamo, a simple engine for converting kinetic into electric energy that naturally gives rise to magnetism (Solanki et al., 2006). However, based on recent global-dynamical observations of brown dwarfs that lack a core but still exhibit the same magnetic patterns as the Sun, that mechanism could also be purely convectional (Route, 2016). At the same time, based on sunspot activity observed in the past as able to shut down entirely per hemisphere, the generation mechanism could also be hemispherical (Grote & Busse, 2000). Furthermore, early global computations do not support the premise that dynamo models could account for global vibrations of the Sun on the order of years (Stenflo & Vogel, 1986). The Ulysses mission, as the only one ever to have scanned the hemispheric magnetic fields in flybys over the polar regions, has confirmed that magnetism in the two polar regions of the Sun differs significantly both between them and from the equatorial magnetism. Ulysses has also directly observed the flipping of the solar magnetism polarity every ~11 yr.

Heliophysics today faces two still unresolved problems of large-scale dynamics. One is **solar abundance**, i.e., a disagreement of standard stellar models (SSM) of evolution/interior for the Sun with the latest helioseismology data from the early 2000s (Bergemann & Serenelli, 2014). This discrepancy is likely due to some physical process not yet accounted for in standard solar models and could entail mixing by mechanical waves below the convection zone and down into the radiative interior and the core (Asplund et al., 2009). At the same time, although the current state of solar magnetic fields is the primary indicator of the Sun's global activity, SSMs, unfortunately, do not account for internal magnetic fields at all (Bergemann & Serenelli, 2014). The second problem is the **million-degree corona**, i.e., the overheating of the outer envelope of the Sun, to >$10^6$ °C (as opposed to a relatively much cooler photospheric/surface temperature, at 5.5·$10^3$ °C), involving millions of small-scale magnetic reconnection events of impulsive heating, called nanoflares, of up to several thousand km in size and constantly firing everywhere on the Sun (Parker, 1988).

In theory, this immense heat surplus in the corona could be explained as the dissipative energy from a buildup of a large-amplitude resonance vibration of small-scale waves whenever the global wave frequency equals the local Alfvén wave frequency through the process of resonance absorption (Davila, 1987). The essential feature of that theory of the Sun and other late-type stars — under which each magnetic tube is a driven, high-quality resonator — is the existence of the global mode and its subsequent coupling to the small-scale dissipative waves (*ibid.*). While spectra of the Sun's p-mode (pressure-force-restoring-) and g-mode (gravity-force-restoring-) vibrations with periods of the order of 1-h or longer have been extracted entirely in heliophysics, the remaining a-mode (global; acoustically-and-magnetic-force-restoring-) vibrations spectra have not. Crude estimates indicate its periods are much longer than those of the p or g modes, typically in the range of years and inexplicable by dynamo theory (Stenflo & Vogel, 1986). In addition, since the acceleration of the (fast) solar wind and the heating of the solar corona occur in essentially the same region, the underlying mechanisms of the two phenomena may be strongly linked (Grail et al., 1996) and due to the damping of Alfvén waves (Markovskii et al., 2009). As mentioned earlier, this parallel is especially significant because the solar wind is a physical system characterized by multi-scale evolution. Since all SSMs are calibrated on the Sun and are routinely used to interpret any other star or stellar populations, completing our global knowledge of the Sun by explaining the above problems would have very large implications for our understanding of stars and galaxies in general (Bergemann & Serenelli, 2014).

Fig. 3 revealed the signature of the Sun's complete and incessant resonance in the 0.01–1.60 ZeV ($\sim$1.6·10$^7$ –2.5·10$^9$ erg) band of the solar wind's extreme (mechanical flapping) energies. This result was as expected of a magnetoalternator engine at work, with the separation of the *solar engine* (left frame) from the *Rieger resonance offshoot* subband (right frame) in which mostly macroscale phenomena of relevance for planetary geodynamics occur, see, e.g., Omerbashich (2023b). While the northerly wind preserves the shaking modality the best, exhibiting AR even at L1, as seen in the ≲50-nT 2004–2021 WIND data, the equatorially mixed (mainly slow) wind carries only a faint AR signature. Thus, the northerly polar fast wind reveals its $P_S$ global driver the most accurately and precisely. At the same time, the Rieger period in MMF turns out to be swamped by turbulence down to the $\sim$67%-significance, making $P_{Rg}$ a feature of the northerly wind primarily and, to a lesser degree, the southerly wind, which winds then transmit the $P_{Rg}$ as folded and into the heliosphere. Besides, AR becomes complete (preceded all the time by companion antiresonance modes) only when the northerly polar wind data alone are analyzed, Fig. 3–e. In addition, the RR band begins (AR band ends) at or around the antiresonances termination frequency, which on the Sun turns out to be around the Earth-annual periodicity. That frequency is when the resonating system becomes decoupled but still not relaxed, and turbulences can give in to external planetary gravitation and magnetism so that, already at the lowest-frequencies-RR, the globally prominent anisotropy (seen as spectral peak splitting) yields as well. As the energy breakdown point of a coupled system, the anti-resonances termination at $\sim$1-yr and its ½ harmonic are also sources of confusion, i.e., seasonal (annual and semiannual) periodicities often seen inexplicably in various records of planetary including terrestrial data, which empirically is why the 180–365-day subband is best to ignore — an approach adopted in the present study as well.

As indicated in Fig. 2–d & e and confirmed in Fig. 3–d & e, AR modulates anisotropy instead of the other way around. The modulation is across all AR harmonics, so the entire Sun (characterized by anisotropy due to turbulence and overall instability of its magnetic fields) — and not just some of its belts or layers — is kept under the AR regime. Since a vibration magnification occurs in such closed and damped vibrating systems naturally via frequency demultiplication that upsurges the energy injected resonantly into such a physical system by 100s of times, e.g., Den Hartog (1985), we can expect a tremendous increase in dissipated heat to follow as well. For example, previously observed high-frequency ($\sim$12–42 mHz) torsional oscillations in the quiet Sun were claimed by Srivastava et al. (2017), based on modeling, to be torsional Alfvén waves that transfer $\sim$10$^3$ W·m$^{-2}$ energy into the overlying corona, thereby heating it and facilitating the creation of the solar wind. However, those authors failed to provide an overlying mechanism for the global sustainment of such localized energy transfers, as different kinds of modeling reveal that such local transfers are insufficient to balance the Sun energy loss, e.g., by Soler et al. (2021). At the same time, the $\sim$1.6·10$^7$–2.5·10$^9$ erg base energies of the global mechanism found herein can multiply $\sim$10$^2$-fold or more via the said frequency demultiplication and thereby supply the minimum input energy required to balance the chromospheric and coronal losses to radiation and supersonic wind, of $\sim$10$^2$–10$^4$ W·m$^{-2}$ (Withbroe & Noyes, 1977). All these characteristics of how resonance–antiresonance (constructive vs. destructive waves) couplings dominate turbulence help discern the northerly and southerly fast polar as the original ("released as intended") winds, i.e., the most faithful representation of the Sun's internal engine at work. Thus, the polar winds (of which mostly are fast) overall reflect the operational regime of the Sun under which it emits them. Note statistical fidelity favoring the northerly polar wind amongst all characteristic data sets examined, Fig. 5.

The above-discussed moderation of anisotropy by AR is not the only reformatting of global-scale dynamics. As seen in Fig. 3-b, besides signal purity breaking down temporally beyond/above the antiresonance train termination point at $\sim$1-yr periodicity when the global coupling ceases, the Sun vibration does not relax either. It also breaks down spatially beyond L1, leaving RR open to external influences, primarily planetary constellations, and fields (gravitational and magnetic). Such interplay has been suggested previously, e.g., by Kurochkin (1998) and Abreu et al. (2012), who proposed such gravitational and orbital-resonance effects on the Rieger process. Earlier, Pap et al. (1990) offered an intuitive explanation that a transient 154±13-day (Rieger) period was related to an emerging strong magnetic field.

As can be seen from Fig. 3, after decades of debates, $P_{Rg}$ turned out to be a genuine Sun period that experiences modulations in the heliosphere, while most, if not all, of Rieger-type and longer periodicities, as modified externally, have been reported in various types of data in heliophysics; see, e.g., Forgacs-Dajka and Borkovits (2007). The secret to strength, stability, and the very presence of the Rieger period in most solar indices lies in it being globally (as sensed here in both the northerly and southerly polar winds) the tailing ≥99%-significant harmonic of the Sun antiresonance twice (N and S simultaneously). While all those reports were demonstrably correct, Fig. 3 and Table 1, planetary fields can reformat the Rieger harmonics, somewhat disturbing the widely reported values listed in the Introduction, and as seen from the steadiness of ⅕$P_{Rg}$ as the one they do not, Fig. 5–callout. As mentioned, these disturbances primarily come from our Solar system's most dominant gravitational and magnetic fields. Besides, global resonances in closed spherical systems can arise externally and internally. However, in stars with a non-uniform rotation like our Sun, no external triggering is possible, so a continuous spectrum of modes can be expected that can undergo amplification when subjected to a disturbing force of the appropriate frequency and hence lead to enhanced dissipation (Papaloizou & Pringle, 1978). Furthermore, even in the opposite case, the main external factors would be closely related to variations in orbital parameters; however, the Sun's most relevant of those parameters — precession and obliquity

— are related practically entirely to Jupiter's gravitational influence and can be safely ruled out since creating negligible effects which on the Sun amount to <1 mm.

Thus, the primary candidate for the AR triggering mechanism lies in couplings amongst the normal modes of vibration of the Sun's latitudinally varied belts, e.g., Cole (2008), here extracted as ~11-, ~10-, and ~9-yr. Specifically, these varying but mutually superimposed global modes found are 9.9-yr for both MMF (by WSO) and N–S polar fields combined (by Ulysses), 11.2-yr both for the IMF at L1 (by WIND) and the IMF above the northern region (by Ulysses), and 8.8-yr for the southern field (by Ulysses), Fig. 3. The Sun–Heliosphere congruency in periods of the AR lead modes is as expected from a global resonance propagated to L1 and beyond. These excellent matches from datasets significantly disparate in length, density, field strength, and sampled epoch reflect the northerly polar (mostly fast) wind's dominance over the heliosphere as far as ~L1 distances. The matchings also reveal both clarity of the signal and GVSA superiority over the Lomb-Scargle and wavelets spectral techniques as applied by Stenflo and Vogel (1986) and Knaack and Stenflo (2005) for their claimed partial recovery of r- and R-mode fragments of the continuous a-mode global resonance.

The discovery that the solar wind gets emitted in the form of a very low-frequency mechanical resonance (that gets maintained in the form of coherent structures as far as L1 and likely beyond) is good news for the Earth since this find enables more concentrated efforts at modeling the solar wind's impact on Earth. Thus, as found for the solar cycles examined, isolating the northerly polar wind as the physically most vigorous pusher of solar ejecta has important implications for modeling solar-terrestrial interactions and forecasting space weather.

In summary, the present spectral study of the B<30 µT IMF in the polar (mostly fast) solar winds at or near their presumed sources has investigated the winds' global dynamics in the 1-month–13-years band. For this purpose, I separated the wind data into the predominantly equatorial (mostly slow-winds) component, represented in the above band of global resonances by MMF observations from the WSO telescope, and the mostly-fast component, emitted by the polar fields and observed by the Ulysses spacecraft magnetometer and also represented by PF data from the WSO telescope. The polar-wind data were separated further — into the northerly and southerly polar data. Since decadal samplings of the near solar wind spanning around one Hale cycle or two $P_S$= ~11-yr (Schwabe) cycles are now available for the same interplanetary sector, at L1, those magnetometer measurements from the WIND mission also were used to verify the result. The multi-mission comparison of solar wind and its components' sources revealed with a ≥99% confidence that the northern polar region drives the wind at $P_S$, with the slow solar wind emitted or mixed mainly equatorially at the $P_S$ degenerated into a ~10-yr global mode, Fig. 3–a. Due to turbulent and wandering local fields, the southern polar region emits the solar wind along a $P_S$ further degenerated into a ~9-yr global mode and under progressive anisotropy towards lower frequencies so that $P_S$ harmonics recover entirely. The differences between the original (northern) global mode and its equatorial and southern degenerations trigger a perfect (integer-ordered) 3D mechanical resonance in the wind waving at least up to degree m=131. Thus, the ~11-yr Schwabe cycle is the guide period among all known magnetism-related solar periods. At the same time, longer periodicities reflect the differentially triggered resonance so that the next longer, ~88-yr Gleissberg, such period is the first superimposition reflection of the 11–10–9-yr coupling. Acting globally as an asymmetrically vibrating (and thereby resonating) magnetic alternator, the Sun fully exerts control over the entire heliosphere domain, i.e., of both solar wind's resonances (near winds) and turbulences (near and far winds). This successful spectral separation of very-low-frequency (down to $P_S$) equatorial (mainly slow) vs. polar (mainly fast) solar winds via extraction of their unique but comparable spectral signatures paves the way for global and differential studies of heliosphere using AR and, generally, for all future modeling of stellar wind and planetary geodynamics as well.

The heliosphere's hypothetical magnetic structure, believed by some to be composed of random flux tubes, is superseded by a centrally and virtually completely guided mechanical resonance of the alternating fast-slow solar wind blanketing and flapping resonantly quasiperiodically (locally transiently) about the ecliptic.

As found by Stenflo and Vogel (1986) in their partial recovery of AR from Sun (remote) data, the global magnetic field has a patterned (modal) structure characterized by sharp global resonances decoupled from each other for the modes of odd and even parity, thus indicating the existence of an underlying selection rule. While failing to extract the complete AR information, including antiresonances and the Rieger resonance, those authors concluded that the new emission-line spectrum of solar magnetic fields is, as mentioned above, not explicable within the framework of current concepts like dynamo theory. They also concluded that the new spectrum should contain potentially powerful diagnostic information on the interior magnetic structure of the Sun and the origins of solar activity. Those conclusions of theirs are in agreement with what was arrived at here, Figs. 2–4 and Tables 1–3, and what then was also followed by various crosschecks: a computational examination of heliospheric preservation of the Sun's global resonance to at least L1, Fig. 3-b, a computational verification against WSO polar-field data, Fig. 6, and a comparison of the agreement with experiment, Fig. 7. The present study has thus ignored not just simplistic dynamos but more ambitious dynamo models as well that, in their root, have polar field reversals, e.g., Babcock-Leighton dynamos, as those now are all shown based on data as the final judge to be nonsensical attempts at blending physically entirely unrelated and thus incompatible approaches. Instead, the present study humbly re-attempted at the main result as reported previously by (*ibid*.) but is now surprisingly able to report **absolute improvement** after using an approach and methodology previously never used for the same task or in heliophysics in general.

As a star with a possibly differentially rotating core, the Sun as a whole is a globally self-resonating ring system of belts and layers that both contrarily vibrate and differentially rotate, and where then no particular layer or field (toroidal or spheroidal, i.e., poloidal), is singly responsible for the a-mode global resonance. That such a mechanism is at play and responsible for the solar wind creation also follows from the recent find by Omerbashich (2023b) that the solar wind causes seismicity on solid bodies but independently of solar activity, which indicates

71

that the Sun indeed releases the wind coherently instead of randomly. This mechanism, which produces not strictly Alfvén resonance but involves both the Alfvén waves and the shear mode, is easily extended to the entire normal-stars class to which our Sun belongs. On the other hand, based on theoretical modeling, the more massive stars beyond the Sun class were previously speculated to release their stellar wind due to radiation pressure on the stellar atmospheric dust, e.g., by Mattsson et al. (2010). However, disentangling the various feedback mechanisms in massive stars, including stellar wind, only becomes possible with observations spanning a significant range of environments, as this allows probing dependences on metallicity, size, and stellar and dust contents (McLeod et al., 2019). Based on the results of the present study, extracted in the global dynamics (highest) energy ranges, the polar (mainly fast) solar wind indeed appears to be the 'main' or 'normal' solar wind in the solar cycles examined, with the northerly polar wind reflecting the Sun's decadal global dynamics the best.

Separating the polar winds data further still — into strictly fast vs. slow polar winds — resulted in the virtually theoretically (perfectly) recovered AR (of the ideal Sun, i.e., slow winds absent) from the fast polar winds alone. This final separation has confirmed conclusively the fast wind and its highly organized global vibrational structure as a genuine image of the Sun's internal engine at work at all times. This result was then corroborated based on the solar activity (sunspot and calcium) data, whose symmetric spectra in the band of interest indeed were found to correctly reflect the operation of the Sun as a global rotating alternator known from mechanical engineering.

Furthermore, basic knowledge in the vibrational analysis of rotating machinery helped discern, based on vibrational modes (extracted from historical solar activity records) and trends and shapes of spectral envelopes of the fast-wind spectra (from the Ulysses in situ polar-wind data), that the Sun contains a rigid or virtually solid core. The core appears offset away from the apex as the Sun orbits the Galactic Center and tags along the core as its carry-on internal engine that globally is not just an oversized nuclear reactor, as believed previously and simplistically, but also a misaligned mechanical rotator with its macroscopic dynamics and the Sun's counter-physics. The eccentricity thus results in the core wobble with a ~2-yr return period, as well as in AR. AR gets complicated further by dynamics of the Sun's depth-stratified layers, but also by antiresonances from differential motions and chemical-physical properties of the polar regions vs. equatorial belt. Consequently, AR overwhelms the Sun and its antiresonances at each equilibrium vibration every ~11 yr, resulting in the mechanically-resonantly induced mechanical flipping of the solar core and thus inverting the global magnetic polarity resonantly (quasi-regularly).

Namely, as it turns out, our star changes its polarity as any ordinary magnetic alternator — by mechanically flipping its core (rotor) once every ~11 years. Here, the solar core appears offset southwards as it lags behind the Sun on its journey through the Milky Way galaxy towards the apex in the direction of star Vega (northern constellation Lyra, $\delta \approx 40°$ $\alpha \approx 19h$). This tugging motion causes natural preferentiality for the northern hemisphere, in which magnetism behaves more regularly than in the southern, in which the core's relative proximity causes significantly more interferences. (In fact, the rest of the Sun creates the interference — by constantly attempting to reduce the wobbling core's inherent tendency to throw the whole Sun into destructive vibration — resulting globally in both damping and turbulence.)

Specifically, with each completed wobble, the solar core forces an AR train so that the whole Sun, as a closed spherical dynamical system, magnetically gives in, resulting in an equilibrium damping state of its magnetic variations every ~11 yr. At first, the Sun attempts to thermally compensate for the (dynamically) core-induced mechanical global resonance and thus restructure the star, i.e., self-adjust the global dynamics magnetically, which we observe as the solar maxima and minima. Then, after all the thermal compensation attempts to restore the equilibrium always fail as the subsequent successive wobbles commence regularly, the Sun finally gives in dynamically, and the core itself flips, thus changing the global magnetic polarity as a non-exclusive property of the solar core. The same as one sunspot cycle can only end after one whole wobble (accompanied by electromechanical traction and the resulting sparking we observe as nanoflares, CMEs, and sunspots), one solar cycle can only end after an integer number of core wobbles has completed since the last flip until the vibration equilibrium ~11 yr since the previous vibration equilibrium. Therefore, the here extracted value for the core wobble, of 2.2 yr, indeed agrees well with the absolute value of the $P_S \in [9$ yr, 13 yr] Schwabe (1844) range of sunspot counts of $|\pm 2|$ yr. Importantly, I also detected the same recurrence rate from a historical record of group sunspot numbers, Fig. 12. These cross-agreements give great credibility not only to the value of the core wobble period herein extracted, Fig. 12 (and then verified, see Fig. 13), but also to the core wobble–AR mechanism as the forcer of mechanical reversals of the global magnetic polarity via flipping the core mechanically.

A southerly offset solar core was already indicated with the finding early on in the present study: on gradual damping of the global vibration in its highest energies, from ~11-yr northwardly to ~10-yr equatorially to ~9-yr southwardly, Fig. 3. Furthermore, we know from the example of the much explored and indirectly observed inner core of the Earth that, while direct detection of core eigenmodes is very unlikely (Triana et al., 2022), the discovery and identification of even a few undertone periods provide valuable gross constraints on the core stability profile and thermal regime (Crossley & Rochester, 1980). By analogy, the extraordinary (conclusive) success of the present global study of the Sun corroborates and justifies the initial physical hypothesis: that the fast winds are the original winds that represent the physics of the Sun better than the slow winds alone or mixed winds ever could. The solar core offset towards the south causes great disturbance to the southern hemisphere and the equatorial belt, causing interference with global decadal vibration. Namely, the interference then propagates down to the shortest scales and thus translates into the overall highest level of turbulence anywhere on the Sun. The southern polar region succumbs to turbulence more than any other region as the relatively nearby solar core wobbles and thus triggers AR. The core makes the magnetic field lines push each other apart, acting globally destructively, and in the southern hemisphere as its preferred domicile — the most vigorously so.

In short, galactic-orbital dynamics behind the magneto-alternator Sun are simple: as the Sun orbits about the Galactic Center, the core gets dragged by a thrice as massive gaseous

72

stellar envelope and thereby lags somewhat, thus creating the offset that causes the core to wobble. In turn, the wobble causes the Sun to resonate globally, and the resulting global Alfven resonance and antiresonance cause the Sun to (differentially) rotate and thus release its excess mass in the form of the (fast) solar winds. This constant interplay between galactic-orbital and inner dynamics of the Sun causes sparking seen in most unshielded rotating machinery, thus giving rise to the impulsive surface sparking, seen as occasional coronal mass ejections (CME), and, by extension, incessant nanoflares (Parker, 1988), as well as to the standing interior sparking that manifests itself on the surface as sunspots.

The above first-ever conclusive detection of the solar core and the newly gained comprehensive understanding of its dynamics and the resulting global vibration of the Sun for the first time paint a complete picture of our star's macroscopic dynamics. These results help explain some standing problems of global heliophysics, such as why there are more emissions tangentially to the photosphere than away from the Sun (Miller & Ramaty, 1989), why flares with emissions >10 MeV are visible only near the solar limb (Rieger, 1989), and why magnetic polarity reversals are preceded by the moving of sunspot regions towards the equator as a solar cycle progresses.

7. CONCLUSIONS

The inner workings of the whole Sun — here revealed acting as a classical revolving-field magnetoalternator and not dynamo or a proverbial engine, Figs. 2–4 vs. Fig. 7 — were extracted for the first time entirely and to an unprecedented (absolute) accuracy and precision from the Ulysses mission scans of the interplanetary magnetic field above the northern and, separately, southern polar regions of the Sun. This virtually complete extraction rationalizes previous remarks by others according to whom, should such a total extraction ever become possible, it would mean that the solar wind originates at highly coherent, discrete wave modes in the Sun, which get released and then effortlessly transported to distances beyond L1. This remarkable result confirms that sources of the slow solar wind, including open field lines from the quiet region, can be ignored from global considerations, so to fundamentally understand the Sun, only global decadal scales matter — quite like with solar-type stars in general.

The rigorous Gauss–Vaníček spectral analysis (GVSA) method, based on strict least-squares Fourier-type-fitting to sinusoids and physical and statistical criteria for the significance of spectral contents, was used to extract the complete and incessant global decadal vibration (resonance and antiresonance) of the Sun as imprinted in the polar solar winds. With its unique statistical-physical rigor, this method has thus proven itself apt for dynamics problems in natural sciences and by far superior to any Fourier, approximate least-squares fitting like Lomb-Scargle or wavelets techniques used in the past to claim (partial) extractions of Sun global resonances. Particularly, GVSA revolutionizes physical sciences by enabling direct computations of nonlinear global dynamics, rendering spherical approximation obsolete. It can also cut costs in planetary and space sciences by using segments of a data record to describe traversals of a probe perfectly as though achieved constantly without interruption (from complete orbits), thus rigorously simulating multiple simultaneously operating spacecraft of identical performance or even fleet formations from the operation of a single spacecraft.

Specifically, spectral analysis of separated northerly and southerly polar (fast) solar wind, both taken to be the original solar wind where the former recreates global decadal vibrations more faithfully, has thus revealed the signature of a Sun as a real magnetic alternator engine that naturally both vibrates and resonates, resembling the theoretical concept of Alfvén resonance. The data further showed that, as with any mechanical resonance, the Sun's global resonance (here termed AR to pay homage to Alfvén) is far more complex than just the Alfvén waves. Thus, AR gets preceded by antiresonances and does not lend itself to theoretical simplifications, so I invoked empirical considerations showing that the result also agrees with disparate (including remote) data and experimental evidence from mechanical engineering. The Sun engine is a globally dynamical multipart system of separately vibrating and rotating conveyor belts, with an added complexity of differentially rotating, contrarily (out-of-phase) vibrating, and vertically stratified layers. Because of this, while working in unison with solar cycles as the common-denominator pulsation driver that periodically emerges from all the mutually struggling possibilities within the frequency space, the Sun then releases the solar wind axially as a result of structural instability and in a shake-off similar to that of a motor engine trying to rid itself of its (fixed) casing while experiencing a global mechanical resonance itself.

Thus, the present study, as a reproducible computation from all and only *in situ* data ever collected above the Sun's poles continuously over several decades, has revealed an entirely ≥99% significant and (both computationally and experimentally) reproducible ideal mode of systematic global decadal stellar dynamics of the entire Sun. Again, the computation also showed that the Sun exhausts the solar wind in an axial shake-off at highly coherent, discrete wave modes generated internally. This finding is remarkable because any study in the energy band of global stellar (closed-physical-system-) dynamics that independently reproduces a previously reported regular dynamic is beyond doubt correct (by definition). Such a confirmation (at the highest system energies) simultaneously overrides, completes, and redefines all considerations at lower energy levels, including those disagreeing with such an independently corroborated and firmly established result. As only global decadal scales matter for a complete understanding of ours and solar-type stars, the recent missions like Parker or SOHO aimed at relatively minute scales mostly were indeed a waste, regardless of how unpopular such a conclusion might be, especially from the point of view of a statistical physicist.

The data also showed that gravitation could couple spontaneously with the well-known ~150–160-day Rieger period, known to dominate planetary geodynamics, resulting in the Rieger resonance (including the Rieger-type periodicities) whose existence, in turn, demonstrates the potential viability of gravitomagnetics on stellar-system scales. The Rieger period arises as the offshoot of the global Alfvén resonance at the antiresonance-termination (couplings-cessation) point. The double-power of this period, stemming from its simultaneous emission from both northern and southern polar regions, gets channeled through the heliosphere unobstructed and forms a planets-modi-

73

fied Rieger resonance — physical waves of flapping solar ejecta pushed outwards and thereby sped up. The Rieger period gets its most common value of ~154 days from being the triple (tri-band) resonance response of the Sun to its three significantly contrary global vibrations (~11, ~10, ~9-yr) around the global mode, i.e., $P_{Rg}=P_S/3/3/3$ to ≤1‰. Here, the heliosphere's hypothetical magnetic structure, imagined by some as composed of turbulently random flux tubes or ropes, is superseded by a guided resonating flux of alternating fast-slow jets blanketing and flapping quasiperiodically (locally transiently) about the ecliptic.

The Sun contains a rigid or virtually solid inner core offset from the apex (towards the south pole), creating the northerly preferentiality in global magnetism and, inversely, heavy turbulence in the south. As expected, the eccentric core also wobbles, with a ~2.2-yr period, thus continuously triggering an incessant global mechanical resonance (AR), i.e., on global decadal scales, maintained then by the whole Sun and emitted equally congruently in the form of solar wind to the heliosphere to L1 and beyond. Every ~11 yr, i.e., each time the Sun's self-sustained vibration attains equilibrium, AR flips the solar core mechanically-resonantly and thus reverts the global magnetic polarity. The conclusion on a naturally lagging, offset, wobbling, (differential) spin-forcing, and global resonance-inducing core is easily extendable to other astrophysical bodies.

While the notion of the whole Sun as a global magneto-alternator (engine) is readily extendable to solar-type and other stars, that of the solar wind as simply the Sun's global-resonant shake-off is probably extendable not only to the stellar wind but, as a generally astrophysical global resonance phenomenon, to planetary atmospheric winds and Earth mantle wind as well. Under such a scenario, thermally differentiated masses get stirred mechanically-resonantly in an attempt of the host body to equilibrate by ridding itself as quickly as possible of the weakest link — the topmost least dense layer alongside the contact interface with the adjacent (~zero-density) layer radially outwards (above). Within its energy budget that includes up to 100s-of-times electromechanically-resonantly magnified energies, this mechanism allows for the solar abundance and million-degree corona global phenomena, which thus turned out to be a pressing direction for future research in global heliophysics. Now completely (across all mesoscales and macroscopic scales) predictable global solar vibration, including the here deciphered Rieger process crucial for the Solar system's planetary dynamics, such as physics-based space weather and seismicity forecasting, pave the way for proper detailed studies of the Sun and heliosphere. The results of the present study also bring a new fundamental understanding of the estimated >100 billion trillions of solar analogs (as the most common star type discarding dwarfs) that so far have mainly remained an enigma themselves, as well as of their stellar systems, host galaxies, and other objects and dynamic phenomena of our observable universe.

## Acknowledgments

I am grateful to the reviewer Academician Jingxiu Wang (National Astronomical Observatories – NAOC, Chinese Academy of Sciences) for meticulously reading the manuscript and for providing thorough corrections and suggestions. Teimuraz Zaqarashvili (Ilia State University, Georgia) has endorsed the arXiv preprint. The scientific software LSSA v.5, based on the only known rigorous method of spectral analysis (by Vaníček, 1969; 1971), was used to compute the spectra and is available at www2.unb.ca/gge/Research/GRL/LSSA/sourceCode.html.

## Availability of Data

WIND data at: https://wind.nasa.gov/data_sources.php#MFI_Data. WSO MMF data at: http://wso.stanford.edu/#MeanField. WSO PF data at: http://wso.stanford.edu/Polar.html. Ulysses data at the UK National Space Science Data Center, Imperial College London: https://spdf.gsfc.nasa.gov/pub/data/ulysses/mag/interplanetary/hour/. The analyzed main ("real Sun") data are in a Supplement at the IEEE *DataPort* repository: https://dx.doi.org/10.21227/dqkp-3040. Historical records of solar-cycle lengths and sunspot/calcium numbers, used to verify the main result, are at the US NOAA Solar Indices database: https://www.ngdc.noaa.gov/stp/solar/solar-indices.html.

74